\documentclass[12pt]{article}

\usepackage{mheck}
\allowdisplaybreaks[4]

\usepackage{float}
\usepackage{wrapfig}
\def\ii{{\,{\rm i}\,}}
\def\e{{\,\rm e}\,}
\def\FOmega{\overline{\underline{\Omega}}}

\institution{LISBOA}{\  Center for Mathematical Analysis, Geometry and Dynamical Systems, \cr
Instituto Superior T\'ecnico, Universidade de Lisboa, \cr
Av. Rovisco Pais, 1049-001 Lisboa, Portugal }
\institution{SCGP}{\ Simons Center for Geometry and Physics, SUNY, Stony Brook, NY, 11794-3636 USA}

\title{Discrete Integrable Systems, \\ \vspace{-1.3cm} Supersymmetric Quantum Mechanics, \\ and Framed BPS States -- I}
\authors{Michele Cirafici\worksat{\LISBOA}\footnote{e-mail: {\tt michelecirafici@gmail.com}} and Michele Del Zotto \worksat{\SCGP}\footnote{e-mail: {\tt mdelzotto@scgp.stonybrook.edu}}}

\abstract{It is possible to understand whether a given BPS spectrum is generated by a relevant deformation of a 4D $\cn=2$ SCFT or of an asymptotically free theory from the periodicity properties of the corresponding quantum monodromy. With the aim of giving a better understanding of the above conjecture, in this paper we revisit the description of framed BPS states of four-dimensional relativistic quantum field theories with eight conserved supercharges in terms of supersymmetric quantum mechanics. We unveil aspects of the deep interrelationship in between the Seiberg-dualities of the latter, the discrete symmetries of the theory in the bulk, and quantum discrete integrable systems.}

\begin{document}

\maketitle

\tableofcontents

\section{Introduction}

In the recent years, enormous progress in the analysis of the BPS properties and spectra of 4D $\cn=2$ quantum field theories has been achieved. There are essentially two approaches for solving 
the BPS spectral problem of a given 4D $\cn=2$ model. One is geometrical, relying upon spectral 
networks \cite{Gaiotto:2009hg,Gaiotto:2012rg,Gaiotto:2012Db,Galakhov:2013oja,Maruyoshi:2013fwa,Galakhov:2014xba,Longhi:2016rjt,Hollands:2016kgm,Longhi:2016wtv,Longhi:2016bte}, the other is algebraic and it is based on BPS quivers, motivated by 
geometric engineering in Type II superstrings \cite{Denef:2002ru,Cecotti:2010fi,Cecotti:2011rv,Cecotti:2011gu,Alim:2011ae,DelZotto:2011an,Alim:2011kw,Xie:2012Dw,Cecotti:2012va,Cecotti:2012sf,Xie:2012jd,Xie:2012gd,Chuang:2013wt,Cecotti:2013lda,Cecotti:2013sza,Cordova:2014oxa,Cordova:2015vma,Cecotti:2015qha,Caorsi:2016ebt}. One of the interesting aspects that emerged from these studies is that there are deep interconnections in between the spectrum of BPS states and the properties of the theory at the origin of the Coulomb branch. In particular, the quantity
\begin{equation}
\ci(q) = (q)^{2r}_{\infty} \,\,\text{Tr }\left( \overrightarrow{\prod_{\gamma \in \text{ BPS}}} \ci_{\tfrac{1}{2}H}(q; \mathsf{Y}_\gamma)\right),
\end{equation}
has been conjectured to agree with the Schur limit of the superconformal index for SCFTs \cite{Iqbal:2012xm,Cordova:2015nma,Cecotti:2015lab,Cordova:2016uwk,Cordova:2017ohl}. Here 
\begin{equation}
(q)_\infty = \prod_{n\geq1}(1-q^n), \qquad \ci_{\tfrac{1}{2}H}(q; z) = \prod_{n\geq0} \left(1+z q^{n+\tfrac{1}{2}}\right)^{-1},
\end{equation}
and the $\mathsf{Y}_\gamma$ are valued in a quantum torus \cite{Cecotti:2010fi}\begin{equation} \mathsf{Y}_\gamma \mathsf{Y}_{\gamma^\prime} \equiv y^{{\langle\gamma,\gamma^\prime\rangle\over2}} \mathsf{Y}_{\gamma+\gamma^\prime} \, ,\end{equation}
and we have identified the fugacity parameter $q$ with the quantum deformation parameter $y$. This motivates the question: given a BPS spectrum at a given point of the Coulomb branch, is it possible to understand whether it corresponds to an SCFT or to an asymptotically free theory? From the above conjecture, it seems that the answer is affirmative: computing $\ci(q)$ and checking that it gives rise to the vacuum character of a 2D chiral algebra \cite{Beem:2013sza} can be interpreted as a signal of superconformal symmetry for the 4D theory. This fact is closely related to another conjecture to which we now turn. Consider the quantum monodromy operator  \cite{Cecotti:2010fi},\footnote{ The operator $M(q) $ defined in \cite{Cecotti:2010fi} is the inverse of the one we are going to use in this paper.}
\begin{equation}
M(q) \equiv \overrightarrow{\prod_{\gamma \in \text{ BPS}}} \ci_{\tfrac{1}{2}H}(q; \mathsf{Y}_\gamma).
\end{equation}
This operator gives rise to an inner automorphism of the quantum torus algebra
\begin{equation}
\mathsf{Y}_\gamma \mapsto \mathsf{Y}_\gamma^\prime \equiv \text{Ad}_{M(q)}\left( \mathsf{Y}_\gamma\right),
\end{equation}
that in the limit $q\to 1$ gives rise to a $Y$-system. More precisely, it can happen that the $M(q)$ admits a $1/2s$ root, a fractional quantum monodromy \cite{Cecotti:2010fi}. In that case $M(q) = N(q)^{2s}$. Notice that every system admits a $1/2$ monodromy corresponding to the action of CPT on the BPS spectrum. Whenever a model admits a BPS chamber with a $\mathbb{Z}_{2s}$ symmetry, it automatically has a $1/2s$ fractional monodromy. Choosing a basis $\gamma_i$ of $\Gamma$, and setting
\begin{equation}
\mathsf{Y}_i \equiv \mathsf{Y}_{\gamma_i}
\end{equation}
we can define a discrete time evolution as follows
\begin{equation}
\mathsf{Y}_i(t\pm1) \equiv  \text{Ad}_{N(q)^{\pm1}} \mathsf{Y}_i(t).
\end{equation}
In the $q\to1$ limit, the $\mathsf{Y}_i$ become just ordinary commutative variables $Y_i$ and the above action reduces to a rational function
\begin{equation}
Y_i(t\pm1) \equiv R^{(\pm)}[ \{Y_i(t)\}].
\end{equation}
This auxiliary dynamical system satisfies the following conjectures:
\begin{enumerate}
\item The $Y$-system above is periodic $\Leftrightarrow$ the theory at the origin of the Coulomb branch is superconformal \cite{Cecotti:2010fi}. Moreover, if the $\cn=2$ theory has $U(1)_R$ charges for Coulomb branch operators of the form
\begin{equation}
{q_1 \over p_1}\,,\,\frac{q_2}{p_2}\,,\,\dots\,,\,\frac{q_r}{p_r}
\end{equation}
where $q_i$ and $p_i$ are coprime, the period of the $Y$-system is given by
\begin{equation}
\ell \equiv \text{ lcm}(p_i) \qquad Y_i(t+ 2 s \ell) = Y_i(t)
\end{equation}
In particular, for Lagrangian theories or for class $\mathcal{S}$ theories with regular punctures, $\ell=1$.
\item The $Y$-system above is non periodic but integrable (it has a number of
independent commuting hamiltonians, that equals the number of independent
$Y$-system variables) $\Leftrightarrow$ the theory at the origin of the Coulomb branch is asymptotically free \cite{Cecotti:2014zga}.
\end{enumerate}
These conjectures were formulated based on a number of explicit computations. We refer our readers to  \cite{Cecotti:2010fi} and to \cite{Cecotti:2014zga} for a review.

\bigskip

Let us address the above conjecture by probing the 4D $\cn=2$ theory of interest with BPS line operators $\mathfrak{L}_\zeta$ 
\cite{Kapustin:2005py,Drukker:2009tz,Gaiotto:2010be}. One can argue for both 1.) and 2.) above just using the Witten effect \cite{Witten:1979ey} on BPS line operators \cite{Henningson:2006hp}. 

\bigskip

More in details, following \cite{Gaiotto:2008cd} we consider the 4D $\cn=2$ theory of interest on a circle $S^1$ of finite radius and we obtain an effective 3d $\cn=4$ theory with a moduli space of vacua $\mathcal{M}_H$. For theories of class $\mathcal{S}$ such a moduli space is identified with the Hitchin moduli space parametrizing harmonic bundles on a curve $\mathcal{C}$ and in particular has the structure of a torus fibration over the four dimensional Coulomb branch $\mathcal{M}_C$. Such moduli space can be given TBA coordinates $X_\gamma(\xi)$ and the $Y$-system above can be considered as a $Y$-system for the TBA equations of the $X_\gamma(\xi)$. Techniques based on spectral networks can be used to employ line defects as probes of the geometry of $\mathcal{M}_H$, as in \cite{Hollands:2013qza,Gabella:2016zxu,Longhi:2016wtv,Xie:2013lca,Gaiotto:2014bza}. Another approach using semiclassical methods is discussed in \cite{Ito:2011ea,Lee:2011ph,Moore:2015szp,Moore:2015qyu,Moore:2014gua,Moore:2014jfa,Brennan:2016znk}. The connection with cluster algebras is further investigated in \cite{Cirafici:2013bha,Williams:2014efa,Allegretti:2015nxa}. For a review see \cite{Okuda:2014fja}.

Now consider wrapping the $S^1$ with a BPS line operator \cite{Gaiotto:2010be}. This gives rise to a chiral operator in 3d that we denote by $\widehat{\mathfrak{L}}_\zeta$. Being a chiral operator it can be given a vev that must be a function of the coordinates on $\mathcal{M}_H$, schematically
\begin{equation}
\langle \widehat{\mathfrak{L}}_\zeta \rangle = F(X_\gamma) = \sum_{\gamma \in \Gamma_L} \FOmega(\gamma ; \widehat{\mathfrak{L}}_\zeta ) X_\gamma(\zeta),
\end{equation}
where the $\FOmega(\gamma ; \widehat{\mathfrak{L}}_\zeta )$ are protected spin characters (PSCs) for framed BPS states\footnote{ Recall that a PSC for framed BPS states is defined as 
\begin{equation}
\FOmega(\gamma ; \mathfrak{L}_\zeta) \equiv \text{Tr}_{\mathcal{H}^{BPS}_{\mathfrak{L}_\zeta}(\gamma)} y^{2 J_3} (-y)^{2 I_3},
\end{equation}
where $\mathcal{H}^{BPS}_{\mathfrak{L}_\zeta}(\gamma)$ is the Hilbert space of framed BPS states of charge $\gamma$ bound to the defect $\mathfrak{L}_\zeta$, $J_3$ is the generator of spin, and $I_3$ is a generator of $SU(2)_R$.}. When it is clear from the context we will omit the dependence of the PSC on the line operator and write simply $\FOmega (\gamma)$. The key for our argument is that while on one hand 
\begin{equation}\label{Ysystemladder}
\langle \widehat{\mathfrak{L}}_{\e^{\pm2\pi i}\zeta} \rangle = \left(\underbrace{\,\,R^{(\pm)}\circ \cdots \circ R^{(\pm)}\,\,}_{2s \text{ times}} \,\,\circ \,\,F\right) (X_\gamma)
\end{equation}
on the other,
\begin{equation}
\mathfrak{L}_{\e^{\pm2\pi i}\zeta} = W^{\mp} \, \mathfrak{L}_{\zeta},
\end{equation}
where $W^{\pm}$ is an operator implementing the Witten effect on dyonic BPS lines. Here it is crucial that the phase $\zeta$ labeling the line defects is not valued on $S^1$ but in a covering of it \cite{Gaiotto:2010be}. From Eqn.\eqref{Ysystemladder} it is easy to reformulate the conjecture 1.) in terms of the structure of such covering: 1.) is equivalent to the requirement that the covering is finite, which is the statement we are going to argue. Also 2.) can be reformulated in terms of BPS line defects: it becomes equivalent to the statement that there are enough Wilson lines, which are BPS lines $\mathfrak{w}$ such that $W^{\pm}\mathfrak{w}_\zeta = \mathfrak{w}_\zeta$.

\bigskip

For Lagrangian theories this can be understood as follows. For theories without defects, the $U(1)_R
$ symmetry of the 4D $\cn=2$ algebra is unbroken at the conformal point, and it gets broken 
by flowing to a generic point along the Coulomb branch to a $\mathbb{Z}_2$ subgroup that 
corresponds to the CPT symmetry. At non-generic points these symmetries can enhance to 
larger cyclic subgroups of $U(1)_R$ commuting with CPT, $\mathbb{Z}_{2s}$. Even if the model is 
asymptotically free, often certain discrete subgroups of the $U(1)_R$ symmetry turn out to be 
still symmetries of the theory (e.g. CPT again), and still survive at special points along the 
Coulomb branch. Indeed, in this case an anomalous $U(1)_R$ rotation of angle $\alpha$ corresponds to a shift in the effective action of 
\begin{equation}
-\frac{2 b\alpha }{32\pi^2} \int F\wedge F,
\end{equation}
where $b$ is the coefficient of the 1-loop beta function. This rotation can be compensated by shifting the theta angle by $\theta_0 = 2 b \, \alpha $. In particular, for theories with integer $b$, this breaks the $U(1)_R$ to a discrete subgroup $\mathbb{Z}_{2b}$ determined by the integrality property of $\tfrac{1}{32\pi^2} \int F\wedge F$, and for $\alpha = 2 \pi (n/2b)$ we are shifting the $\theta$ angle of a multiple of its period. For a $U(1)_R$ rotation of angle $\alpha$, the central charge rotates $Z \to e^{i 2 \alpha} Z$. In particular, a $2 \pi$ rotation of the $Z$-plane, corresponds to a shifting of $\theta_0 = 2 \pi b$. 

Of course these discrete symmetries are broken by coupling the model to a BPS line defect $\mathfrak{L}_\zeta$. Hence they have a non-trivial action on the space of line defects. A $\pm 2\pi$ rotation of the $Z$-plane corresponds to a $\mp2\pi$ rotation of the argument $\zeta$ of the BPS line. Now, for gauge theories with a non-zero beta function, such rotation is a symmetry for the theory in the bulk corresponding to the Witten effect: shifting the theta angle $\theta \to \theta\pm\theta_0$ a dyon of charge $(m_i,e_i)$, where $m_i$ are magnetic charges while $e_i$ are electric ones, is mapped to a dyon of charge $(m_i,e_i^\prime)$, where
\begin{equation}
(m_i,e_i) \xrightarrow{\quad W^{\pm}\quad} (m_i,e_i^\prime) = (m_i,e_i \pm m_i \, \theta_0/2 \pi).
\end{equation}
The same effect takes place for the BPS dyonic lines: the theory $\ct$ coupled to a defect $\mathfrak{L}_\zeta$ is mapped to itself, but now coupled to a different defect $\mathfrak{L}_{e^{\pm 2 \pi i}\zeta} = W^{\mp} \mathfrak{L}_\zeta$. This gives a proof of the conjectures above in the case of Lagrangian theories. Indeed, if $b=0$, there is no Witten effect and this entails the fact that for Lagrangian SCFTs $\ell=1$. If instead $b\neq 0$, we have that the Wilson BPS lines in a representation $\rho$ have null magnetic charges $m_i = 0$ and therefore are such that
\begin{equation}
\mathfrak{w}^\rho_{\e^{\pm2\pi i}\zeta} = \mathfrak{w}^\rho
\end{equation} 
and hence give rise to constants of motion for the Y-system:
\begin{equation}
w^\rho (X_\gamma) \equiv \langle \, \widehat{\mathfrak{w}^\rho}_\zeta \,\rangle = \langle \, \widehat{\mathfrak{w}^\rho}_{\e^{\pm2\pi i}\zeta} \, \rangle = (R^{(\pm)} \circ w^\rho) (X_\gamma)  
\end{equation}
To complete the proof of integrability, one has to show that the $Y$-system depends on a number of independent $Y$-variables which equals the number of Wilson lines in the fundamental representations. Using the results of \cite{Cecotti:2014zga} it is not hard to see that this is the case for Lagrangian gauge theories with known BPS quivers as well as for non-Lagrangian theories.

\bigskip

In this paper we consider the consequences of the above effect on the framed quiver SQM characterizing framed BPS states \cite{Chuang:2013wt,Cordova:2013bza}.\footnote{ See also \cite{Cirafici:2013bha}, where a 
different approach is given for class $\cs[A_1]$ models.} A BPS line operator with $\zeta \in S^1$ extended along the time direction can be interpreted as an infinitely 
heavy BPS particle sitting steady at a point in space. In 
presence of such an external heavy object, the 4D $\cn=2$ system develops extra sectors of BPS 
excitations bound to the defect, the framed BPS states \cite{Gaiotto:2010be}. The quiver SQM 
description we alluded to above is easily obtained from this picture \cite{Chuang:2013wt,Cordova:2013bza}.\footnote{ See \cite{Andriyash:2010qv,Andriyash:2010yf} for similar results in the context of BPS black hole physics.} The 
framed BPS degeneracies of 4D $\cn=2$ theories which have a quiver regime are captured by the representation theory of framed BPS quivers.\footnote{ We address a minor puzzle of \cite{Cordova:2013bza} in Appendix \ref{failarmy}.}  In particular, we find that rotations of the $Z$-plane are mapped to sequences of framed BPS mutations and using this method the argument for showing the validity of the conjectures above becomes extremely simple, combinatorial and graphical.

\bigskip


This paper is organized as follows: in Section \ref{solitabrodazza}, in order to establish the notation we are going to use throughout this paper, we briefly review certain aspects of the BPS quiver theory; in Section \ref{framedda} we revisit the description of framed BPS states by means of quiver SQM; in Section \ref{su2onceandforall} we study the case of the pure SYM $\mathfrak{su}_2$ theory in detail. In Section \ref{Qsystems} we introduce certain discrete integrable systems and discuss the identification of their conserved charges with the fundamental Wilson line defects. Sections \ref{SU3cluster}, \ref{SU4cluster} and \ref{SU5cluster} contain several computations of vevs of line defects using the techniques developed in the first part of the paper. We end with the Conclusions. Two Appendices contain technical results.


%
\section{BPS Quivers and Mutations \emph{ad Usum Delphini}}\label{solitabrodazza}

\subsection{BPS Quiver Quantum Mechanics}

Often the BPS states of a 4D $\cn=2$ theory of rank $r$ in its 
Coulomb phase can be described as boundstates of certain elementary BPS monopoles and dyons, hypermultiplets that can become massless along the Coulomb branch \cite{Seiberg:1994rs,Seiberg:1994aj}. For a given model with such  property, let us denote $\gamma_{1},...,\gamma_{D} \in \Gamma$  the charges of the corresponding elementary BPS particles, taking values in the lattice of charges $\Gamma$. In a Coulomb phase 
a central charge $Z$ for the 4D $\cn=2$ supersymmetry algebra is generated 
dynamically \cite{Witten:1978mh}, which gives a map $Z : p\in \mathcal{P} \mapsto Z(p,-)\in\text{Hom}(\Gamma,\C)$, where $\mathcal{P}$ is the parameter space of the model (the space of all couplings, mass deformations and vevs). In the regions of $\mathcal{P}$ such that the central 
charges $Z(p,\gamma_i)$ became all almost aligned, these BPS bound-states can be characterized in terms of a quiver supersymmetric quantum mechanics with four 
supercharges (SQM) \cite{Denef:2002ru}. We call this a quiver regime. Rotating the $Z$-plane if necessary, we can choose the $Z(\gamma_i)$ to be all almost aligned with the purely imaginary 
axis. From now on let us assume that we have done so. In absence of exotics BPS states \cite{Gaiotto:2010be}, there are no walls of the third kind  \cite{DelZotto:2014bga}, and the validity of the 
quiver SQM description can be extended to the whole upper half $Z$-plane $\text{Im } Z > 0$: let us assume that this is the case for the models at hand.\footnote{ This is indeed the case for all the quiver  theories obtained from SYM theories by adjoint higgsing as was shown in \cite{DelZotto:2014bga} as well as from all theories obtained from toric CY three-folds as would follow from generalizing the argument in \cite{Chuang:2013wt}. Also, we refer the interested readers to the seminar by Thomas T. Dumitrescu, \emph{Current Algebra Constraints on BPS Particles}, \href{http://pirsa.org/displayFlash.php?id=16030131}{PIRSA 16030131} about his result with C. Cordova on the absence of exotic BPS particles, which is an even stronger motivation for our assumption.} Consider a boundstate consisting of $N_i$ copies of each elementary $\gamma_i$ BPS particle, 
where $N_i\geq 0$. This corresponds to a BPS state of charge $\gamma = \sum_i N_i 
\gamma_i$. The quantum mechanics governing such boundstate has gauge group $
\prod_i U(N_i)$. Let us denote by $\xi_i$ the corresponding F-I terms. At a fixed point $p\in \mathcal{P}$, for any given $
\gamma$, the $\xi_i$ are determined from the central charge of the 4D $\cn=2$ 
supersymmetry algebra as follows
\begin{equation}\label{FISQM}
\xi_i \equiv \text{Im}(Z(p\,;\gamma_i)/Z(p\,;\gamma)).
\end{equation}
In addition to the gauge group, whenever the DSZ elecromagnetic pairing $\langle\gamma_i \,,\,\gamma_j \rangle$ is positive, there are $B_{ij}\equiv \langle\gamma_i \,,\,\gamma_j \rangle$ bifundamental fields $X_\alpha$ transforming in the representation $(\widebar{N}_i,N_j)$. This matter content can be conveniently repackaged into a quiver, the BPS quiver of the theory \cite{Cecotti:2010fi,Cecotti:2011rv,Alim:2011kw}, with intersection matrix $B_{ij}$. Whenever a BPS quiver has an oriented loop, $\ell$, one can write down a corresponding (single trace) superpotential term $\cw_\ell = tr(\prod_{\alpha \in \ell}X_\alpha)$. The resulting potential is 
\begin{equation}\label{pototo}
\cw = \sum_\ell c_\ell \cw_\ell
\end{equation} for $c_\ell \in \C$. The boundstate with charge $\gamma$ is stable provided the corresponding SQM has a non-empty moduli space of vacua, $\cm(\gamma,\zeta_i)$, determined by solving the F-term and the D-term equations for the corresponding SQM. The quantum numbers of the BPS state are encoded in the Dolbeault cohomology of $\cm(\gamma,\zeta_i)$. In particular, the protected spin character \cite{Gaiotto:2010be} is given by:\footnote{ The protected spin character $\Omega(\gamma,p;y)$ is defined as  \begin{equation} \Omega(\gamma,p;y) \equiv ( y-1/y)^{-1}  \text{ Tr}_{\mathcal{H}_{BPS,p}(\gamma)} (2 J_3) (-1)^{2J_3} (-y)^{2(J_3 + I_3)} ,\end{equation}where $J_3$ (resp. $I_3$) is an $SU(2)_{spin}$ (resp. $SU(2)_R$) generator and the trace is taken over the single particle BPS sector of charge $\gamma$ at $p\in\mathcal{P}$ \cite{Gaiotto:2010be}.}
\begin{equation}
\Omega(\gamma,p;y) = \sum_{m,n}(-1)^{m-n} y^{2m - d} h^{m,n}(\cm(\gamma,\zeta_i)).
\end{equation}
Assuming the absence of exotics BPS particles the above simplifies to
\begin{equation}
\Omega(\gamma,p;y) = \sum_m y^{2m - d} h^{m,m}(\cm(\gamma,\zeta_i)).
\end{equation}
There are several different methods to compute the protected spin characters $\Omega(\gamma,p;y)$. From SQM, one of the simplest is given by a standard geometric invariant theory argument: the moduli spaces $\cm(\gamma,\zeta_i)$ are obtained trading the D-terms for a complexification of the gauge groups and a stability condition (dictated by the F-I terms), and solving the F-terms $\partial \cw = 0$. In this case, computing  $\cm(\gamma,\zeta_i)$ boils down to a well-posed problem in the representation theory of the jacobian algebra underlying the BPS quiver for the theory. There are other methods to determine the $\Omega(\gamma,p;y)$ from SQM: one is the MPS Coulomb branch formula \cite{Manschot:2010qz,Manschot:2011xc,Sen:2011aa,Manschot:2012rx,Manschot:2013sya,Manschot:2013dua,Manschot:2014fua}, another is a direct evaluation of the corresponding index via a localization formula, which boils this down to a matrix model like path integral \cite{Hori:2014tda,Kim:2015fba} (see also \cite{Cordova:2014oxa,Hwang:2014uwa}). In particular, from the representation theory perspective, it follows a remarkable property of these BPS quiver SQMs: the category of representations of a quiver with potential does not depend on the choice of the coefficients $c_\ell$ in the definition of $\cw$, provided $\cw$ is DWZ-generic \cite{DWZ} (see below for a definition).

\subsection{Quiver Mutations and Discrete Symmetries}

For the sake of CPT symmetry, any BPS spectrum must be symmetric under $\gamma \to - \gamma$. In particular, it is enough to determine the spectrum of stable BPS particles to recover the whole BPS spectrum: if the BPS spectrum contains a given particle with charge $\gamma$, a corresponding antiparticle with charge $-\gamma$ is automatically part of the spectrum as well. 

Any given quiver regime is selecting a preferred basis of elementary BPS particles (denoted $\gamma_i$ above). Moving around the moduli space the $Z_{i} \equiv Z(p\, ; \gamma_i)$ spread over the $Z$-plane, and it can happen that at a given $p\in \mathcal{P}$ one of the $Z_{i}$ crosses the real axis of the $Z$-plane, thus sorting out of the upper half plane. Whenever this occurs, the BPS particles which looked elementary in the original quiver regime can no longer be interpreted as elementary: this region of the parameter space is connected to a different regime, corresponding to a novel set of elementary constituents, with charges $\gamma_i^\prime$. 

These transitions occur at codimension one walls along $\cp$, the walls of the second kind. Notice that transitions of this sort which connects different quiver regimes always occur when the charge $\gamma_i$ that is being ``rotated out'' of the $Z$-plane is the charge of a (elementary) hyper and the charges $\gamma_i^\prime$ of the new regime are charges of (elementary) hypers as well. That is the case because in a quiver regime all charges of BPS particles are expressed as linear combinations of the $\gamma_i$ (or of the $\gamma_i^\prime$) with non-negative integer coefficients, and therefore their image in the $Z$-plane is a convex cone, by linearity of $Z$. 

In particular, in a quiver regime the two boundaries of such a cone are always basis elements, which are elementary by definition. Say that the charge $\gamma_j$ of the original quiver basis exits the $Z$-plane along the negative (resp. positive) real axis, the novel quiver basis of elementary BPS constituents $\gamma_i^\prime$ is determined from the former by a right (resp. left) mutation $\gamma_i^\prime \equiv \mu^{-}_j (\gamma_i)$ (resp. $\gamma_i^\prime \equiv \mu^{+}_j (\gamma_i)$), defined as follows:\footnote{ Here $[x]_+\equiv \text{max}(0,x)$}
\begin{equation} \label{quivermutation}
\gamma_i^\prime \equiv \mu_j^\mp(\gamma_i) \equiv \begin{cases} - \gamma_i & \text{if } i = j \\ \gamma_i + [\pm\langle \gamma_j, \gamma_i \rangle]_+ \gamma_j & \text{otherwise}\end{cases}
\end{equation}
Notice that $\mu^+_j\mu^-_j = \text{id}$ and $\mu^-_j\mu^+_j = \text{id}$. The mutations acts on the corresponding quiver SQMs as Seiberg dualities for the correspoding $U(N_j)$ gauge group, which also transform the SQM superpotential $\cw$ (see e.g.\cite{DWZ}). 

We say that a superpotential is good whenever it allows to integrate out all the quadratic terms generated upon mutation, otherwise it is bad. A given superpotential is DWZ-generic iff it is good for all quivers that can be obtained from a given one by sequences of mutations. Only in this case the representations of the quiver are independent from the choice of $c_\ell$ in Eqn.\eqref{pototo}.

\begin{figure}
\begin{center}
\includegraphics[scale=0.5]{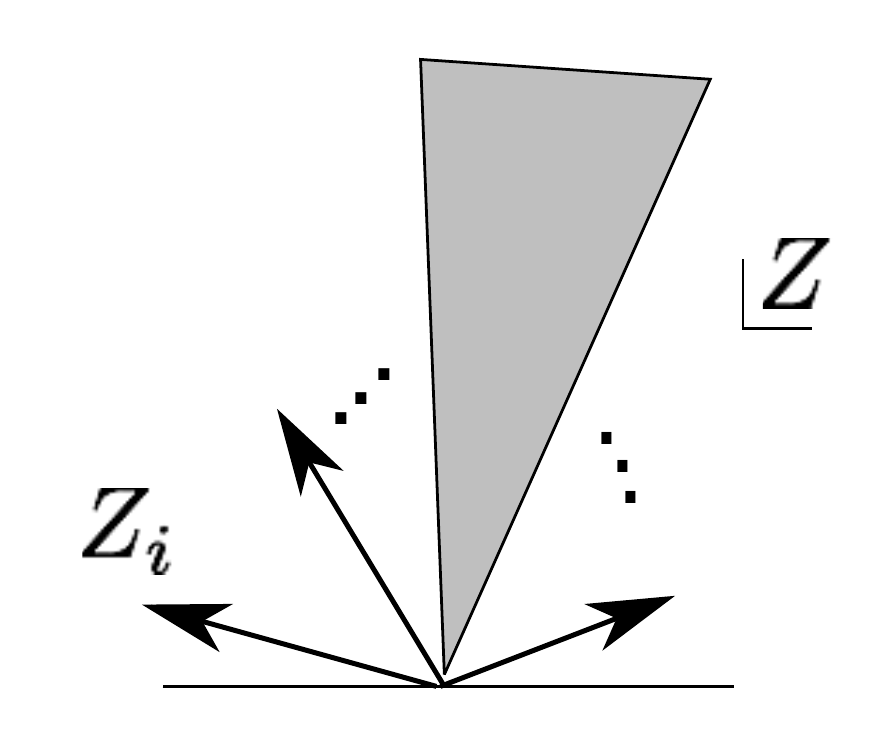}
\end{center}
\caption{Infinite BPS spectrum: generic structure. In the grey shaded region a dense subset with higher spin BPS particles. The dots stand for an accumulating sequence of hypers.}\label{obstruct}
\end{figure}

It is important to stress that not all the walls of the second kind along $\mathcal{P}$ can be described in the way discussed above. That is the case only for the walls corresponding to hypers that are rotated out of the $Z$-plane. Whenever a higher spin particle lies on the boundary of a BPS cone of particles, that obstructs a quiver description in the sense above --- see Figure \ref{obstruct}: the boundaries of the grey shaded region represent BPS particles with spin $\geq 1$, tilting the $Z$-plane the quiver description breaks down when we are hitting the boundary of such cones.\footnote{ This suggest that quivers are not fundamental and another description in terms of triangulated categories is much more adequate. This is beyond the scope of the present note and will be discussed elsewhere.} By contrast, finite BPS spectra of stable hypers correspond to finite mutation sequences 
\begin{equation}\label{halfmonodromy}
\mathbf{m}\equiv \mu_{i_\ell}  \cdots \mu_{i_2} \, \mu_{i_1}
\end{equation}
such that $i_j\neq i_{j+1},$ $\mathbf{m}(Q,\cw) = (Q,\cw)$ and $\mathbf{m}(\gamma_i) = - \gamma_{\pi(i)}$ where $\pi \in \mathsf{S}_D$ is an elementary permutation of the $D$ nodes. This can be seen as follows, if at a point $p\in\mathcal{P}$ one such finite BPS spectrum is found, just tilt the $Z$-plane $180^\circ$ clockwise or counterclockwise to produce the sequence of elementary mutations stated above. 

In particular, if the finite BPS chamber above has a $\mathbb{Z}_{2k}$ discrete symmetry, this can be interpreted at the level of the corresponding mutation sequences as the following fact: there is a shorter mutation sequence 
\begin{equation}\label{fractionalmonodromy}
\mathbf{s} \equiv \mu_{i_s} \cdots \mu_{i_1}
\end{equation} 
and a permutation of the nodes $\sigma$ such that $\mathbf{s}(Q,\cw)=\sigma(Q,\cw)$ and $\mathbf{m} = (\sigma^{-1}\mathbf{s})^k$. If that is the case, $\mathbb{Z}_{2k}$ can be interpreted as a finite discrete subgroup of the $U(1)_R$ symmetry of the $\cn=2$ supersymmetry algebra. Indeed, we have that
\begin{equation}
\{Q_\alpha^i,Q_\beta^j\} = \varepsilon^{ij}\varepsilon_{\alpha\beta} Z,
\end{equation}
and by assigning $R[Q] = 1$, we have that $R[Z]=2$: a $\pi/2$ $U(1)_R$ rotation acts on $Z$ as the CPT map, which corresponds to the mutation sequence $\mathbf{m}$. If a discrete subgroup of $U(1)_R$ is left unbroken by $Z$, then by compatibility with CPT its order must be even, and its action on the $Z$-plane corresponds to the mutation sequence $\sigma^{-1}\mathbf{s}$.

Whenever two mutually non-local BPS states with charges $\gamma$ and $\gamma^\prime$ are such that $Z(\gamma, p)/Z(\gamma^\prime, p) \in \mathbb{R}^+$, the BPS spectrum changes discontinuously, giving rise to a wall-crossing phase transition, characterized by the value of $\langle\gamma, \gamma^\prime\rangle$.\footnote{ The change in the BPS spectrum is controlled by the KS wall-crossing formulas, as have been argued in \cite{Gaiotto:2008cd,Cecotti:2009uf, Gaiotto:2010be} (see also \cite{Dimofte:2009bv,Dimofte:2009tm}).} The loci 
\begin{equation}
\mathrm{WMS}_{\gamma,\gamma^\prime} \equiv \left\{p\in \mathcal{P} \, \colon \, Z(\gamma, p)/Z(\gamma^\prime, p) \in \mathbb{R}^+ \right\} \subset \mathcal{P}
\end{equation}
where wall-crossing phase transitions occur are codimension one in $\mathcal{P}$ and are called walls of the first kind or wall of marginal stability. The space $\mathcal{P}$ splits into BPS chambers, regions bounded by the walls of the first kind corresponding to the pairs of stable mutually non-local BPS excitations in the spectrum.

\subsection{Quantum monodromies and rational transformations} \label{quantummonodromies} \label{rationaltransf}

When a finite BPS chamber is associated to a sequence of mutations $\mathbf{m}$ or to a fractional sequence $\mathbf{s}$, the quantum monodromy $M(q)$ and the fractional quantum monodromy $N(q)$ can be decomposed as a sequence of quantum mutations \cite{keller:2012de}. Quantum mutations are compositions of an ordinary mutation \eqref{quivermutation} with the adjoint action of the function $\mathcal{I}_{\frac12 H} (q ; \mathsf{Y}_\gamma)$.

Denote by $\mathbb{T}_\Gamma$ the quantum torus algebra spanned by the $q$-commutative variables $\mathsf{Y}_\gamma$. The quantum monodromy can be written as an element of $\mathbb{T}_\Gamma$ as follows. Consider a BPS chamber $\mathscr{C}$ with a finite number of BPS states. The operator $M (q)$ defines the inner automorphisms of $\mathbb{T}_\Gamma$
\begin{equation}
\mathsf{Y}_\gamma \longrightarrow  \mathrm{Ad}_ {M (q)^{\pm 1}} \, \mathsf{Y}_\gamma
\end{equation}
and is a wall crossing invariant. The equivalence of different decompositions of $M (q)$ in different chambers is the KS wall-crossing formula \cite{Kontsevich:2008fj}. 

If our chamber $\mathscr{C}$ has a discrete symmetry we can define finer monodromies, precisely as we have done with quiver mutations. Assume that the spectrum within the chamber $\mathscr{C}$ admits a $\mathbb{Z}_{2 s}$ involution. This involution can be understood as the action of a kinematic operator acting on the charges. This operator lifts to $\mathbb{T}_\Gamma$ and allows to define the $1/(2 s)$ fractional quantum monodromy 
\begin{equation} \label{actionN}
\mathsf{Y}_\gamma \longrightarrow  \mathrm{Ad}_ {N (q)^{\pm 1}} \, \mathsf{Y}_\gamma
\end{equation}
which has the property that
\begin{equation}
M (q) = N(q)^{2 s} \, .
\end{equation}
In particular the $1/2$ monodromy is always well defined due to CPT.

In the $q \longrightarrow 1$ limit all the monodromy operators reduce to rational transformations on the function $Y_\gamma$. In particular we can consider rational transformations arising from the limit of the half monodromy or of any fractional monodromy. Consider a sequence of mutations $\mathbf{s} = \mu_{i_s} \cdots \mu_{i_1}$  corresponding to the finest fractional monodromy, within a finite chamber in an $\mathcal{N}=2$ model. Let $\sigma$ be the associated permutation. We denote by $\gamma_i$ the charges of the nodes of the quiver and employ the usual notation $Y_{\gamma_i} = Y_i$. The $q \longrightarrow 1$ limit of \eqref{actionN} defines the rational transformations
\begin{equation} \label{Rpm-transf}
Y_{i}  (t \pm 1) =R^{( \pm ) }_i [ \{ Y_1 , \dots , Y_n \} ]  \, .
\end{equation}
This recursive relation defines a \textit{discrete integrable system} associated to the finite chamber $\mathscr{C}$ \cite{Cecotti:2014zga}.  After an evolution given by one unit of time, the underlying quiver is back to itself, since we are taking into account the permutation $\sigma$, while all the $Y_i$ variables have evolved by a sequence of mutations.

The rational transformation \eqref{Rpm-transf} can be decomposed according to elementary rational transformations, which can be seen as the commutative limit of quantum mutations
\begin{align}
R_j^{ (-)} [ \{ Y_1 , \dots , Y_n \} ] & = \sigma^{-1}  R_{j , i_s}^{(-)} \ R_{j , i_{k-1}}^{(-)} \ \cdots R_{j , i_1}^{(-)} [\{ Y_1 , \cdots , Y_n \}] \, , \cr
R_j^{ (+)} [ \{ Y_1 , \dots , Y_n \} ] & =  \sigma^{-1} R_{j , i_s}^{(+)} \ R_{j , i_{k-1}}^{(+)} \ \cdots R_{j , i_1}^{(+)} [\{ Y_1 , \cdots , Y_n \}] \, ,
\end{align}
where by partial abuse of notation we use the same symbol $\sigma$ to denote the action of the permutation on the quiver nodes and on the $Y_\gamma$ functions. The elementary rational transformations are given by cluster transformations \cite{FominZelevinsky,FG1}
\begin{equation} \label{clusterR}
R_{i , k}^{\pm} [\{ Y_1 , \dots , Y_n \} ]=  \left\{ 
\begin{array}{cc}
Y_i^{-1} & \text{if} \ i = k \\
Y_j \, \left( 1 + Y_k^{\mp \mathrm{sign} \langle \gamma_k , \gamma_i \rangle} \right)^{\mp \langle \gamma_k , \gamma_i \rangle} & \text{if} \ i \neq k 
\end{array}
\right. \, .
\end{equation}
These sets of rational transformations endow the moduli space $\mathcal{M}_H$ with the structure of a cluster variety, at least locally.\footnote{ Some of these $Y$-systems and their properties are well known and important in the literature. In particular, we find extremely interesting connections with the works \cite{Faddeev:1999fe,Faddeev:2007uv,Faddeev:2008xy,Faddeev:2012zu,Derkachov:2013cqa,Faddeev:2014qya,Faddeev:2014kra} that we are going to report about in part II of this work. In particular, the evolution operator can be generalized by means of the Faddeev Modular Double leading to a unitary discrete integrable system.}

\section{Framed SQM Revised}\label{framedda}

\subsection{4D Defect Groups}

As we have reviewed in the Introduction, a 4D $\cn=2$ theory in presence of a BPS line defect 
$L$ develops a novel subsector of BPS excitations, describing BPS states which are bound to 
the defect, the framed BPS states. For a 4D $\cn=2$ theory of rank $r$ in its Coulomb phase, 
a BPS line defect can be described with a mild extension of the ideas outlined in Section 
\ref{solitabrodazza} \cite{Cordova:2013bza}. Fix a point $p \in \mathcal{P}$. As we reviewed in the Introduction, any BPS line 
defect with $\zeta \in U(1)$ can be thought of as an extra elementary BPS particle which is very heavy, sitting at the 
origin of space. One of the differences among ordinary BPS particles and line defects is that the charges of line defects are valued in the dual of the lattice of charges
\begin{equation}
\Gamma^* \equiv \{\alpha \in \mathbb{R}^D \colon \langle \alpha, \gamma\rangle \in \mathbb{Z} \quad \forall \, \gamma \, \in \Gamma\}.
\end{equation}
In particular, a framed BPS state can be modeled as bound states among a heavy BPS particle with charge $\alpha$ valued in $\Gamma^*$ and the various BPS 
particles in the bulk, with charges valued in $\Gamma$. Notice that the charges of the framed BPS states are always valued in a torsor of $\Gamma$ of the form $\alpha + \Gamma \subseteq \Gamma^*$. The charge $\alpha \in \Gamma^*$ is the core charge of the defect, which can jump discontinuously across walls that are analogue of the walls of the second kind (the GMN anti-walls), see Section \ref{corazonespinado} for a review in the context of quiver SQM. Consistency with the Dirac quantization condition requires that for a given model, only core charges such that
\begin{equation}\label{diracco}
\langle \a , \a^\prime \rangle \in \mathbb{Z}
\end{equation}
are allowed. Maximal sub-lattices of $\Gamma^*$ that satisfy \eqref{diracco} characterize different models which have the same content of local operators, but are distinguished by the respective spectra of non-local operators. Example of models of this kind are e.g. the pure SYM theories with gauge groups $SU(N)/\mathbb{Z}_N$ and $SU(N)$ (see \cite{Aharony:2013hda} for a detailed discussion). To a given lattice $\Gamma$ with DSZ pairing $\langle - , - \rangle$ therefore correspond at least one maximal local sublattice 
\begin{equation}
\Gamma \subset \Gamma_L \subseteq \Gamma^*,
\end{equation}
but there can be more $\Gamma_{L^{'}}$, $\Gamma_{L^{''}}$ such that while $\langle \Gamma_L,\Gamma_L \rangle \in \mathbb{Z}$, in general $\langle \Gamma_L,\Gamma_{L'} \rangle \in \mathbb{Q}$. It is interesting to consider the group $\Lambda_L \equiv \Gamma_L/\Gamma$, consisting of those defect charges that cannot be screened by charges of particles in the bulk.\footnote{ This discussion suggests a possible refinement of the arguments of \cite{DelZotto:2015isa} in the context of 6D theories, which, of course, goes beyond the scope of this note.} It is clear that whenever two core charges have the same class in $\Lambda_L$, they are valued in a torsor of $\Gamma$ of the form $\a + \Gamma$, where $\alpha \in \Gamma^* \setminus \Gamma$. While the core charge $\alpha$ undergoes (anti-)wall-crossings, its class $[\alpha]\in \Lambda_L$ does not jump, it is an (anti-)wall-crossing invariant and, as such, a property of the model that is independent of the boundary conditions (a UV property). It is tempting to identify it with an IR analogue of the 't Hooft flux of the corresponding BPS line defect.

\subsection{Framed BPS quiver quantum mechanics}

\begin{figure}
\begin{center}
\includegraphics[scale=0.9]{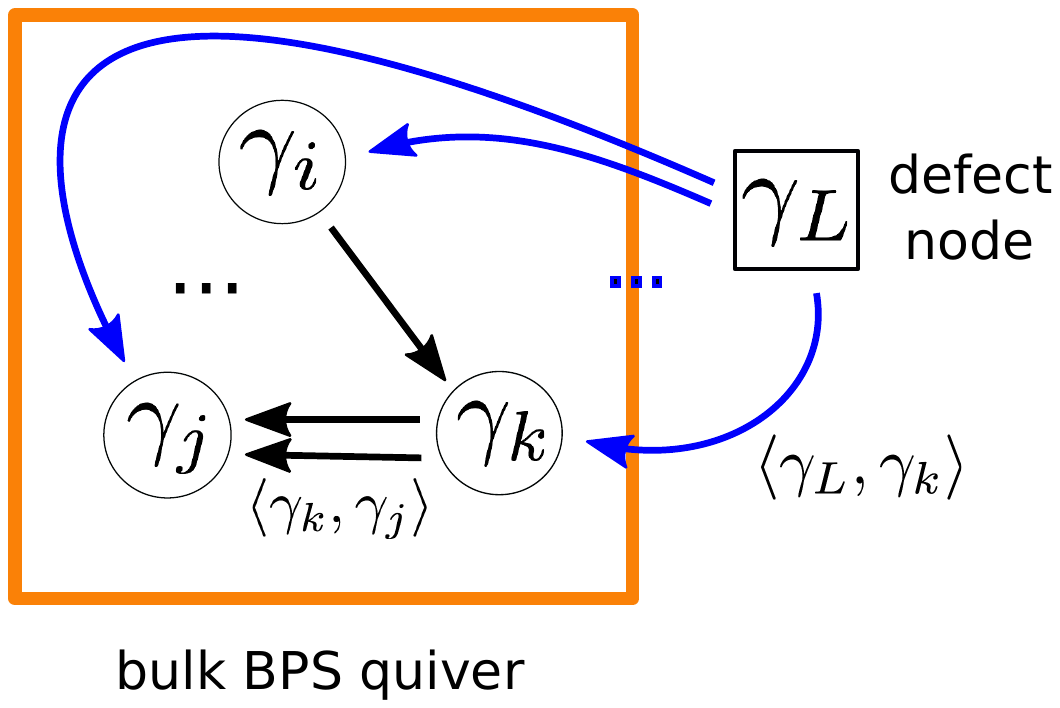}
\end{center}
\caption{A framed BPS quiver.}\label{maserve?}
\end{figure}

To obtain the SQM describing the framed 
BPS states, the SQM governing the BPS states for the theory in the bulk can be modified by 
adding an extra elementary constituent with charge
\begin{equation}
\gamma_L \equiv \alpha + \gamma_F
\end{equation}
where $\alpha \in \Gamma^*$ is the core charge of the BPS line defect, while $\gamma_F$ is an additional flavor charge
\begin{equation}
\langle \gamma_F, \gamma_i\rangle = 0 \qquad \forall \, i = 1, ..., D.
\end{equation}
extending the charge 
lattice with the addition of an extra flavor charge $\gamma_F$. In particular, notice that $\alpha$ has an expansion 
\begin{equation}
\alpha = \sum_i \alpha_i \gamma_i \qquad \alpha_i \in \mathbb{Q},
\end{equation}
where the coefficients $\alpha_i$ can be rational numbers as long as $\alpha\in\Gamma^*$. The central charge can be extended by linearity setting
\begin{equation}
Z(\gamma_F) \equiv m \zeta \quad\text{for}\quad \zeta \in U(1).
\end{equation}
The fact that the $\gamma_L$ particle is very heavy is modeled by studying the regime 
\begin{equation}
m \ggg |Z_i| \qquad \forall \, i = 1, ..., D.
\end{equation}
Now, if the bulk theory admits a quiver basis, the same framework we described in Section \ref{solitabrodazza} can be applied to characterize the framed BPS states. We view the framed BPS states as bound states among some elementary BPS particles in the bulk with charge $\gamma_i\in\Gamma$ and the heavy particle with charge $\gamma_L$  which 
models the defect. Notice that the charge
\begin{equation}
\gamma = \sum_i N_i \gamma_i \in \Gamma
\end{equation}
can be viewed as the charge of a bulk BPS particle. Of course, we always obtain an SQM with gauge group
\begin{equation}
U(1)_L \times \prod_i U(N_i),
\end{equation} 
where the $N_i$ are supported on the nodes of the bulk BPS quiver, and $U(1)_L$ corresponds to an extra node. The latter group decouples in the strict infinite mass limit. The matter content of the SQM is determined again by the DSZ pairing: if  $\langle\gamma_L, \gamma_i\rangle$  is  positive (resp. negative) it gives rise to a fundamental (resp. antifundamental) of $U(N_i)$ with positive (resp. negative) $U(1)_L$ charge --- see Figure \ref{maserve?}. The F-I terms for such SQM are determined by \eqref{FISQM}. In particular \cite{Gaiotto:2010be}, 
\begin{equation}
M(\gamma_L + \gamma) \geq |Z(\alpha+\gamma) + m \zeta| \sim m + \text{Re}(Z(\alpha+\gamma)/\zeta) + O(1/m),
\end{equation}
where the charge $\widehat{\gamma} =  \alpha + \gamma$, renormalized by subtracting $m$, gives the framed BPS bound $M(\widehat{\gamma}) \geq \text{Re}(Z(\widehat{\gamma})/\zeta)$ in the limit $m\to \infty$.\footnote{ Our conventions 
differs from those of \cite{Gaiotto:2010be,Moore:2015szp} as far as the particle/antiparticle 
splitting is concerned.} For our purposes, it is more natural 
to keep $m$ very large but finite, and to study the problem from the SQM perspective.

\subsection{Framed BPS Quivers and Mutations}\label{corazonespinado}

\begin{figure}
\begin{center}
\includegraphics[scale=1.5]{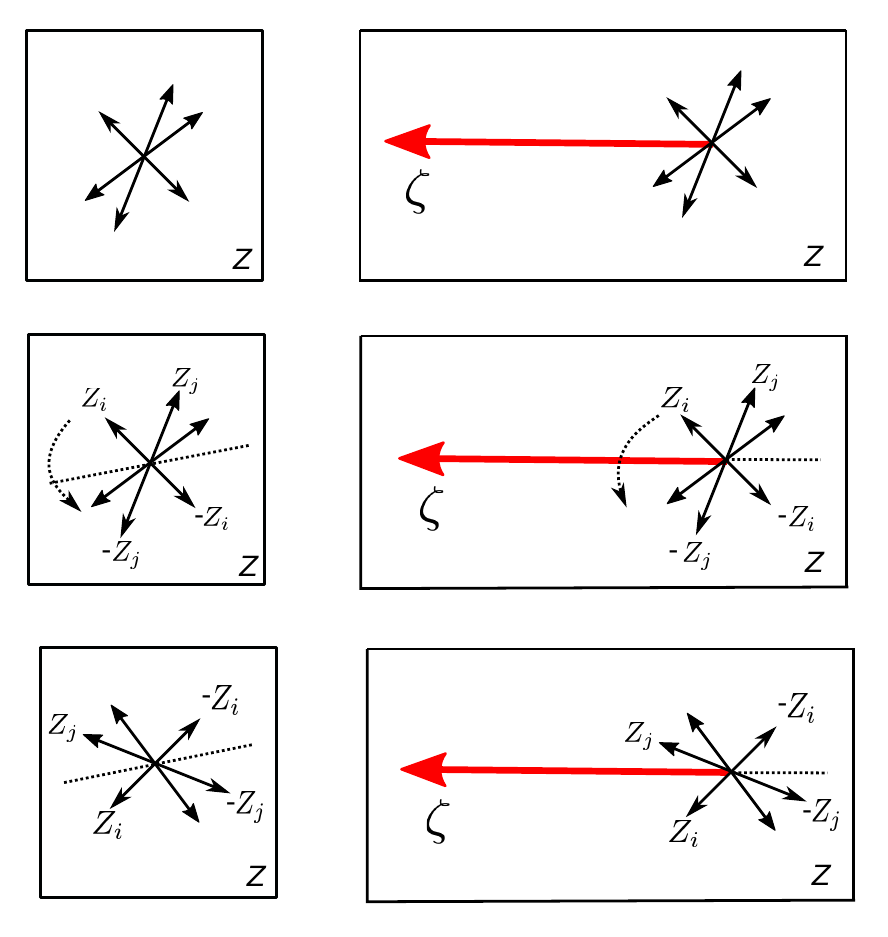}\\
\end{center}
\caption{Quiver mutation VS framed quiver mutation. \textsc{up left:} mapping a BPS spectrum to the $Z$-plane.   \textsc{up right:} BPS spectrum in the $Z$-plane in presence of a BPS line defect with phase $\zeta$. \textsc{mid left:} choice of a quiver basis, splitting among particles and anti-particles. \textsc{mid right:} our choice of conventions: anti-walls and walls of the second kind coincide. \textsc{down left:} rotating the $Z$-plane or equivalently changing convention about the particle/antiparticle splitting thus leading to a quiver mutation. \textsc{down right:} framed BPS quiver mutation: the $Z_i$ rotate, while $\zeta$ is held fixed. Notice that this implies also a framed BPS wall-crossing.}\label{notyet}
\end{figure}

In the construction of the framed BPS quiver SQM we are still free to choose a quiver basis of particles (see Figure \ref{notyet}).\footnote{ Such a choice was labeled right adapted in \cite{Cordova:2013bza}.} The behavior under mutations of a framed BPS quiver is rather different with respect to that of an ordinary BPS quiver. This can be understood from the following remark.\footnote{ See Section 12 of \cite{Gaiotto:2010be}. Notice that we are working with a different set of conventions: in the language of that paper, it is natural for us to choose a triangulation of the curve for which the anti-walls coincide with the walls of the second kind.\label{conventia}} Whenever one of the $Z_i$ aligns with $\zeta$ a boundstate among the defect and the BPS particle of 
charge $\gamma_i$ can be formed if $\langle \gamma_i, \gamma_L\rangle \neq 0$. This has 
the effect of screening the charge $\gamma_L$, by shifting the core charge $\alpha$. The precise change of the core charge is encoded in a mutation
\begin{equation}\label{FRAMUTAH}
\alpha^\prime \equiv \mu_i^-(\alpha) \equiv \alpha + [\langle \gamma_i, \alpha \rangle]_+ \gamma_i.
\end{equation}
where $\gamma_i$ is the charge of the BPS hyper such that (anti-wall)
\begin{equation}
Z(\gamma_i)/\zeta \in \mathbb{R}^+.
\end{equation}
The arrows connecting the defect node $\gamma_L$ to the rest of the bulk BPS quiver are completely determined by the core charge
\begin{equation}
\langle\gamma_i,\gamma_L\rangle \equiv \langle\gamma_i,\alpha\rangle,
\end{equation}
therefore changing it via a mutation of the type in Eqn.\eqref{FRAMUTAH} has the effect of changing the framed BPS quiver. One of the main features of our choice of conventions is that whenever this occurs we are simultaneously changing the spectrum of framed BPS states. The walls of the second kind for a framed BPS quiver encode not only loci where a mutation occurs for the bulk BPS quiver, they are simultaneously walls of marginal stability for the framed BPS particles and anti-walls for the core charges. In particular, the jump in the framed BPS spectrum is accounted for by the fact that whenever a mutation for the bulk BPS quiver occurs, the stability condition governing the structure of the moduli spaces for the framed BPS SQMs changes as we are holding $\zeta$ fixed while the $Z_i$ are rotating. The corresponding framed BPS quiver mutates following the ordinary rules we have reviewed in Section \ref{solitabrodazza}. There is however a main difference: the framing node $\gamma_L$ never mutates. This can be understood from the fact that while the central charges $Z(p,\cdot)$ does change as we vary $p\in \mathcal{P}$, the phase $\zeta$ corresponding to the node $\gamma_L$ is held fixed as it is part of the definition of the defect we are considering, hence it never exits the upper half plane.\footnote{ Of course one can still use the mutation method to determine the framed BPS states, as was done e.g. in \cite{Cordova:2013bza,Cordova:2016uwk}.} In the language of Cluster Algebras, the node corresponding to the BPS defect is frozen. This corresponds to a framing of the $Z$-plane. Considering for simplicity a model with a finite BPS chamber, we compare the two cases in Figure \ref{notyet}.

\begin{figure}
\begin{center}
\includegraphics[scale=1.5]{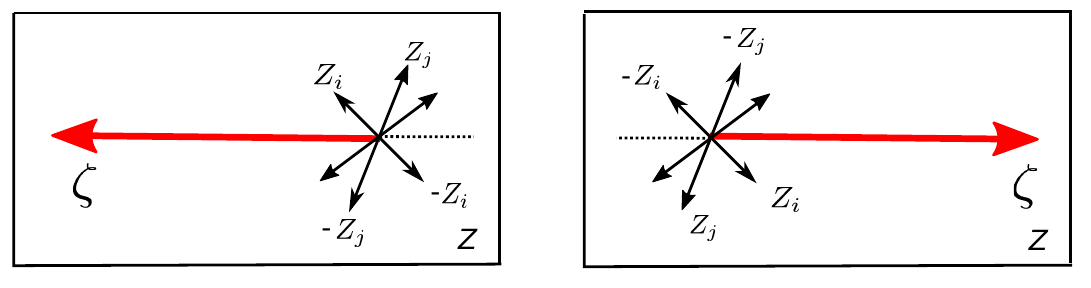}
\end{center}
\caption{Two possible choices of conventions for framed BPS quivers such that walls of the second kind, anti-walls and walls of marginal stability for framed BPS states coincide. Notice that they are CPT conjugates.}\label{theotherone}
\end{figure}

Let us notice that there is another possible choice of conventions leading to the same mechanisms we illustrated above, it is the CPT transform of ours (see Figure \ref{theotherone}), obtained by choosing a quiver basis with charges $\gamma_i^\prime = - \gamma_i$. If one use this other basis, the mutation corresponding to an anti-wall is given by
\begin{equation}\label{FRAMUTAHB}
\alpha^\prime \equiv \mu_i^+(\alpha) \equiv \alpha + [\langle \alpha,\gamma_i \rangle]_+ \gamma_i.
\end{equation}
These two choices were called respectively right adapted or left adapted in \cite{Cordova:2013bza}. 
These two choices have a mathematical description in terms of the representation of the jacobian algebra of the underlying quiver: they correspond to co-cyclic or cyclic stability conditions. Imposing cyclic stability conditions amounts to consider only cyclic modules over the jacobian algebra, that is modules that are generated by a vector. 

From the above discussion, a key difference in between a quiver SQM and a framed quiver SQM emerges: while mutations are symmetries for the modeling of BPS states in the absence of BPS line defects, this is no longer the case in their presence. This can be understood from the idea that rotations of the $Z$-plane correspond on one hand to mutations while on the other to $U(1)_R$ rotations: in presence of a BPS line defect, no non-trivial subgroup of the $U(1)_R$ survives by definition. The SQM description accounts for this fact by combining two physical facts. The first is that the core charge mutates as we have reviewed above, the second is that the spectrum of framed BPS states jumps across walls of marginal stability. From our perspective this is essentially analogous to the ordinary wall-crossing as captured e.g. from quiver representation theory, however again there is a difference here because the change in the stability condition for the framed BPS states coincides with a wall of the second kind and not of the first kind. Let us proceed by explaining this fact in more detail below.

\subsection{Framed BPS chambers and mutations of framed BPS quivers}

\begin{figure}
\begin{center}
\includegraphics[scale=1.5]{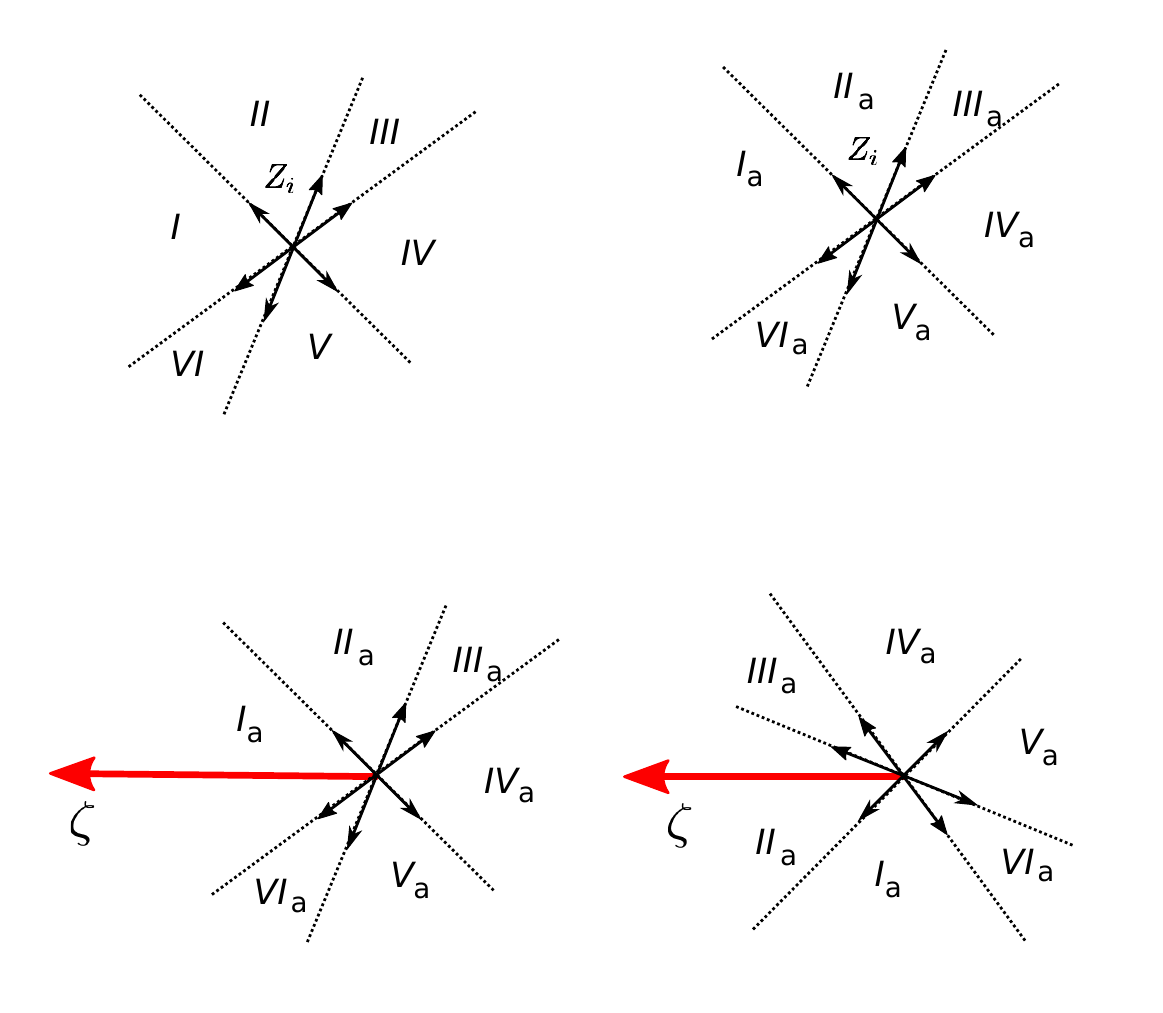}
\end{center}
\caption{Structure of framed BPS chambers in the $Z$-plane. \textsc{up left:} structure of chambers for an SCFT with integer valued $R$-charges. \textsc{up right:} generic structure of chambers: for SCFTs the index $a$ is defined modulo $\ell$, whenever the $R$-charges are valued in $\mathbb{N}/\ell$, for AF theories $a$ runs over the integers. \textsc{down left/right:} chambers related by crossing a wall of the second kind, corresponding to rotating out the charge $\gamma_{(1)}$ on the $a$-th sheet of the covering.}\label{squallors}
\end{figure}

Consider a theory $\ct$ with parameter space $\mathcal{P}_\ct$. Given a point $p\in \mathcal{P}_\ct$, consider the BPS chamber at $p$:
\begin{equation}
\mathfrak{S}_p \equiv \{\gamma \in \Gamma \colon \mathcal{H}^{BPS}_p(\gamma) \text{ is not empty}\} \subset \Gamma.
\end{equation}
Notice that $\mathfrak{S}_p$ can be given an ordering mod $2 \pi$ from the ordering inherited mapping the charges on $S^1$ with $\arg Z(p, -)$.

In this paper we focus on theories such that they admit at least a $p\in \mathcal{P}$ such that $\mathfrak{S}_p$ is finite and $\Omega(\gamma)=1$ for all $\gamma \in \mathfrak{S}_p$. These are models that have a finite chamber of BPS hypermultiplets. Let us denote one such finite chamber as $N_\mathfrak{S}$ where 
\begin{equation}
N \equiv \# (\mathfrak{S}_p) /2.
\end{equation}
In this case, let us label the charges of the BPS states $\gamma_{(1)},...,\gamma_{(2N)}$ in decreasing order of $\arg Z(p, \gamma_{(I)})$ chosen from $\pi$ to $-\pi$, notice that $\gamma_{(N+J)} = - \gamma_{(J)}$. Famous examples of $\cn=2$ theories of this sort are provided e.g. by the pure $G$ SYM theories, which have finite BPS chambers with $N=r_G h_G$.

Now couple the theory $\ct$ to a BPS line defect $\mathfrak{L}_\zeta$. Consider a point $p\in\mathcal{P}_\ct$ such that we have a finite chamber of hypermultiplets with charges $N_\mathfrak{S}$. Corresponding to the $\gamma_{(I)}\in N_\mathfrak{S}$ we have a family of framed BPS chambers at the point $p\in\mathcal{P}_\mathcal{T}$. The stability of the corresponding framed BPS particles is controlled by the position of $\arg \zeta$ relative to the $\arg Z(\gamma_{(I)})$. We refer to the $I$-th framed BPS chamber as the one for which
\begin{equation}
\arg Z(\gamma_{(I)}) < \zeta < \arg Z(\gamma_{(I+1)})
\end{equation}
From this discussion it is evident that for every finite BPS chamber of hypers of the theory in the bulk, there are inequivalent framed BPS chambers in a number that is at least equal to twice the number of BPS states. Indeed, as we shall see below this is the case for superconformal $\cn=2$ theories that have the feature of having purely integer $R$-charges and a finite BPS chamber of hypermultiplets, e.g. $SU(2)$ $N_f=4$. In general, however, this turns out to be too naive. As was remarked already in \cite{Gaiotto:2010be} the parameter $\zeta$ is not valued on $S^1$ but in a covering of the circle. We claim that a necessary condition for a given BPS spectrum to give rise to an SCFT is that such a covering is finite. This result an many other concerning SCFT will be discussed in a forthcoming paper. As we shall see below, this claim can be interpreted as a different version of an old conjecture relating superconformal invariance to the periodicity of the quantum monodromy \cite{Cecotti:2010fi,Cecotti:2011gu,Cecotti:2014zga}. 

For superconformal $\cn=2$ theories that have $R$-charges valued over the rationals, of the form $q_i/p_i$ where $q_i$ and $p_i$ are coprime, let $\ell\equiv\text{lcm}(p_i)$. The naive number of inequivalent framed BPS chambers corresponding to a finite chamber $N_\mathfrak{S}$ in that case is $2N\ell$. We label the corresponding framed BPS chambers with roman numerals from $1$ to $2N$ with an extra index taking values $\text{mod } \ell$. For instance the chamber $III^{(a)}$ denotes the one for which
\begin{equation}
\arg Z(\gamma_{(3)}) < \zeta < \arg Z(\gamma_{(4)})
\end{equation}
and $\zeta$ belong to the $a$-th sheet of a covering of $S^1$ with $\ell$ sheets.

For asymptotically free theories, instead, the covering turns out to have an infinite number of sheets and there are infinitely many inequivalent framed BPS chambers. In such a case we are going to label them in the same way, where $(a)$ takes integer values. 

Inequivalent framed BPS chambers correspond to inequivalent framed BPS quivers, which is related to the fact that we have chosen conventions such that the wall of the second kind for the bulk BPS quiver correspond to the anti-walls and to marginal stability walls for framed BPS states. 

From this perspective, under the assumptions above, the framed BPS chamber $(K-1)^{(a)}$ correspond to a framed BPS quiver such that $\gamma_{(K)}$ and $\gamma_{(K + N - 1)}$ are part of a quiver basis for the finite BPS chamber $N_\mathfrak{S}$. Such chamber is related to the framed BPS quiver for chamber $(K)^{(a)}$ by a mutation corresponding to rotating out $\gamma_{(K)}$ along the negative real axis of the $Z$-plane, this gives right to a mutation $\mu_{\gamma_{(K)}}^-$ for the corresponding bulk quiver, while the core charge shifts from $\alpha$ to $\alpha^\prime$ determined from Eqn.\eqref{FRAMUTAH}:
\begin{equation}
\alpha^\prime = \alpha + [\langle \gamma_{(K)}, \alpha \rangle]_+ \gamma_{(K)}.
\end{equation}
Similarly to shift from the framed BPS chamber $(K-1)^{(a)}$ to the chamber $(K-2)^{(a)}$ one needs to rotate out $\gamma_{(K + N - 1)}$ along the positive real axis of the $Z$-plane. This gives right to a mutation $\mu_{\gamma_{(K+ N - 1)}}^+$ for the corresponding bulk quiver, while the core charge still shifts according to Eqn.\eqref{FRAMUTAH} with the difference that now it is $-\gamma_{(K+ N - 1)}$ that aligns with $\alpha$, therefore
\begin{equation}
\alpha^\prime = \alpha - [\langle \alpha,\gamma_{(K+N-1)} \rangle]_+ \gamma_{(K+N-1)}
\end{equation}
is the framed BPS charge corresponding to the defect in such chamber.

\section{The $\mathfrak{su}_2$ case}\label{su2onceandforall}
\subsection{Example: a 't Hooft BPS line for pure $SU(2)$ SYM}\label{THOFTOH}
To illustrate the above phenomenon it is more convenient to start with an example. Consider the $SU(2)$ SYM theory. The corresponding quiver SQM is given by the Kronecker quiver $\widehat{A}(1,1)$
\begin{equation}
\xymatrix{\gamma_\circ \ar@<-0.5ex>[rr]\ar@<+0.5ex>[rr] &&\gamma_\bullet}
\end{equation}
Choose a 't Hooft defect with core charge $\alpha = \gamma_\bullet$. The corresponding framed BPS quiver is given by
\begin{equation}\label{thofty}
\begin{gathered}
\xymatrix{\gamma_\circ \ar@<-0.5ex>[rr]\ar@<+0.5ex>[rr] \ar@{..>}@<-0.5ex>[dr]\ar@{..>}@<+0.5ex>[dr] && \gamma_\bullet\\
&\framebox{$\alpha$}&}
\end{gathered}
\end{equation}
With our choice of conventions, this quiver gives rise to a single framed BPS state of charge $\alpha$. If we consider the strongly coupled chamber of $SU(2)$, corresponding to the choice of central charges
\begin{equation}
0 < \arg Z_\circ < \arg Z_\bullet < \pi,
\end{equation}
there are only two stable hypermultiplets: one has charge $\gamma_\circ$ and the other has charge $\gamma_\bullet$. Consider varying the central charge in such a way that the $\gamma_\bullet$ particle exits the upper half plane from the negative imaginary axis without crossing any walls of marginal stability. This corresponds to a mutation $\mu_\bullet^-$. Then we obtain a framed BPS quiver
\begin{equation}
\begin{gathered}
\xymatrix{\gamma_\circ\ar@{..>}@<-0.5ex>[dr]\ar@{..>}@<+0.5ex>[dr]  && - \gamma_\bullet\ar@<-0.5ex>[ll] \ar@<+0.5ex>[ll]  \\
&\framebox{$\alpha$}&}
\end{gathered}
\end{equation}
Relabeling the basis $\gamma_\circ^\prime = - \gamma_\bullet$ and $\gamma_\bullet^\prime = \gamma_\circ$, we obtain
\begin{equation}
\begin{gathered}
\xymatrix{\gamma_\bullet^\prime\ar@{..>}@<-0.5ex>[ddr]_{\psi_1} \ar@{..>}@<+0.5ex>[ddr]^{\psi_2}  && \gamma_\circ^\prime\ar@<-0.5ex>[ll]_{A_1} \ar@<+0.5ex>[ll]^{A_2}  \\
&&\\
&\framebox{$\alpha^\prime$}&}
\end{gathered}
\end{equation}
In the new basis $\alpha^\prime = \gamma_\bullet = - \gamma_\circ^\prime$. In this case no bound state was formed among the defect and the BPS particle with charge $\gamma_\bullet$ because $\langle\alpha, \gamma_\bullet\rangle = 0$. Notice also that we had no wall crossing on the framed BPS spectrum for the same reason. The mutated central charges satisfy
\begin{equation}
0 < \arg Z_{\gamma_\circ^\prime} < \arg Z_{\gamma_\bullet^\prime} < \pi.
\end{equation}
Let us proceed and rotate the charge $\gamma_\bullet^\prime$ out along the imaginary axis. In this case we expect the occurence of both phenomena because $\langle \alpha^\prime, \gamma_\bullet^\prime\rangle \neq 0$: the core charge will change and the framed BPS spectrum as well. We obtain
\begin{equation}
\begin{gathered}
\xymatrix{-\gamma_\bullet^\prime\ar@{<..}@<-0.5ex>[dr]_{\psi_j^*}\ar@{<..}@<+0.5ex>[dr]  && \gamma_\circ^\prime\ar@{<-}@<-0.5ex>[ll]_{A_i^*} \ar@{<-}@<+0.5ex>[ll] \ar@{..>}@<+0.5ex>[dl]^{[A_i \psi_j]}_4   \\
&\framebox{$\alpha$}&}
\end{gathered}\qquad \begin{matrix} i=1,2 \\ j = 1,2\end{matrix}
\end{equation}
We can relabel the basis as we did before obtaining
\begin{equation}\label{cooperA}
\gamma^{\prime\prime}_{\circ} = -\gamma_\bullet^\prime = - \gamma_\circ, \qquad \gamma^{\prime\prime}_{\bullet} = \gamma_\circ^\prime = - \gamma_\bullet, \qquad
\alpha^{\prime\prime} = -\gamma^\prime_\circ + 2 \gamma^\prime_\bullet = - (\gamma^{\prime\prime}_{\bullet} + 2 \gamma^{\prime\prime}_{\circ})
\end{equation} 
\begin{equation}
\begin{gathered}
\xymatrix{\gamma_\circ^{\prime\prime}\ar@{<..}@<-0.5ex>[dr]_{\psi_j^*}\ar@{<..}@<+0.5ex>[dr]  && \gamma_\bullet^{\prime\prime}\ar@{<-}@<-0.5ex>[ll]_{A_i^*} \ar@{<-}@<+0.5ex>[ll] \ar@{..>}@<+0.5ex>[dl]^{[A_i \psi_j]}_4   \\
&\framebox{$\alpha^{\prime\prime}$}&}
\end{gathered}\qquad \begin{matrix} i=1,2 \\ j = 1,2\end{matrix}
\end{equation}
In this case a superpotential is generated
\begin{equation}
\cw = \sum_{i,j=1}^2 [\psi_j A_i ] A_i^* \psi_j^* 
\end{equation}
The jacobian relations entail
\begin{equation}
A_i^* \psi_j^*  =[\psi_j A_i ] A_i^* = \psi_j^*  [A_i \psi_j] = 0
\end{equation}
And we obtain 3 framed BPS states with charges and PSCs
\begin{equation}\label{doperA}
\begin{aligned}
&\beta_1 = \alpha^{\prime\prime} = - (\gamma^{\prime\prime}_{\bullet} + 2 \gamma^{\prime\prime}_{\circ}) && \FOmega(\beta_1) = 1,\\
&\beta_2 = \alpha^{\prime\prime} + \gamma_{\circ}^{\prime\prime} = - (\gamma^{\prime\prime}_{\bullet} + \gamma^{\prime\prime}_{\circ})&& \FOmega(\beta_2) = y + 1/y ,\\
&\beta_3 = \alpha^{\prime\prime} + 2 \gamma_{\circ}^{\prime\prime} = -\gamma^{\prime\prime}_{\bullet}&& \FOmega(\beta_3) = 1.
\end{aligned}
\end{equation}

Now consider instead starting from the quiver in Eqn.\eqref{thofty} and start rotating the $Z$-plane in the opposite direction with the same stability condition. This leads to rotating $\gamma_\circ$ out of the upper half-plane going in the opposite direction. Simultaneously, the bulk BPS quiver undergoes a left mutation $\mu^+_\circ$, and $\alpha$ changes as dictated by Eqn.\eqref{FRAMUTAH}:
\begin{equation}
\alpha^\prime \equiv \alpha - [\langle - \gamma_\circ, \alpha \rangle]_+ \gamma_\circ = \alpha - [\langle  \alpha, \gamma_\circ \rangle]_+ \gamma_\circ = \alpha
\end{equation}
because the argument of $\alpha$ aligns with that of $-\gamma_\circ$ and $\langle - \gamma_\circ, \alpha \rangle < 0$ in the quiver of Eqn.\eqref{thofty}. We obtain
\begin{equation}
\begin{gathered}
\xymatrix{\gamma_\bullet^\prime \ar@{<-}@<-0.5ex>[rr]\ar@{<-}@<+0.5ex>[rr] \ar@{<..}@<-0.5ex>[dr]\ar@{<..}@<+0.5ex>[dr] && \gamma_\circ^\prime\\
&\framebox{$\alpha^\prime$}&}
\end{gathered}
\end{equation}
where we have relabeled the basis as follows:
\begin{equation}\label{cooper}
\gamma_\bullet^\prime = -\gamma_\circ \qquad \gamma_\circ^\prime = \gamma_\bullet \qquad \alpha^\prime = \gamma_\bullet = \gamma_\circ^\prime.
\end{equation}
There are again 3 framed BPS states with charges and PSCs
\begin{equation}\label{duper}
\begin{aligned}
&\beta_1^\prime = \alpha^{\prime} = \gamma_\circ^\prime&& \FOmega(\beta_1) = 1,\\
&\beta_2^\prime = \alpha^{\prime} + \gamma_{\bullet}^{\prime} = \gamma_\circ^\prime + \gamma_\bullet^\prime && \FOmega(\beta_2) = y + 1/y ,\\
&\beta_3^\prime = \alpha^{\prime} + 2 \gamma_{\bullet}^{\prime} = \gamma_\circ^\prime + 2\gamma_\bullet^\prime&& \FOmega(\beta_3) = 1.
\end{aligned}
\end{equation}

\begin{figure}
\begin{center}
\includegraphics[scale=1.5]{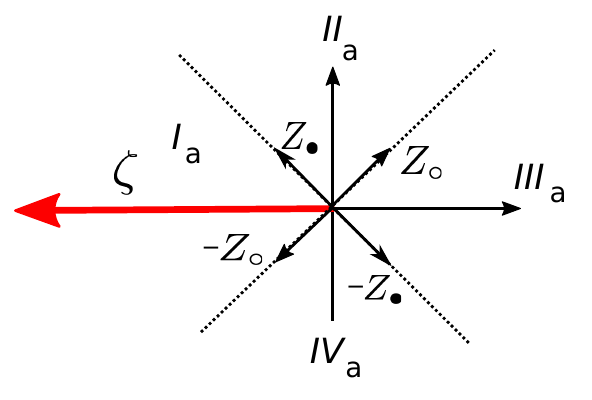}
\end{center}
\caption{The $\mathbb{Z}_4$ point of $\mathfrak{su}_2$ at strong coupling.}\label{ZU2}
\end{figure}

Let us proceed tilting the plane clockwise. $Z(\gamma_\circ^\prime)$ gets rotated out and we obtain 
\begin{equation}
\begin{gathered}
\xymatrix{\gamma_\circ^{\prime\prime} \ar@{->}@<-0.5ex>[rr]\ar@{->}@<+0.5ex>[rr] \ar@{<..}@<-0.5ex>[dr]\ar@{<..}@<+0.5ex>[dr] && \gamma_\bullet^{\prime\prime}\\
&\framebox{$\alpha^{\prime\prime}$}&}
\end{gathered}
\end{equation}
where we have relabeled the basis as follows:
\begin{equation}
\gamma_\bullet^{\prime\prime} = -\gamma_\circ^\prime \qquad \gamma_\circ^{\prime\prime} = \gamma_\bullet^\prime \qquad \alpha^{\prime\prime} = -\gamma_\bullet^{\prime\prime}.
\end{equation}
It is a simple exercise in combinatorics to see that in this case we obtain the following framed BPS spectrum:\footnote{ We have explicitly checked these results also match with the MPS Coulomb branch formula.}
\begin{equation}\label{mavaff...}
\begin{aligned}
&\beta_{2,1} = \alpha^{\prime\prime} = -\gamma_\bullet^{\prime\prime} &&\FOmega (\beta_{2,1})= 1\\
&\beta_{2,2} = \alpha^{\prime\prime} + \gamma_\circ^{\prime\prime} = -\gamma_\bullet^{\prime\prime} + \gamma_\circ^{\prime\prime} && \FOmega (\beta_{2,2}) = y + 1/y\\
&\beta_{2,3} = \alpha^{\prime\prime} + 2 \gamma_\circ^{\prime\prime} = -\gamma_\bullet^{\prime\prime} + 2 \gamma_\circ^{\prime\prime} && \FOmega (\beta_{2,3}) = 1\\
&\beta_{2,4} =  \alpha^{\prime\prime} + \gamma_\circ^{\prime\prime} + \gamma_\bullet^{\prime\prime} = \gamma_\circ^{\prime\prime} && \FOmega (\beta_{2,4}) = 2 + y^2 + 1/y^2 \\
&\beta_{2,5} = \alpha^{\prime\prime} + \gamma_\circ^{\prime\prime} + 2\gamma_\bullet^{\prime\prime} = \gamma_\circ^{\prime\prime} + \gamma_\bullet^{\prime\prime} && \FOmega (\beta_{2,5}) = y + 1/y\\
&\beta_{2,6} = \alpha^{\prime\prime} + 2 \gamma_\circ^{\prime\prime} + \gamma_\bullet^{\prime\prime} = 2\gamma_\circ^{\prime\prime} && \FOmega (\beta_{2,6}) = 1/y^3 + 1/y + y + y^3\\
&\beta_{2,7} = \alpha^{\prime\prime} + 2 \gamma_\circ^{\prime\prime} + 2 \gamma_\bullet^{\prime\prime} = 2\gamma_\circ^{\prime\prime} + \gamma_\bullet^{\prime\prime} && \FOmega (\beta_{2,7}) = 2 + 1/y^4 + 1/y^2 + y^2 + y^4\\
&\beta_{2,8} = \alpha^{\prime\prime} + 2 \gamma_\circ^{\prime\prime} + 3 \gamma_\bullet^{\prime\prime} = 2\gamma_\circ^{\prime\prime} + 2\gamma_\bullet^{\prime\prime} && \FOmega (\beta_{2,8}) = 1/y^3 + 1/y + y + y^3\\
&\beta_{2,9} =  \alpha^{\prime\prime} + 2 \gamma_\circ^{\prime\prime} + 4 \gamma_\bullet^{\prime\prime} = 2\gamma_\circ^{\prime\prime} + 3\gamma_\bullet^{\prime\prime} && \FOmega (\beta_{2,9})= 1
\end{aligned}
\end{equation}
One can continue this exercise \textit{ad libitum}, but we prefer to stop here. Notice that for a special tuning this chamber can preserve a $\mathbb{Z}_4$ symmetry (cfr. Figure \ref{ZU2}).

\subsection{The $\mathfrak{su}_2$ $Y$-system}
The $\mathfrak{su}_2$ $Y$-system can be obtained from the BPS spectum of the $\mathfrak{su}_2$ gauge theory at strong coupling easily with the method discussed in \cite{Cecotti:2014zga}. It is given by
\begin{equation}
\begin{aligned}
&R^{(+)} \colon \left(Y_\bullet , Y_\circ \right) \mapsto \left({1 \over Y_\circ} , {Y_\bullet Y_\circ^2 \over (1 + Y_\circ)^2} \right)\\
&R^{(-)} \colon\left(Y_\bullet , Y_\circ \right) \mapsto \left(Y_\circ (1 + Y_\bullet)^2,{1 \over Y_\bullet }\right)\\
\end{aligned}
\end{equation}
Consider the example of the previous section. To start with we have
\begin{equation}
\langle \, \widehat{\mathfrak{L}}_\zeta \, \rangle = Y_\bullet.
\end{equation}
Now tilting the $Z$-plane counterclockwise, we had to mutate on $\bullet$, which in this case corresponds to $R^{(+)}$, indeed
\begin{equation}
\langle \, \widehat{\mathfrak{L}}_{e^{i \pi/4} \zeta} \, \rangle = 1/Y_\circ = R^{(+)}(Y_\bullet).
\end{equation}
Now, rotating again, we had to mutate on $\gamma_{\bullet}^\prime$, which at the $\mathbb{Z}_4$ symmetric point, is identified with $\gamma_\bullet$. We obtain
\begin{equation}
\langle \, \widehat{\mathfrak{L}}_{e^{i \pi/2} \zeta} \, \rangle = {(1 + Y_\circ)^2 \over Y_\bullet Y_\circ^2} = {1 \over Y_\bullet Y_\circ^2} + {2 \over Y_\bullet Y_\circ} + {1 \over Y_\bullet}.
\end{equation}
Notice the agreement in between this expression, and the framed BPS degeneracies as caputerd by the framed quiver SQM in Eqn.\eqref{doperA}. The effect of the $\mathbb{Z}_4$ symmetry is encoded in the base change (dropping the primes). Consider now a rotation in the opposite direction. We obtain
\begin{equation}
\langle \, \widehat{\mathfrak{L}}_{e^{- i \pi /4}  \zeta} \, \rangle = Y_\circ + 2 Y_\circ Y_\bullet +  Y_\circ Y_\bullet^2 = R^{(-)} (Y_\bullet),
\end{equation}
that matches precisely with the spectrum obtained using the BPS quiver in Eqn.\eqref{duper}. A further iteration of $R^{(-)}$ leads to
\begin{equation}
\langle \, \widehat{\mathfrak{L}}_{e^{- i \pi /2}  \zeta} \, \rangle = {1 \over Y_\bullet} + 4 Y_\circ + {2 Y_\circ \over Y_\bullet} + 2 Y_\bullet Y_\circ + 4 Y_\circ^2 + { Y_\circ^2 \over Y_\bullet}+ 6 Y_\bullet Y_\circ^2 + 4 Y_\bullet^2 Y_\circ^2 + 
 Y_\bullet^3 Y_\circ^2. \end{equation}
 Notice that it matches exactly with the $y\to 1$ limit of  Eqn.\eqref{mavaff...}.
 
\subsection{Wilson lines} \label{SU2-Wilson-lines}

The $SU(2)$ $W$--boson is represented by the representation with dimension vector $\d \equiv (1,1)$, therefore the unit electric charge and the corresponding magnetic charge are given by \cite{Cecotti:2012va}
\begin{equation}
\mathfrak{q} \equiv \frac{1}{2} \d \quad\text{and}\quad m(X) = \dim X_\circ - \dim X_\bullet \, .
\end{equation}
Consider the Wilson line in the $\mathbf{n+1}$ representation of $SU(2)$. The corresponding core charge is:
\be
\alpha[\mathfrak{w}^{\mathbf{n+1}}_{\zeta}] \equiv - n \, \mathfrak{q}.
\ee
This determines the corresponding framed BPS quiver, but not the corresponding superpotential. The framed BPS quiver is:
\begin{equation}\label{wSu2q}
\begin{gathered}
Q^{SU(2)}_{\mathfrak{w}^{\mathbf{n+1}}_{\zeta}} \equiv
\begin{gathered}
\xymatrix@R=1.0pc@C=1.5pc{\bullet\ar@/_0.9pc/@{..>}[ddrr]_{\a_1}\ar@{}[ddrr]|{{\tiny\vdots}}\ar@/^0.7pc/@{..>}[ddrr]^{\a_n} \\
\\
&& \ast \ar@/^0.9pc/@{..>}[ddll]^{\b_n}\ar@{}[ddll]|{{\tiny\vdots}}\ar@/_0.7pc/@{..>}[ddll]_{\b_1} \\
\\
\circ \ar@/^0.5pc/@<-0.5ex>[uuuu]_A  \ar@/^0.5pc/@<0.5ex>[uuuu]^B}
\end{gathered}
\end{gathered}
\end{equation}

Notice that this quiver has the following property:
\begin{equation}\label{wowiezowie}
\mu_\bullet(Q^{SU(2)}_{\mathfrak{w}^{\mathbf{n+1}}_{\zeta}}) = \sigma(Q^{SU(2)}_{\mathfrak{w}^{\mathbf{n+1}}_{\zeta}}) \qquad\mu_\circ(Q^{SU(2)}_{\mathfrak{w}^{\mathbf{n+1}}_{\zeta}}) = \hat{\sigma}(Q^{SU(2)}_{\mathfrak{w}^{\mathbf{n+1}}_{\zeta}}).
\end{equation}
As we have discussed in the Introduction, the Wilson lines are constants along the $Y$-system evolution. This, in particular, entails that the corresponding superpotentials have to satisfy:
\be\label{SUKA}
\mu_\bullet \cw |_\text{on shell} = \cw \qquad \text{up to relabeling}.
\ee
To show this fact, it is sufficient to consider the vev of the Wilson line in the fundamental, corresponding to the case $n=1$ for the quiver above. In this case there is a \emph{unique} superpotential that is linear in the fields and satisfies \eqref{SUKA}, it is
\be
\cw_{1} = \b \a A,
\ee
where unicity follows from the fact that $\hat{A}(2,1)$ is an acyclic quiver. 

To discuss the relevant representation theory of this quiver it is useful to introduce a diagrammatic tool to keep track of the cyclic modules. Recall that for left-adapted stability conditions the relevant modules over the jacobian algebra which describe stable states are generated by a vector $v$. This in practice means that the vector spaces of the representation have a basis whose element are all of the form of jacobian algebra elements acting on the vector $v$, for example $\{ v , \beta_i \, v , A \, \beta_i \, v  , \cdots \}$. Such a collection of vectors has to be linearly independent over the jacobian algebra, which means upon imposing the F-term relations $\partial \, \cw = 0$. Once this is done, we pick a representative from each equivalence class and connect them with arrows which mimic the way the module is generated. We call this a skeleton diagram.

For example, in the case $n=1$, the skeleton diagram for cyclic modules is given by
\be
\xymatrix{v \ar[r]^\beta & \beta v \ar[r]^B& B \beta v \ar[r]^{\a\quad}& \a B \b v = b v } \qquad b\in\C,
\ee
as $\partial_A\cw = \b \a = 0$, either $\b v =0$ or $b = 0$ in this case. Then, the representation theory reduces to that of the $A_3$ Dynkin quiver, and the allowed representations are $(1,0,0)$, $(1,1,0)$, and $(1,1,1)$. From this follows that
\be\label{wsu22vev}
\langle \, \widehat{\mathfrak{w}}^{\mathbf{2}}_\zeta \, \rangle = \frac{1 +  Y_\circ + Y_\circ Y_\bullet}{(Y_\circ Y_\bullet)^{1/2}} = \left[ \frac{1}{(Y_\circ Y_\bullet)^{1/2}} + (Y_\circ Y_\bullet)^{1/2} \right] + \frac{Y_\circ^{1/2}}{Y_\bullet^{1/2}},
\ee
in perfect agreement with \cite{Gaiotto:2010be}. This vev is indeed invariant with respect to the $Y$-system evolution we described above.

Using the fact that the OPE algebra descends by construction to the 3d vev's of BPS lines, this property has to be shared by all other Wilson line vev's for $n>1$, as these are completely determined using simply the vev above and the OPE relations.\footnote{ For instance, from the OPE relations $\mathfrak{w}^{\mathbf{2}}_{\zeta} \ast \mathfrak{w}^{\mathbf{2}}_{\zeta} = 1 + \mathfrak{w}^{\mathbf{3}}_{\zeta}$ we obtain that $\langle \widehat{\mathfrak{w}}^{\mathbf{3}}_{\zeta} \rangle = (\langle \widehat{\mathfrak{w}}^{\mathbf{2}}_{\zeta}\rangle)^2 -1$, and so on using the multiplication table of $SU(2)$ representations.} The uniqueness of the superpotential $\cw_1$ subject to the constraint Eqn.\eqref{SUKA} is therefore enough to \emph{completely} fix the values of the vev's of all the tower of Wilson lines from their well known OPEs.


A nice consistency check for the superpotentials for $n>1$ is that there is always more than one solution to eqn.\eqref{SUKA} that is linear in the fields, and moreover, higher order terms are allowed. There are however several universal properties for this class of superpotentials that are worth to continue this discussion. The stability of a module $X$ of the quiver $Q^{SU(2)}_{W_{\mathbf{n+1}}}$ for left adapted stability conditions entails that
\be
\text{ker } \b_i \cap \text{ker } \b_j = 0 \qquad\text{for all pairs} \, i \neq j \in \{1,...,n\}
\ee
As $\dim X_\ast = 1$ by construction, this is enough to constrain
\be
\dim X_\circ \leq n
\ee
independently of the choice of $\cw$. Without using the OPE this implies that the corresponding generating function of cyclic modules satisfies
\be
w_n(Y_\circ,Y_\bullet) = \frac{1}{(Y_\circ Y_\bullet)^{n/2}}\sum_{0 \leq m_1 \leq n \atop m_2 \geq 0} c_{m_1,m_2} \, Y_\circ^{\,m_1} Y_\bullet^{\,m_2},
\ee
where $c_{m_1,m_2}$ are some positive integers depending on $\cw$ that we have to determine. Notice that the quiver of eqn.\eqref{wSu2q} is invariant along the $Y$-system evolution by construction, and therefore the corresponding $w_n(Y_\circ,Y_\bullet)$ have to be constants as well. As we will see, this is essentially enough to fix almost all of the $c_{m_1,m_2}$'s. 


Consider the class of cubic superpotentials. Imposing the constraint above it follows that the only allowed superpotentials are of the form
\be
\cw_{n,k} \equiv  \sum_{i=1}^{k} \b_i \a_i A +  \sum_{j=k+1}^n \b_j \a_j B \qquad k = 1,...,n
\ee
Indeed, a mutation at the node $\bullet$ gives
\be
\mu_\bullet \cw_{n,k} = \sum_{i=1}^{k} \b_i [\a_i A] +  \sum_{j=k+1}^n \b_j [\a_j B] + \sum_{\ell=1}^n [\a_\ell A] A^* \a_\ell ^* + [\a_\ell B] B^* \a_\ell^*.
\ee
Eliminating the quadratic terms, the partial derivatives with respect to the $\b_i$ arrows set the arrows $[\a_i A]$ for $i=1,...,k$ and $[\a_j B]$ for $j=k+1,...,n$ to zero, while the partial derivatives with respect to the arrows $[\a_i A]$ and $[\a_j B]$ set $\b_i = A^* \a_i^*$ and $\b_j = B^* \a^*_j$, as these arrows are fixed by the eom's we can safely neglect them in the mutated quiver, we obtain:
\be
\mu_\bullet \cw_{n,k} |_\text{on shell} =  \sum_{j=k+1}^n [\a_j A] A^* \a_j ^* + \sum_{i=1}^k[\a_i B] B^* \a_i^*.
\ee
Therefore, the resulting quiver with superpotential on shell is exactly the one we started with. The relations from $\cw_{n,k}$ are
\bea
&\a_i A = A \b_i  = 0\qquad i =1,\dots,k\\
&\a_j B = B \b_j = 0\qquad j =k+1,\dots,n\\
& \sum_{i=1}^k \b_i \a_i =0\qquad\qquad  \sum_{j=k+1}^n \b_j \a_j =0. \\
\eea

\begin{figure}
\begin{equation}
\begin{gathered}\label{skelwsu2}{\footnotesize
\xymatrix@R=0.7pc@C=0.7pc{
&&&\a_1 B \b_1 v&\cdots\\
&\b_1 v \ar[r] &B \b_1 v\ar[ur]\ar[dr]&\vdots\\
&&&\a_k B \b_1 v&\cdots\\
&\vdots &\\
&&&\a_1 B \b_k v&\cdots\\
&\b_k v \ar[r] &B \b_k v\ar[ur]\ar[dr]&\vdots\\
&&&\a_k B \b_k v&\cdots\\
v\ar[uur]\ar[uuuuuur]\ar[ddr]\ar[ddddddr]&\\\
&&&\a_{k+1} A \b_{k+1} v&\cdots\\
&\b_{k+1} v\ar[r] &A \b_{k+1} v\ar[ur]\ar[dr]&\vdots\\
&&&\a_n A \b_{k+1} v&\cdots\\
&\vdots&\\
&&&\a_{k+1} A \b_n v&\cdots\\
&\b_n v\ar[r] &A \b_n v\ar[ur]\ar[dr]&\vdots\\
&&&\a_n A \b_n v&\cdots\\
\\
X_\ast&X_\circ&X_\bullet&X_\ast
}
}
\end{gathered}
\end{equation}
\caption{Skeleton diagram for a Wilson line in the $n+1$ dimensional representation of $SU(2)$}\label{MACOSATIFUMI}
\end{figure}
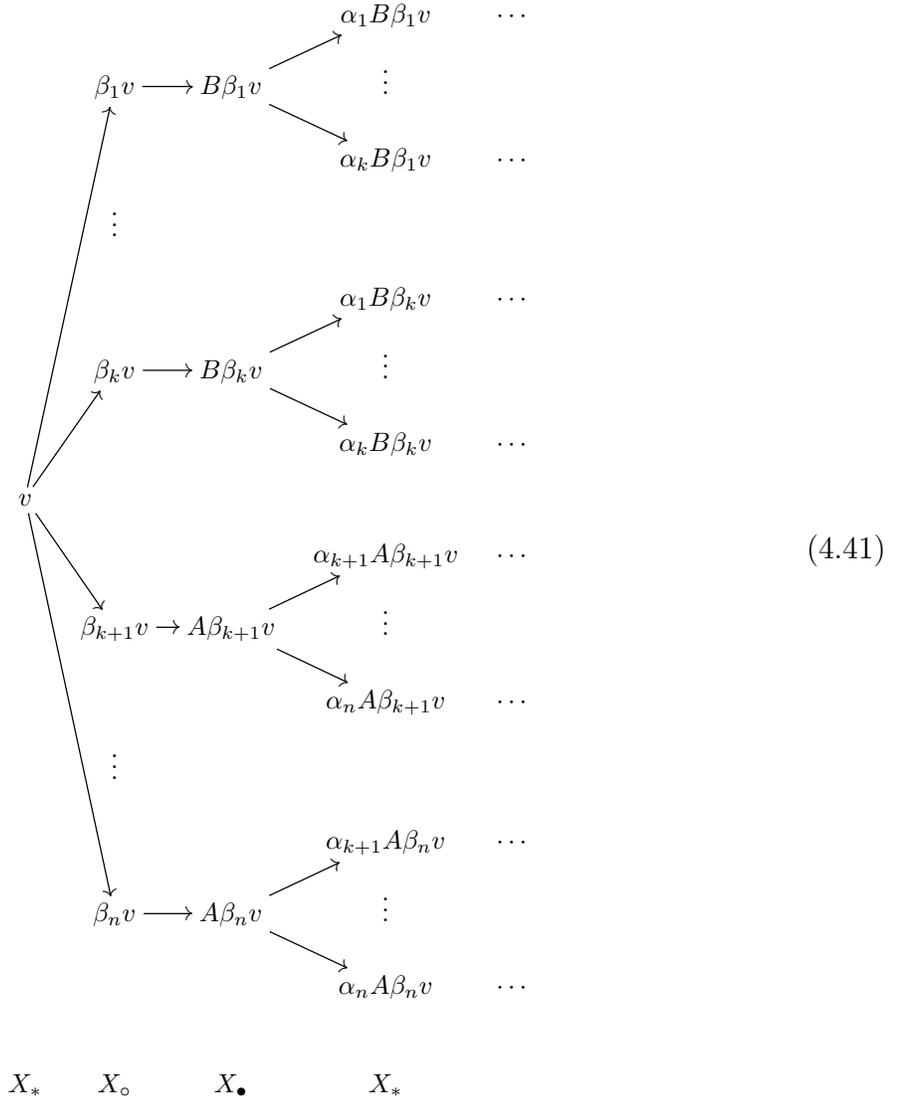
To keep track of the structures for cyclic modules, we draw the relevant skeleton diagram in Figure \ref{MACOSATIFUMI}. Notice that, as $\dim X_\ast = 1$,
\be
\a_{i_1} B \b_{i_2} v \equiv b_{i_1\,,i\,_2} v \qquad\qquad\a_{j_1} A \b_{j_2} \equiv a_{j_1-k \,,\, j_2-k} v, \qquad\qquad b_{i_1\,,i\,_2}, a_{j_1-k \,,\, j_2-k} \in \C,
\ee
for $i_1,i_2 = 1 ,...,k$ and $j_1,j_2 = k+1,...,n$. In principle, the skeleton diagram would replicate itself at each of the nodes of the last column in Figure \ref{MACOSATIFUMI}, times the appropriate complex constant. However, we have the relations
\bea\label{wsu2rels}
&\sum_{i=1}^k \b_i \a_i = 0 \quad \Longrightarrow \quad \sum_{i=1}^{k} b_{i\ell} \b_i v = 0 \quad\forall\quad \ell=1,...,k\\
&\sum_{j=k+1}^{n} \b_j \a_j = 0 \quad \Longrightarrow \quad  \sum_{j=1}^{n-k} a_{j\ell} \b_{j+k} v = 0 \quad\forall\quad \ell=1,...,n-k
\eea
From these equations it follows that whenever some $\a_i$'s are non-zero, they introduce some linear relations in between the otherwise linearly independent vectors $\b_i v \in X_\circ$: therefore, whenever the diagram has the chance to replicate, this also lowers the allowed dimensions on the node $X_\circ$. From the skeleton diagram we see that the dimension of $X_\bullet$  is constrained to be less than or equal to that of $X_\circ$. Therefore,  the biggest possible dimension vector for a cyclic module corresponds to the case in which all $\a_i$'s are zero: only in that case there are no additional relations in between the vectors $\beta_i v \in X_\circ$, and we can choose them to be linearly independent. Therefore, the dimension vectors of the cyclic modules of the quiver $Q^{SU(2)}_{\mathfrak{w}^{\mathbf{n}}_\zeta}$ with superpotential $\cw_{k,n}$ have to obey the following constraint (independent from $k$):
\be
0 \leq \dim X_\bullet \leq \dim X_\circ \leq n.
\ee
From this simple analysis it follows that a BPS Wilson line in the $\mathbf{n+1}$ representation carries $(n+1)(n+2)/2$ distinct framed BPS states, and
\be
\langle \widehat{\mathfrak{w}}^{\mathbf{n+1}}_{\zeta}\rangle = \frac{1}{(Y_\circ Y_\bullet)^{n/2}}\sum_{0\leq m_1\leq m_2 \leq n} c_{m_1,m_2} \, Y_\circ^{\,m_1} Y_\bullet^{\,m_2},
\ee
where $c_{m_1,m_2}$ are some positive integers that we have to determine. These coefficients however should depend from the specific choice of the superpotential $\cw_{k,n}$, and this is problematic as only one element of the class $\cw_{k,n}$ correctly describes the SQM of the BPS excitations framed by the BPS Wilson line in the $\mathbf{n+1}$ of $SU(2)$. There are several ways to determine these coefficients. The most  direct approach is to compute the allowed cyclic modules. In the case $n=1$, $k=1$ was discussed above. If we consider the case $n=2$ there are two possibilities to be considered:
\be
\cw_{2,1} = \b_1 \a_1 A + \b_2 \a_2 B \quad\text{and}\quad\cw_{2,2} = (\b_1 \a_1+\b_2 \a_2)A.
\ee
Before doing that, however, notice that the quiver of eqn.\eqref{wSu2q} with superpotential $\cw_{k,n}$ is invariant along the $Y$-system evolution by construction, and therefore the corresponding vev's have to be constants of motions. This is essentially enough to fix almost all of the $c_{m_1,m_2}$'s. First of all, notice that with respect to $R^{(+)}$
\be
\frac{1}{(Y_\bullet Y_\circ)^{n/2}} \to \frac{(1+Y_\circ)^n}{(Y_\bullet Y_\circ)^{n/2}}\qquad
Y_\circ^{\,m_1} Y_\bullet^{\,m_2} \to \frac{Y_\bullet^{\, m_1} Y_\circ^{\,2m_1 - m_2}}{(1+Y_\circ)^{2 m_1}}.
\ee
So, the invariance of the Wilson lines amounts to the condition:
\be
\sum_{0\leq m_1\leq m_2 \leq n} c_{m_1,m_2} Y_\circ^{\,m_1} Y_\bullet^{\,m_2} = \sum_{0\leq m_1\leq m_2 \leq n} c_{m_1,m_2} Y_\bullet^{\, m_1} Y_\circ^{\,2m_1 - m_2} (1+Y_\circ)^{n - 2 m_1}.
\ee
Consider for example the case $n=1$: we obtain
\bea
c_{00} + c_{10} Y_\circ + c_{11} Y_\circ Y_\bullet &= c_{00}(1+ Y_0) + c_{10} \frac{Y_\bullet Y_\circ^{\,2}}{(1+Y_\circ)} + c_{11} \frac{Y_\bullet Y_\circ}{(1+Y_\circ)}\\
&=c_{00}(1+ Y_0) + \frac{Y_\bullet Y_\circ}{(1+Y_\circ)}(c_{10}Y_\circ + c_{11})\\
\eea
This equality can be satisfied only if $c_{00} = c_{10} = c_{11}$. Moreover, $c_{00}=1$ since it corresponds to the rigid simple module with support on the framing node. This correctly reproduces eqn.\eqref{wsu22vev} above. Consider the case $n=2$: we obtain
\bea
&c_{00} + c_{10} Y_\circ + c_{11} Y_\circ Y_\bullet  + c_{20} Y_\circ^2 + c_{21} Y_\circ^2 Y_\bullet + c_{22}Y_\circ^2 Y_\bullet^2=\\
&=c_{00}(1+Y_\circ)^2 + c_{10} Y_\bullet Y_\circ^2 + c_{11} Y_\bullet Y_\circ + c_{20} \frac{Y_\bullet^2 Y_\circ^4}{(1+Y_\circ)^2} + c_{21} \frac{Y_\bullet^2 Y_\circ^3}{(1+Y_\circ)^2} + c_{22} \frac{Y_\bullet^2 Y_\circ^2}{(1+Y_\circ)^2}\\
&=c_{00}(1+Y_\circ)^2 + c_{10} Y_\bullet Y_\circ^2 + c_{11} Y_\bullet Y_\circ + \frac{Y_\bullet^2 Y_\circ^2}{(1+Y_\circ)^2} \left( c_{20} Y_\circ^2 + c_{21} Y_\circ + c_{22}\right)
\eea
Again this equation can be satisfied only if $c_{00} = c_{20} = c_{22} = c_{21} / 2 = c_{10} / 2$. Moreover $c_{00}=1$ as previously. There is only one ambiguity to be fixed: the coefficient $c_{11}$ is not determined in this case. This is indeed a constant contribution to the vev and the fact that this argument is not able to fix this term uniquely is related to the fact that there is more than one superpotential that remains invariant w.r.t. to the mutation on $\bullet$. This is very interesting: the choice in between $\cw_{2,1}$ and $\cw_{2,2}$ can only affect $c_{11}$ in this case. We will discuss these issues further in appendix \ref{Wsu23vevk}, where we carry out the corresponding representation theory. In particular one finds that with a superpotential $\cw_{2,1}$ we obtain $c_{11}=1$, while with a superpotential $\cw_{2,2}$ we obtain $c_{11}\neq1$, and all other coefficients are exactly equal. The consistency with the OPE
\be
\mathfrak{w}^{\mathbf{2}} \ast \mathfrak{w}^{\mathbf{2}} = 1 + \mathfrak{w}^{\mathbf{3}} 
\ee
entails that the correct superpotential for the case $n=2$ is $\cw_{2,1}$, indeed in that case we obtain:
\bea
\langle \widehat{\mathfrak{w}}_{\zeta}^{\mathbf{3}} \rangle &= \frac{1 + 2 Y_\circ + Y_\circ^2 + 2 Y_\circ^2 Y_\bullet  + Y_\circ Y_\bullet  +  Y_\circ^2 Y_\bullet^2 }{Y_\circ Y_\bullet}\\
&= 
\left[ Y_\bullet Y_\circ+1+\frac{1}{Y_\bullet
   Y_\circ} \right]+ \frac{Y_\circ}{Y_\bullet}+\frac{2}{Y_\bullet}+2 Y_\circ
\eea
that is the correct result. The Coulomb branch formula should account for both possibilities as well: the case in which all single centered degeneracies are set to zero corresponding to the superpotential $\cw_{2,1}$.

Let us proceed by considering the case $n=3$. We have only two inequivalent superpotentials, namely $\cw_{3,1}$ and $\cw_{3,3}$, as $\cw_{3,2}$ give rise to the same relations of $\cw_{3,1}$ up to relabeling:
\be
\cw_{3,1} =  \b_1 \a_1 A + (\b_2 \a_2+\b_3 \a_3)B \quad\text{and}\quad\cw_{3,3} =(\b_1 \a_1+ \b_2 \a_2+\b_3 \a_3)A.
\ee
Proceeding as before we obtain the following system of equations:
\be
\begin{cases}
c_{00} + c_{10} Y_\circ+ c_{20} Y_\circ^2 + c_{30} Y_\circ^3 = c_{00}(1+ Y_\circ)^3\\
c_{11} + c_{21} Y_\circ + c_{31} Y_\circ^2 = (1+Y_\circ)(c_{11}+c_{10} Y_\circ)\\
c_{22} + c_{32} Y_\circ = (c_{22} + c_{21} Y_\circ + c_{20} Y_\circ^2 )/(1+Y_\circ)\\
c_{33} = (c_{33} + c_{32} Y_\circ + c_{31} Y_\circ^2 + c_{30} Y_\circ^3)/(1+Y_\circ)^3
\end{cases}
\ee
Combining the first two equations with the last one gives:
\be
1=c_{00} = c_{10}/3 = c_{20}/3 = c_{30} = c_{31}/3 = c_{32}/3=c_{33} \qquad c_{21} = 3  + c_{11}
\ee
From the relations above the third equation admits a family of solutions:
\be
c_{22} = a + 3 \qquad c_{11} = a + 3 \qquad c_{21} = a+6 \qquad a\in\mathbb{Z}_{\geq -3}.
\ee
This in turn corresponds to the following family of rational functions
\bea
f_a(Y_\circ,Y_\bullet) &\equiv \Big(1+3Y_\circ + 3 Y_\circ^2 + Y_\circ^3 + (a+3) Y_\circ Y_\bullet  + (a+6 )Y_\circ^2Y_\bullet  + 3 Y_\circ^3Y_\bullet \\
& + (a+3) (Y_\circ Y_ \bullet)^2 +  3 Y_\circ^3 Y_\bullet^2  + (Y_\bullet Y_\circ)^3 \Big)/ (Y_\circ Y_\bullet)^{3/2}
\eea
all constant along the $Y$-system evolution. From the representation theory above, it follows that the minimal possible allowed choice for $a$ is $a=-2$ (indeed, there are always cyclic modules with the corresponding dimension vectors). With this choice we obtain:
\bea
f_{-2}(Y_\circ,Y_\bullet) = & \left[
(Y_\bullet  Y_\circ)^{3/2}+ (Y_{\bullet} Y_{\circ})^{1/2} + \frac{1}{(Y_{\bullet} Y_{\circ})^{1/2}}+\frac{1}{(Y_\bullet
   Y_\circ)^{3/2}} \right]
   \cr &
   +3 \, \frac{ Y_{\circ}^{1/2}}{Y_\bullet^{3/2}} +\frac{Y_\circ^{3/2}}{Y_\bullet^{3/2}} +\frac{3}{Y_\bullet^{3/2}
   Y_{\circ}^{1/2}}+3 \, Y_{\bullet}^{1/2} Y_\circ^{3/2}+\frac{3
   Y_\circ^{3/2}}{Y_{\bullet}^{1/2}} +4 \, \frac{
   Y_{\circ}^{1/2}}{Y_{\bullet}^{1/2}}
\eea
Notice that this is precisely the result obtained in \cite{Cordova:2013bza} for the case $n=3$ using the Coulomb branch formula. The logic of the argument above can be reversed: we know the for the Wilson line in the fundamental there is no ambiguity in the choice of the superpotential. Moreover, we know the OPE of Wilson lines. Therefore we can obtain all the vev's of the Wilson lines starting from the one in the fundamental representation and using the OPE relations.

\subsection{Lines for $\mathfrak{su}_2$ theories}
\begin{figure}
\begin{center}
\begin{tabular}{ccc}
\includegraphics[width=0.3\textwidth]{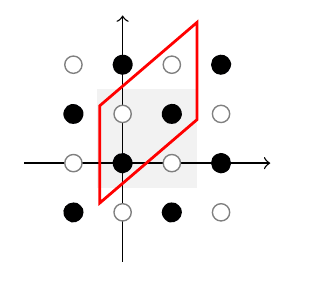}&
\includegraphics[width=0.3\textwidth]{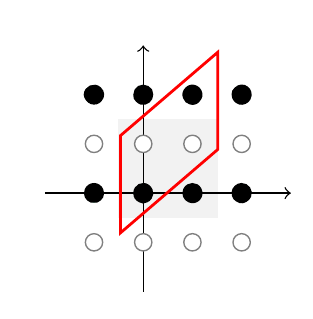}&
\includegraphics[width=0.3\textwidth]{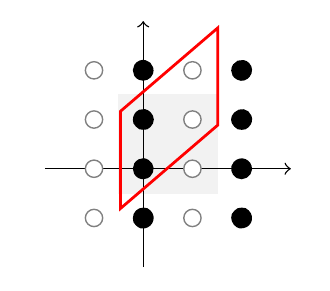}\\
$\Gamma^*_{SU(2)} $&$\Gamma^*_{SO(3)_-}$&$\Gamma^*_{SO(3)_+} $
\end{tabular}
\end{center}
\caption{Here we draw the fundamental cells of the lattices in eqn\eqref{su2latx}. The cartesian grid refers to the $(\gamma_\bullet,\gamma_\circ)$ basis, the $\gamma_\circ$ direction being the vertical one. In red we draw the fundamental cells identified in \cite{Aharony:2013hda}}\label{su2lattices}
\end{figure}

As discussed in \cite{Gaiotto:2010be}, the charge lattice of line operators is determined by a maximal sublattice of the dual lattice
\be
\Gamma^*_{\mathfrak{su}_2} \equiv \Big\{ \alpha \in \mathbb{Q}^n \,\colon \, \langle \alpha, \gamma \rangle_D \in \mathbb{Z}, \, \forall \, \gamma \in \Gamma \Big\}
\ee
compatible with Dirac quantization. Notice that the dual lattice $\Gamma^*$ is determined uniquely by the $\mathfrak{su}_2$ structure, which is reflected in the Dirac pairing defining the BPS quiver. The lattice $\Gamma_{\mathfrak{su}_2}^*$ is $\tfrac{1}{2}\Z \times \tfrac{1}{2}\Z$. To determine the maximal sublattices compatible with Dirac quantization we use the fact that the latter are strongly constrained by the knowledge of the Wilson line operators in the spectrum. We obtain three distinct maximal sublattices of $\Gamma^*_{\mathfrak{su}_2}$ are:
\bea\label{su2latx}
&\Gamma^*_{SO(3)_-} \equiv \{ (\tfrac{n}{2},k),\, n,k\in \Z \} = \tfrac{1}{2}\Z \times \Z \subset \tfrac{1}{2}\Z \times \tfrac{1}{2}\Z\\
&\Gamma^*_{SO(3)_+} \equiv \{ (k,\tfrac{n}{2}),\, n,k\in \Z \} = \Z \times \tfrac{1}{2}\Z \subset \tfrac{1}{2}\Z \times \tfrac{1}{2}\Z\\
&\Gamma^*_{SU(2)} \equiv \{ (\tfrac{n}{2},\tfrac{n}{2}+k),\,n,k\in \Z \} = (0,\Z) + (\tfrac{1}{2},\tfrac{1}{2})\Z \subset \tfrac{1}{2}\Z \times \tfrac{1}{2}\Z
\eea
In figure \ref{su2lattices} we draw them and identify the AST cells \cite{Aharony:2013hda}. In the case of $\mathfrak{su}_2$ theories the defect group consists of one non-trivial element only, as expected.

\subsection{Framed BPS states of $\mathfrak{su}_2$ theories}

\begin{figure}
\begin{center}
\includegraphics[scale=0.65]{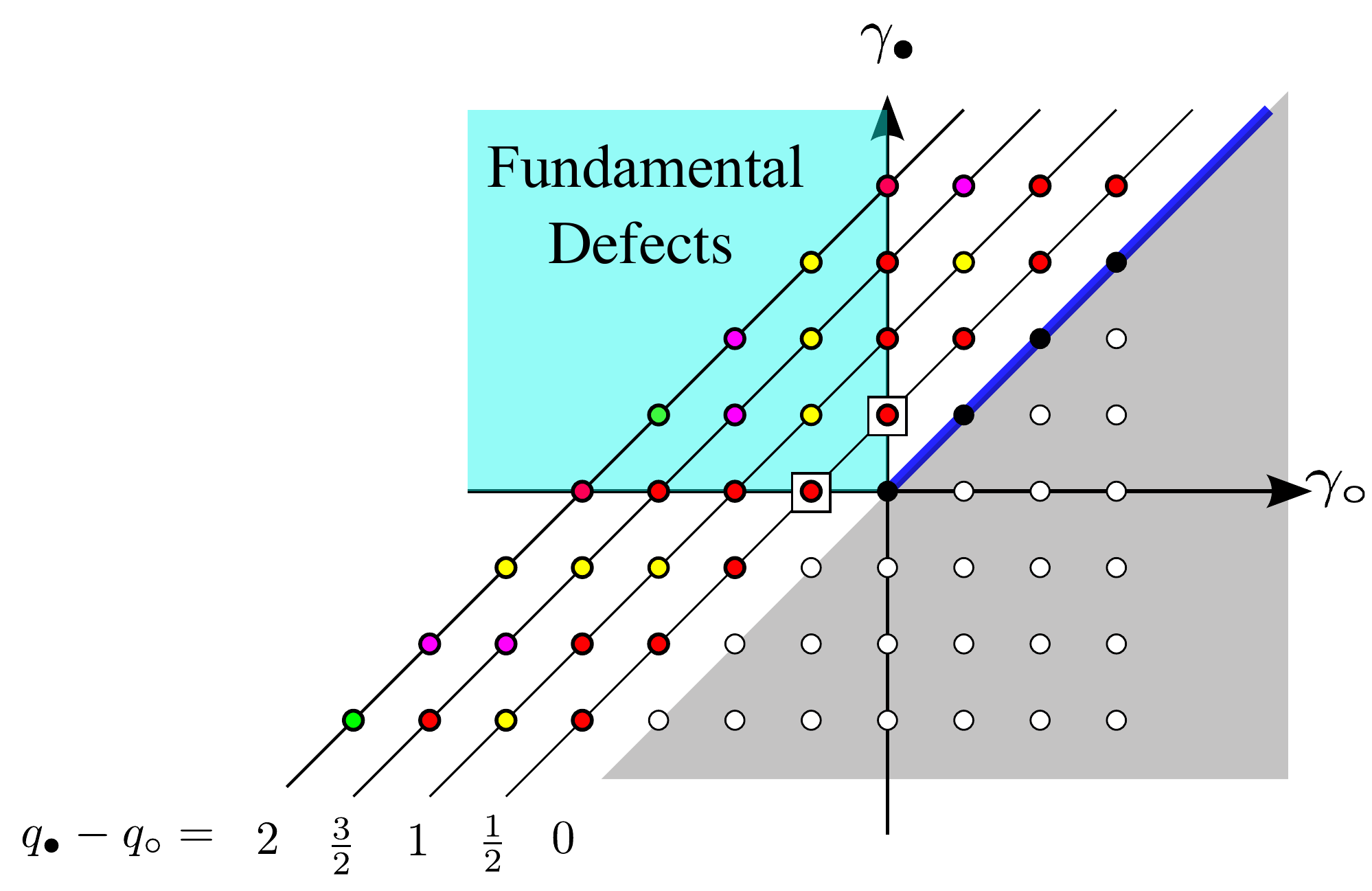}
\end{center}
\caption{Orbits of defects in the $\mathfrak{su}_2$ lattice. Lines of constant magnetic charge contain BPS lines which are part of the same orbit (indicated here using dots of the same colors along the same line). In blue we are drawing the Wilson lines.}\label{ORBITAH}
\end{figure}

We begin by identifying the set of fundamental defects. These are the defects which have a framed BPS spectrum consisting of a single framed BPS state with charge that equals the core charge of the defect. In this case the core charge and the UV charge of the defect are expected to match.\footnote{ We thank Andy Royston for a discussion about this point.}  Let us denote $\alpha = q_\bullet \gamma_\bullet + q_\circ \gamma_\circ \in \Gamma^*_{\mathfrak{su}_2}$, we claim that $\a$ is the core charge of a fundamental defect iff:
\be\label{simpleSU2}
\langle \gamma_\bullet,\alpha \rangle \geq 0 \quad\text{and}\quad \langle \gamma_\circ,\alpha \rangle \geq 0\, \Longrightarrow\, q_\circ \leq 0 \text{ and } q_\bullet \geq 0.
\ee
The set of fundamental defects is closed with respect to the OPE because of charge conservation. Indeed, if $\a_1$ and $\a_2$ are two elements of $\Gamma^*$ which satisfy eqn.\eqref{simpleSU2}, so does $\a_1 + \a_2$. Therefore the set of simple line defects is an associative magma, a semigroup. We denote such a set as $\Gamma^*_{\mathfrak{su}_2,F}$. The framed BPS quiver for a fundamental line defect with charge $\a \in \Gamma^*_{\mathfrak{su}_2,F}$ is
\begin{equation}
\begin{gathered}\label{simpleSU2}
\xymatrix{\bullet \ar@{..>}[dr]^{2 |q_\circ|} \\
&\ast\\
\circ \ar@<-0.5ex>[uu]  \ar@<0.5ex>[uu]\ar@{..>}[ur]_{2 q_\bullet}}
\end{gathered}
\end{equation}
and by construction there is a unique representation of this quiver which satisfies the axioms for a framed BPS state: the simple representation corresponding to the framing node, $S_\ast$. Simple representations being rigid by construction, the corresponding vev is
\be\label{simpleSU2vev}
\langle \mathfrak{f}_{\alpha,\zeta}\rangle = \frac{(Y_\bullet)^{q_\bullet}}{(Y_\circ)^{|q_\circ|}} \qquad\qquad \forall \, \a \in \Gamma^*_{G,s}.
\ee
Notice that the quivers in eqn.\eqref{simpleSU2} are all acyclic, and therefore have null superpotential. From these remarks, we see that the vevs of all fundamental defects in $\Gamma^*_{\mathfrak{su}_2,F}$ are determined from the vevs of two of them, which are
\be
\langle \mathfrak{f}_{1,\zeta} \rangle = (Y_\bullet)^{1/2} \quad \text{and}\quad \langle \mathfrak{f}_{2,\zeta} \rangle = \frac{1}{(Y_\circ)^{1/2}}
\ee
These two are the simple fundamental defects of the $\Gamma^*_{\mathfrak{su}_2}$ lattice. From eqn.\eqref{simpleSU2vev} we have that \cite{Cordova:2013bza}
\be
\mathfrak{f}_{\alpha,\z} = \underbrace{\mathfrak{f}_{1,\zeta}\cdot ... \cdot \mathfrak{f}_{1,\zeta}}_{2 q_\bullet \text{ times}} \cdot \underbrace{\mathfrak{f}_{2,\zeta} \cdot ... \cdot \mathfrak{f}_{2,\zeta} }_{2|q_\circ| \text{ times}}
\ee
for all $\a \in \Gamma^*_{\mathfrak{su}_2,F}$. Not surprisingly, the knowledge of the simple defects uniquely determines all framed BPS spectra of dyonic BPS defects by means of the interplay in between $U(1)_R$ rotations, the Witten effect, and framed wall crossings. With reference to Figure \ref{ORBITAH} moving up/down along a diagonal with constant magnetic charge corresponds precisely to evolving the corresponding fundamental defect using the $Y$-system. The situation is identical to the one we discussed in Section \ref{THOFTOH}.

\subsection{$SU(2)$ vs. $SO(3)$}

To construct the lattice of charges of line defects we use the following procedure. We fix a chamber and we consider certain classes of core charges which are in principle allowed. Then we complete the lattice by considering all the core charges which are mutually consistent, in the sense that they obey the Dirac quantization condition. This is exactly the same procedure used in \cite{Aharony:2013hda,Xie:2013vfa}, but now at the level of the IR defects. We stress that this requires fixing a point $p \in \mathcal{P}$, the parameter space, although the results are independent of this choice by consistency with wall-crossing.

Consider the Lie algebra $\mathfrak{su} (2)$. We know that the Wilson line is present in $SU(2)$, with $\alpha = - {1 \over 2} (\gamma_\circ + \gamma_\bullet)$. Then core charges of the form $n \gamma_\circ + m \gamma_\bullet$ with $m,n \in \mathbb{Z}$ obey the consistency condition because of the pairing $\langle \gamma_\circ, \gamma_\bullet \rangle = 2$ which cancels the $2$ in the denominator. On the other hand $\frac12 \gamma_\circ$ or $\frac12 \gamma_\bullet$ don't have integral pairing with all the Wilson lines, and therefore they cannot be part of the same set of line defects, they must be in different theories, which we will call $SO(3)_{\pm}$ following \cite{Aharony:2013hda}. Note that $\frac12 \gamma_\circ$ and $\frac12 \gamma_\bullet$ have non integer pairing between themselves and therefore they cannot be both part of the same lattice. We will call  $SO(3)_+$ the theory which contains $\frac12 \gamma_\circ$.

The $SU(2)$ and the $SO(3)$ theories have different smallest fractional monodromy available. The reason is that the presence of a fractional monodromy is a consequence of the discrete R-symmetry of the UV theory. The surviving non-anomalous R-symmetry in turns is related to the periodicity of the $\theta$-angle. Let us briefly recall this connection. 

By the index theorem given an instanton background of instanton number $k$ the number of zero modes of a fermion $\psi$ in the representation $\mathbf{r}$ minus the zero modes of $\overline{\psi}$ is given by $2 C_2 (\mathbf{r}) k$, where $C_2 (\mathbf{r})$ is the quadratic Casimir of the representation. Therefore under a phase rotation $\psi \longrightarrow \e^{\ii \delta} \psi$ the integration measure changes by the phase $\e^{2 C_2 (\mathbf{r}) k \ii \delta}$, corresponding to the extra zero modes. This phase shift is a symmetry of the theory if it can be absorbed by a shift of the $\theta$ angle in the instanton term. In the case of $SU(2)$ the periodicity of the $\theta$ angle is $\theta \longrightarrow \theta + 2 \pi$, and therefore only chiral rotations with $\delta = \frac{2 \pi}{2 C_2 (\mathbf{r})}$ are allowed. If we specialize to $SU(2)$, since the gaugino is in the adjoint representation, we have $C_2 (\mathbf{adj}) = 2$ and therefore the allowed phase rotation is $\e^{\pi \ii / 2}$ and the surviving R-symmetry is $\mathbb{Z}_4$. On the other hand if we consider $SO(3)$, the periodicity of the $\theta$ angle is $\theta \longrightarrow \theta + 4 \pi$, which corresponds to the fact that the instanton number on a spin manifold (as is $\mathbb{R}^4$) is $k \in \frac12 \mathbb{Z}$. Now the allowed phase is $\e^{ \ii \delta} =\e^{\ii  \frac{4 \pi}{2 C_2 (\mathbf{r})}} = \e^{\ii \pi}$, since for $SO(3)$ again $C_2 (\mathbf{adj}) = 2$ (for $SO(N)$ is $2N-4$). Therefore the only surviving R-symmetry is $\mathbb{Z}_2$. These arguments play a key role in understanding S-duality for $\mathcal{N}=4$ SYM \cite{Vafa:1994tf}.

This implies that a $SU(2)$ gauge theory admits a chamber with a $\frac14$-fractional monodromy, while an $SO(3)$ theory only has the half-monodromy. Note that this does \textit{not} imply any difference in the spectrum of stable BPS states. However it \textit{does} imply a difference in the spectrum of allowed line defects, since as we have shown these can be generated by the action of the smallest fractional monodromy. 

We will now show this explicitly; as usual we take the following quiver
\begin{equation}
\xymatrix@C=8mm{  
 \gamma_\circ \ar@<-0.5ex>[rr]  \ar@<0.5ex>[rr] & &  \gamma_\bullet  
 } \, .
\end{equation}
Consider the three cases:

\underline{$SU(2)$}.

In this case we have the Wilson lines  $\mathfrak{w}^{\mathbf{n+1}}_\zeta$  which correspond to
\begin{equation}
\xymatrix@C=8mm{  & \framebox{$-\frac{n}{2}  (\gamma_\circ + \gamma_\bullet)$}  \ar@{..>}[dl]_{n} & \\
 \gamma_\circ \ar@<-0.5ex>[rr]  \ar@<0.5ex>[rr]  & &  \gamma_\bullet  \ar@{..>}[ul]_{n}  
 }
\end{equation}
where $n \in \mathbb{Z}$. If we insist that we must have any value of $n \in \mathbb{Z}$, then we must include core charges $\gamma_\circ$ and $\gamma_\bullet$, but not their fractions. We will take cyclic stability conditions. Therefore we have the line operators 
\begin{equation}
\xymatrix@C=8mm{  &\framebox{$ -\gamma_\circ$}  & \\
 \gamma_\circ \ar@<-0.5ex>[rr]  \ar@<0.5ex>[rr]  & &  \gamma_\bullet  \ar@{..>}[ul]_{2}  
 }
\qquad
\xymatrix@C=8mm{  & \framebox{$ \gamma_\bullet $} & \\
 \gamma_\circ \ar@{..>}[ur]_{2}  \ar@<-0.5ex>[rr]  \ar@<0.5ex>[rr]  & &  \gamma_\bullet  
 }
\end{equation}
where we have written explicitly the core charge at the framing node. The corresponding vevs are
\begin{align}
\langle \, \widehat{\mathfrak{L}}_{\gamma_\bullet}\,  \rangle = & Y_{\gamma_\bullet} \, , \\
\langle \, \widehat{\mathfrak{L}}_{- \gamma_\circ} \, \rangle = & 1/ Y_{\gamma_\circ} \, .
\end{align}
For notational simplicity we omit to write explicitly the phase $\zeta$. These operators generate the following cluster families of line defects. Starting from $\mathfrak{L}_{\gamma_\bullet}$, by applying the rational transformation corresponding to the $1/4$ monodromy we find
\begin{align}
\langle \, \widehat{\mathfrak{L}}_{-2 \gamma_\bullet - \gamma_\circ} \, \rangle = & \frac{\left(1+2 Y_{\gamma_\circ}+2 Y_{\gamma_\bullet} Y_{\gamma_\circ}+Y_{\gamma_\circ}^2+3 Y_{\gamma_\bullet} Y_{\gamma_\circ}^2+3 Y_{\gamma_\bullet}^2 Y_{\gamma_\circ}^2+Y_{\gamma_\bullet}^3 Y_{\gamma_\circ}^2\right)^2}{Y_{\gamma_\bullet}^2 Y_{\gamma_\circ}} \, , \\
\langle \, \widehat{\mathfrak{L}}_{-\gamma_\bullet} \rangle = & \frac{\left(1+Y_{\gamma_\circ}+2 Y_{\gamma_\bullet} Y_{\gamma_\circ}+Y_{\gamma_\bullet}^2 Y_{\gamma_\circ}\right)^2}{Y_{\gamma_\bullet}} \, , \\
\langle \, \widehat{\mathfrak{L}}_{\gamma_\circ} \, \rangle = & (1+Y_{\gamma_\bullet})^2 Y_{\gamma_\circ} \, , \\
\langle \, \widehat{\mathfrak{L}}_{\gamma_\bullet} \, \rangle = & Y_{\gamma_\bullet} \, , \\
\langle \, \widehat{\mathfrak{L}}_{-\gamma_\circ} \, \rangle = & \frac{1}{Y_{\gamma_\circ}} \, , \\
\langle \, \widehat{\mathfrak{L}}_{-\gamma_\bullet - 2 \gamma_\circ} \, \rangle = & \frac{(1+Y_{\gamma_\circ})^2}{Y_{\gamma_\bullet} Y_{\gamma_\circ}^2} \, , \\
\langle \, \widehat{\mathfrak{L}}_{-2 \gamma_\bullet - 3 \gamma_\circ} \, \rangle = & \frac{\left(1+2 Y_{\gamma_\circ}+Y_{\gamma_\circ}^2+Y_{\gamma_\bullet} Y_{\gamma_\circ}^2\right)^2}{Y_{\gamma_\bullet}^2 Y_{\gamma_\circ}^3}  \, , \\
\langle  \,  \widehat{\mathfrak{L}}_{-3 \gamma_\bullet - 4 \gamma_\circ} \, \rangle = & \frac{\left(1+3 Y_{\gamma_\circ}+3 Y_{\gamma_\circ}^2+2 Y_{\gamma_\bullet} Y_{\gamma_\circ}^2+Y_{\gamma_\circ}^3+2 Y_{\gamma_\bullet} Y_{\gamma_\circ}^3+Y_{\gamma_\bullet}^2 Y_{\gamma_\circ}^3\right)^2}{Y_{\gamma_\bullet}^3 Y_{\gamma_\circ}^4} \, .
\end{align}
In general we generate $\langle \, \widehat{\mathfrak{L}}_{-(n-1) \gamma_\bullet -n \gamma_\circ} \, \rangle$ and $\langle \, \widehat{\mathfrak{L}}_{- n \gamma_\bullet - (n-1) \gamma_\circ} \, \rangle$, both with $n \ge 0$. The framed quivers are
\begin{equation}
\xymatrix@C=8mm{  &\framebox{$ -(n-1) \gamma_\bullet -n \gamma_\circ$}   \ar@{..>}[dl]_{2 n -2}  & \\
 \gamma_\circ \ar@<-0.5ex>[rr]  \ar@<0.5ex>[rr]  & &  \gamma_\bullet   \ar@{..>}[ul]_{2n}  
 }
\qquad
\xymatrix@C=8mm{  &\framebox{$- n \gamma_\bullet - (n-1) \gamma_\circ$} \ar@{..>}[dl]_{2 n}  & \\
 \gamma_\circ \ar@<-0.5ex>[rr]  \ar@<0.5ex>[rr]  & &  \gamma_\bullet   \ar@{..>}[ul]_{2n-2}  
 }
\end{equation}

\underline{$SO(3)_+$}.

We call $SO(3)_+$ the case with the core charges  $\gamma_\bullet$ and $- \frac12 \gamma_\circ$. They correspond to the two quivers
\begin{equation}
\xymatrix@C=8mm{  & \framebox{$ -\frac12 \gamma_\circ $} & \\
 \gamma_\circ \ar@<-0.5ex>[rr]  \ar@<0.5ex>[rr]  & &  \gamma_\bullet  \ar@{..>}[ul]
 }
\qquad
\xymatrix@C=8mm{  & \framebox{$ \gamma_\bullet  $}& \\
 \gamma_\circ \ar@{..>}[ur]_{2}  \ar@<-0.5ex>[rr]  \ar@<0.5ex>[rr]  & &  \gamma_\bullet  
 }
\end{equation}
and vevs
\begin{align} \label{SO3pvev}
\langle \, \widehat{\mathfrak{L}}_{\gamma_\bullet} \, \rangle = & Y_{\gamma_\bullet} \, , \cr
\langle \, \widehat{\mathfrak{L}}_{- \frac12 \gamma_\circ} \, \rangle = & 1/ Y_{\frac12 \gamma_\circ} = 1 / Y_{\gamma_\circ}^{1/2} \, .
\end{align}
Note that only the Wilson lines $\mathfrak{w}^{\mathbf{n+1}}_\zeta$ with core charge $ -n (\gamma_\circ + \gamma_\bullet)$ where $n \in \mathbb{Z}_{>0}$ are now allowed by the consistency condition. These are precisely the Wilson lines in those representations of $SU(2)$ which are invariant under the action of the center $\mathbb{Z}_2$.

Now we can use the monodromy operators to generate families of line defects starting from the defects above. However here we are in an $SO(3)$ theory and so we have to be careful. In particular the periodicity of the theta angle is changed from $\theta \longrightarrow \theta + 2 \pi$ to $\theta \longrightarrow \theta + 4 \pi$. As a consequence the $R$-symmetry is not anymore $\mathbb{Z}_4$ but only a $\mathbb{Z}_2$ remains. This is a consequence of the fact that the $R$-symmetry of the fermions is related to the shift of the theta angle in an instanton background (and on spin manifolds the instanton number of an $SO(3)$ instanton is in $\frac12 \mathbb{Z}$). Since now we only have $\mathbb{Z}_2$ R-symmetry, the half monodromy is the only fractional monodromy. Therefore starting from the two operators in  (\ref{SO3pvev}) we generate two \textit{distinct} families. Indeed applying the 1/4-monodromy to $ 1/ Y_{\frac12 \gamma_\circ}$ would give $Y_{\frac12 \gamma_\bullet}$ which would violate Dirac consistency condition.

One family is precisely as before, and contains
\begin{align}
\langle \, \widehat{\mathfrak{L}}_{-\gamma_\bullet} \, \rangle = & \frac{\left(1+Y_{\gamma_\circ}+2 Y_{\gamma_\bullet} Y_{\gamma_\circ}+Y_{\gamma_\bullet}^2 Y_{\gamma_\circ}\right)^2}{Y_{\gamma_\bullet}} \, , \\
\langle \, \widehat{\mathfrak{L}}_{\gamma_\bullet} \, \rangle = & Y_{\gamma_\bullet} \, , \\
\langle \, \widehat{\mathfrak{L}}_{-\gamma_\bullet - 2 \gamma_\circ} \, \rangle = & \frac{(1+Y_{\gamma_\circ})^2}{Y_{\gamma_\bullet} Y_{\gamma_\circ}^2} \, , \\
\langle \, \widehat{\mathfrak{L}}_{-3 \gamma_\bullet - 4 \gamma_\circ} \, \rangle = & \frac{\left(1+3 Y_{\gamma_\circ}+3 Y_{\gamma_\circ}^2+2 Y_{\gamma_\bullet} Y_{\gamma_\circ}^2+Y_{\gamma_\circ}^3+2 Y_{\gamma_\bullet} Y_{\gamma_\circ}^3+Y_{\gamma_\bullet}^2 Y_{\gamma_\circ}^3\right)^2}{Y_{\gamma_\bullet}^3 Y_{\gamma_\circ}^4} \, , 
\end{align}
and the other contains the operators
\begin{align} \label{dyonSO3p}
\langle \, \widehat{\mathfrak{L}}_{- \gamma_\bullet - \frac12 \gamma_\circ} \, \rangle = & \frac{\left(1+2 Y_{\gamma_\circ}+2 Y_{\gamma_\bullet} Y_{\gamma_\circ}+Y_{\gamma_\circ}^2+3 Y_{\gamma_\bullet} Y_{\gamma_\circ}^2+3 Y_{\gamma_\bullet}^2 Y_{\gamma_\circ}^2+Y_{\gamma_\bullet}^3 Y_{\gamma_\circ}^2\right)}{Y_{\gamma_\bullet} Y_{\gamma_\circ}^{1/2}} \, , \\
\langle \, \widehat{\mathfrak{L}}_{\frac12 \gamma_\circ} \, \rangle = & (1+Y_{\gamma_\bullet}) Y_{\gamma_\circ}^{1/2} \, , \\
\langle \,\widehat{\mathfrak{L}}_{-\frac12 \gamma_\circ} \, \rangle = & \frac{1}{Y_{\gamma_\circ}^{1/2}} \, , \\
\langle \, \widehat{\mathfrak{L}}_{- \gamma_\bullet - \frac32 \gamma_\circ} \, \rangle = & \frac{\left(1+2 Y_{\gamma_\circ}+Y_{\gamma_\circ}^2+Y_{\gamma_\bullet} Y_{\gamma_\circ}^2\right)}{Y_{\gamma_\bullet} Y_{\gamma_\circ}^{3/2}}  \, , 
\end{align}
which are simply given by the square root of the corresponding operators in $SU(2)$. Note that these $SU(2)$ line operators are still present, since they arise from multiplication of (\ref{dyonSO3p}) with themselves (or equivalently they are generated by iteration of the half-monodromy rational transformation starting from $1/Y_{\gamma_\circ}$).

To summarize the line operators obtained in this way are 
\begin{equation}
\begin{cases}
& \langle \, \widehat{\mathfrak{L}}_{-n \gamma_\bullet - (2n+1)/2 \gamma_\circ} \, \rangle \ , \langle \, \widehat{\mathfrak{L}}_{-n \gamma_\bullet - (2 n-1)/2 \gamma_\circ} \, \rangle  \text{ \ with \ } n\ge 0 \left( \text{\ from \  }  Y_{- \frac12 \gamma_\circ}  \right)\\
& \langle \, \widehat{\mathfrak{L}}_{-(n-1) \gamma_\bullet -n \gamma_\circ}  \, \rangle \text{\ with\ } n \ge 0 \text{\ even and\ } \langle \, \widehat{\mathfrak{L}}_{- n \gamma_\bullet - (n-1) \gamma_\circ} \, \rangle \text{\ with\ } n \ge 0 \text{\ odd} \left( \text{from\ } Y_{\gamma_\bullet} \right) \\
& \langle \, \widehat{\mathfrak{L}}_{-(n-1) \gamma_\bullet -n \gamma_\circ} \, \rangle \text{\ with\ } n \ge 0 \text{\ odd and\ } \langle \, \widehat{\mathfrak{L}}_{- n \gamma_\bullet - (n-1) \gamma_\circ} \, \rangle \text{\ with\ } n \ge 0 \text{\ even} \left( \text{from\ } Y_{- \gamma_\circ}  \right) \\
& \langle \widehat{\mathfrak{w}}_{-n (\gamma_\bullet + \gamma_\circ)} \rangle \text{\ with\ } n>0
\end{cases}
\end{equation}
The third line is redundant, but we include it to clarify that the dyonic operators of $SU(2)$ are always present

\underline{$SO(3)_-$}.

Now the remaining case of $SO(3)_+$ is the one with core charges  $\frac12 \gamma_\bullet$ and $ -\gamma_\circ$. They correspond to the two quivers
\begin{equation}
\xymatrix@C=8mm{  & \framebox{$- \gamma_\circ $} & \\
 \gamma_\circ \ar@<-0.5ex>[rr]  \ar@<0.5ex>[rr]  & &  \gamma_\bullet  \ar@{..>}[ul]_2
 }
\qquad
\xymatrix@C=8mm{  &\framebox{$ \frac12 \gamma_\bullet$} & \\
 \gamma_\circ \ar@{..>}[ur]  \ar@<-0.5ex>[rr]  \ar@<0.5ex>[rr]  & &  \gamma_\bullet  
 }
\end{equation}
and v.e.v.'s
\begin{align} \label{SO3mvev}
\langle \, \widehat{\mathfrak{L}}_{\frac12 \gamma_\bullet}  \, \rangle = & Y_{\frac12 \gamma_\bullet} = Y_{\gamma_\bullet}^{1/2} \,  , \\
\langle \, \widehat{\mathfrak{L}}_{- \gamma_\circ} \, \rangle = & 1/ Y_{\gamma_\circ} \, .
\end{align}
Once again the only consistent Wilson lines have charges $- n (\gamma_\circ + \gamma_\bullet)$. As before applying the half-monodromy we find the consistent set of line operators
\begin{align} \label{dyonSO3m}
\langle \, \widehat{\mathfrak{L}}_{-\frac12 \gamma_\bullet} \, \rangle = & \frac{\left(1+Y_{\gamma_\circ}+2 Y_{\gamma_\bullet} Y_{\gamma_\circ}+Y_{\gamma_\bullet}^2 Y_{\gamma_\circ}\right)}{Y_{\gamma_\bullet}^{1/2}} \, , \cr
\langle \, \widehat{\mathfrak{L}}_{\frac12 \gamma_\bullet} \, \rangle = & Y_{\gamma_\bullet}^{1/2} \, , \cr
\langle \, \widehat{\mathfrak{L}}_{-\frac12 \gamma_\bullet -  \gamma_\circ} \, \rangle = & \frac{(1+Y_{\gamma_\circ})}{Y_{\gamma_\bullet}^{1/2} Y_{\gamma_\circ}} \, , \cr
\langle \, \widehat{\mathfrak{L}}_{-\frac32 \gamma_\bullet - 2 \gamma_\circ} \, \rangle = & \frac{\left(1+3 Y_{\gamma_\circ}+3 Y_{\gamma_\circ}^2+2 Y_{\gamma_\bullet} Y_{\gamma_\circ}^2+Y_{\gamma_\circ}^3+2 Y_{\gamma_\bullet} Y_{\gamma_\circ}^3+Y_{\gamma_\bullet}^2 Y_{\gamma_\circ}^3\right)}{Y_{\gamma_\bullet}^{3/2} Y_{\gamma_\circ}^2} \, ,
\end{align}
and
\begin{align}
\langle \, \widehat{\mathfrak{L}}_{-2 \gamma_\bullet - \gamma_\circ} \, \rangle = & \frac{\left(1+2 Y_{\gamma_\circ}+2 Y_{\gamma_\bullet} Y_{\gamma_\circ}+Y_{\gamma_\circ}^2+3 Y_{\gamma_\bullet} Y_{\gamma_\circ}^2+3 Y_{\gamma_\bullet}^2 Y_{\gamma_\circ}^2+Y_{\gamma_\bullet}^3 Y_{\gamma_\circ}^2\right)^2}{Y_{\gamma_\bullet}^2 Y_{\gamma_\circ}} \, , \cr
\langle \, \widehat{\mathfrak{L}}_{\gamma_\circ}  \, \rangle = & (1+Y_{\gamma_\bullet})^2 Y_{\gamma_\circ} \, , \cr
\langle \, \widehat{\mathfrak{L}}_{-\gamma_\circ} \, \rangle = & \frac{1}{Y_{\gamma_\circ}} \, , \cr
\langle \, \widehat{\mathfrak{L}}_{-2 \gamma_\bullet - 3 \gamma_\circ} \, \rangle = & \frac{\left(1+2 Y_{\gamma_\circ}+Y_{\gamma_\circ}^2+Y_{\gamma_\bullet} Y_{\gamma_\circ}^2\right)^2}{Y_{\gamma_\bullet}^2 Y_{\gamma_\circ}^3} \, .
\end{align}
As before we can generate another family of operators starting from $Y_{\gamma_\bullet}$, which would give the square of (\ref{dyonSO3m}). Summarizing the set of line operators obtained in this way is
\begin{equation}
\begin{cases}
& \langle \, \widehat{\mathfrak{L}}_{-(2n-1)/2 \gamma_\bullet - n \gamma_\circ} \, \rangle \ , \langle \, \widehat{\mathfrak{L}}_{-(2n+1)/2 \gamma_\bullet - n \gamma_\circ} \, \rangle  \text{ \ with \ } n\ge 0 \left( \text{\ from \  }  Y_{\frac12 \gamma_\bullet}  \right)\\
& \langle \, \widehat{\mathfrak{L}}_{-(n-1) \gamma_\bullet -n \gamma_\circ} \, \rangle \text{\ with\ } n \ge 0 \text{\ odd and\ } \langle \, \widehat{\mathfrak{L}}_{- n \gamma_\bullet - (n-1) \gamma_\circ} \, \rangle \text{\ with\ } n \ge 0 \text{\ even} \left( \text{from\ } Y_{ -\gamma_\circ} \right) \\
& \langle \, \widehat{\mathfrak{L}}_{-(n-1) \gamma_\bullet -n \gamma_\circ} \, \rangle \text{\ with\ } n \ge 0 \text{\ even and\ } \langle \, \widehat{\mathfrak{L}}_{- n \gamma_\bullet - (n-1) \gamma_\circ} \, \rangle \text{\ with\ } n \ge 0 \text{\ odd} \left( \text{from\ } Y_{ \gamma_\bullet} \right) \\
& \langle \, \widehat{\mathfrak{w}}_{-n (\gamma_\bullet + \gamma_\circ)} \, \rangle \text{\ with\ } n>0
\end{cases}
\end{equation}
Again the third line is redundant.


Note that if we apply the $1/4$-fractional monodromy to an operator in the first line, we would get precisely an operator in $SO(3)_+$ (simply because the generators of these families are mapped into each other). The fractional monodromy is related to the $\mathbb{Z}_4$ R-symmetry, which is not a symmetry anymore, and its action is compensated by $\theta \longrightarrow \theta + 2 \pi$ (while the physical invariance is now $\theta \longrightarrow \theta + 4 \pi$). This operation exchages the line defects of $SO(3)_+$ with those of $SO(3)_-$: in other words what we are seeing is the statement that $SO(3)_+^{\theta} = SO(3)_-^{\theta + 2 \pi}$ as noted in \cite{Aharony:2013hda}.

%
%
%
%

\section{$Q$-systems and conserved charges}  \label{Qsystems}

Now we develop a more general approach to study Wilson lines in asymptotically free theories. We will mostly focus on $SU(N)$ theories and their respective line defects. For these class of theories the finer fractional monodromy, in the $q \longrightarrow +1$ limit, is a certain rational transformation which coincides with the evolution of a certain integrable system, the so-called $Q$-system. In this Section we will introduce $Q$-systems from the point of view of cluster algebras, and their relation with $\mathcal{N}=2$ models \cite{Cecotti:2014zga,Kedem:2007zz,kedem2}. We will also discuss the conserved charged of these systems which we will identify as the Wilson lines of the gauge theories. We will then discuss an explicit example.

\subsection{$Q$-systems and cluster algebras} \label{sistemidelQ}

The  $A_r$ $Q$-system is defined in terms of the family of commutative variables $\{ \mathcal{Q}_{\alpha,n} \, : \, \alpha \in I_r , n \in \mathbb{Z} \}$, with $I_r = {1 , \dots , r}$ together with the recursion relation
\begin{equation} \label{Qrelations}
\mathcal{Q}_{\alpha , n+1} \, \mathcal{Q}_{\alpha , n-1} = \mathcal{Q}_{\alpha , n}^2 - \mathcal{Q}_{\alpha + 1 , n} \, \mathcal{Q}_{\alpha-1,n} \ , \qquad \mathcal{Q}_{0,n} = \mathcal{Q}_{r+1,n} = 1 \ , \  ( \alpha \in I_r \ , n \in \mathbb{Z} ) \ .
\end{equation}
In this case the variables $\mathcal{Q}_{\alpha , n}$ are related to the $A_r$ Kirillov-Reshetikhin modules \cite{KR}, by imposing the boundary conditions
\begin{equation}
\mathcal{Q}_{\alpha,0} = 1 \ , \qquad \mathcal{Q}_{\alpha,1} = \mathrm{ch} \, V (\omega_{\alpha}) \, , \ \ \alpha \in I_r  \ .
\end{equation}
In the original applications of the $Q$-systems, $\mathcal{Q}_{\alpha, 1}$ is the character of the fundamental Kirillov-Reshetikhin module $V (\omega_{\alpha})$, the fundamental representation of $\frak{sl}_{r+1}$ associated with the highest weight $\omega_{\alpha}$. The relation between $Q$-systems and cluster algebras becomes apparent by relaxing this boundary condition and passing to the following variables
\begin{equation}
\mathcal{R}_{\alpha , n} = \epsilon_\alpha \, \mathcal{Q}_{\alpha , n} \ , \qquad \epsilon_{\alpha} = \e^{i \pi \alpha (r+1-\alpha)/2} \ .
\end{equation}
These variables obey a recursion relation given by
\begin{equation} \label{Rrelations}
\mathcal{R}_{\alpha , n+1} \, \mathcal{R}_{\alpha , n-1} = \mathcal{R}^2_{\alpha , n} + \mathcal{R}_{\alpha+1 , n} \, \mathcal{R}_{\alpha-1 , n} \ , \qquad \mathcal{R}_{0,n} = \mathcal{R}_{r+1 , n} = 1 \ , \qquad  ( \alpha \in I_r \ , n \in \mathbb{Z} ) \ .
\end{equation}
These relations are precisely the exchange relations of cluster algebras. Introduce the seed $\left( {\mathbf{x}} , B \right)$ with
\begin{equation}
\mathbf{x} = \left( \mathcal{R}_{1,0} , \dots , \mathcal{R}_{r,0} ; \mathcal{R}_{1,1} , \dots , \mathcal{R}_{r,1} \right) \, , \qquad B = \left( \begin{matrix} 0 & -C \\ C & 0 \end{matrix}  \right)
\end{equation}
where $C$ is the Cartan matrix of $A_r$.

Time evolution for the $Q$-system takes the form of sequence of mutations of the seed $(\mathbf{x} , B)$. As usual we label mutation by a discrete time variable $t$. Then the recursion relations for the $Q$-system (\ref{Rrelations}) have the form
\begin{equation} \label{xcluster}
x_j [t+1] = x_{j,t+1} = \mu_k (x_{j,t} ) = \left\{ \begin{matrix}
x^{-1}_{j,t} \left( \prod_{i=1}^{2r} \, x_{i,t}^{[B_{ij}^{(t)}]_+} + \prod_{i=1}^{2r} x_{i,t}^{[-B_{ij}^{(t)}]_+} \right) \ , & k=j \\
x_{j,t} & \text{otherwise}
\end{matrix} \right.
\end{equation}
with the identification $x_{\alpha} [t] = \mathcal{R}_{\alpha , 2 t}$ and $x_{\alpha + r} [t] = \mathcal{R}_{\alpha , 2 t +1}$. Similarly the adjacency matrix $B$ evolves in time with 
\begin{equation} \label{mutB}
B'_{ij} = \left\{ \begin{matrix} - B_{ij} & \text{if} \ i=k  \ \text{or} \ j=k \\ B_{ij} + \text{sgn} (B_{ik}) [ B_{ik} \, B_{kj} ]_+ & \text{otherwise} \end{matrix} \right. \ .
\end{equation}
Note that since mutations have inverse, the $Q$-system can evolve forward or backward in time. After a sequence of mutations $\mu_r  \cdots  \mu_1  \, (\mathbf{x} , B)$ or $\mu_{2r}  \cdots  \mu_{r+1} \, (\mathbf{x} , B)$, \textit{all} the variables $\mathcal{R}_{\alpha , n}$ will have evolved forward or backward in time, and the generic seed
\begin{equation} \label{Rseed}
\mathbf{x} = \left( \mathcal{R}_{1,2t} , \dots , \mathcal{R}_{r,2t} ; \mathcal{R}_{1,2t+1} , \dots , \mathcal{R}_{r,2t+1} \right) \ ,  \qquad B = \left( \begin{matrix} 0 & -C \\ C & 0 \end{matrix}  \right) \, ,
\end{equation}
will have evolved to one of the seeds
\begin{eqnarray}
\mathbf{x}' = \left( \mathcal{R}_{1,2t+2} , \dots , \mathcal{R}_{r,2t+2} ; \mathcal{R}_{1,2t+1} , \dots ,  \mathcal{R}_{r,2t+1} \right) \ ,  \qquad B' = \left( \begin{matrix} 0 & C \\ -C & 0 \end{matrix}  \right) \, , \\
\mathbf{x}'' = \left( \mathcal{R}_{1,2t} , \dots , \mathcal{R}_{r,2t} ; \mathcal{R}_{1,2t-1} , \dots ,  \mathcal{R}_{r,2t-1} \right) \ ,  \qquad B'' = \left( \begin{matrix} 0 & C \\ -C & 0 \end{matrix}  \right) \, ,
\end{eqnarray}
respectively \cite[Lemmata 3.6,3.8]{Kedem:2007zz}. Note that the adjacency matrix has changed sign. 

We can express the $Q$-system in terms of the $Y$-variables at the same instant $t$; if we keep track of the ``time" variable by writing $Y_{i,t}$ for the mutated $Y$-seed, we can write
\begin{equation} \label{y-variables}
Y_{i,t} = \prod_i x_{i,t}^{B_{ij}^{(t)}} \, ,
\end{equation}
and the transformations (\ref{xcluster}) coincide now with the $Y$-mutations \eqref{clusterR}. Therefore the time evolution of the $Q$-system corresponds precisely to the fractional monodromy associated with the sequence of mutations $\mathbf{r} = \mu_{r}   \cdots  \mu_1$ and discussed in Section \ref{rationaltransf}. In the following we will usually drop the discrete time dependence.

\subsection{Conserved Charges} \label{conserved}

Consider now a certain line defect $\mathfrak{L}$ with core charge $\alpha$. Assume that the framed quiver is invariant under the finest fractional monodromy transformations, that is under a sequence of mutations whose iteration generates the vanilla BPS spectrum in a certain chamber, eventually up to a permutation. We will argue that this is the case for the Wilson lines. The action of the sequence of mutations on the framed quiver  lifts to an action on the line defect vevs $\langle \, \widehat{\mathfrak{L}}_{\zeta , \alpha} \, \rangle$. In particular we have argued that if a sequence of mutations corresponding to the fractional monodromy maps a framed quiver into another, then the corresponding line defects are related by the corresponding rational transformations $R^{\pm}$ \eqref{Rpm-transf}. On the other hand if a framed quiver is invariant under the sequence of mutations, this translates into a condition for the corresponding line defect generating function $\langle \widehat{\mathfrak{L}}_{\zeta , \alpha} \rangle$: \textit{it must be a fixed point of a series of cluster transformations}. Note that in these arguments one must keep explicitly track of any permutation in the nodes of the quiver.

It is easy to see why Wilson lines should correspond to conserved charges from a graphical point of view. Consider for example a Wilson line given by the following framed BPS quiver
\begin{equation}
\xymatrix@C=8mm{
\framebox{$ f_{\mathbf{N}}$} \ar@{..>}[r] & \circ_1  \ar@<-0.5ex>[d]  \ar@<0.5ex>[d] & & \bullet_2 \ar[rr] \ar[ll] & & \cdots \ar[rr] & & \circ_{N-1} \ar@<-0.5ex>[d]  \ar@<0.5ex>[d] \\
& \ar@{..>}[ul]  \bullet_1 \ar[rr] & & \circ_2  \ar@<-0.5ex>[u]  \ar@<0.5ex>[u] & & \cdots \ar[ll] & & \bullet_{N-1} \ar[ll]
 } \, .
\end{equation}
The unframed BPS quiver describes $\mathfrak{su}_N$ SYM and the framing a Wilson line corresponding to one of the fundamental representations. We would like to show that such a framed quiver is invariant under the sequence of quiver mutations which generates the spectrum in the strong coupling chamber, the finest fractional monodromy. But this simply follows from the fact that the coupling of the fundamental Wilson lines is \textit{local} in the quiver, that is it involves only a Kronecker sub-quiver. Indeed for this class of quivers the sequence of mutations which generates the strong coupling spectrum is such that two adjacent mutations involve nodes with mutually local charges. As a consequence we can analyze separately what happens to the Kronecker sub-quiver involving the framing and what happens to the remaining of the quiver. But we know that both will return to themselves after the sequence of mutations: the framed Kronecker sub-quiver, because we have showed it in Section \ref{SU2-Wilson-lines}, and the rest of the quiver, by construction. Note that this argument involves also the superpotential. 

Assume that a certain line defect $ \langle \, \widehat{\mathfrak{L}}_{\zeta , f} \, \rangle$ is fixed by a  fractional monodromy operator. Using the identification between cluster algebras and $Q$-systems outlined in Section \ref{sistemidelQ}, this translates into the fact that the line defect generating function is invariant under the evolution of the $Q$-system. In other words it is a constant of motion. Therefore we conclude that constructing such defects in supersymmetric quantum field theory is equivalent to the determination of the constants of motions of the associated $Q$-system. There is however a slight difference between the evolution of the $Q$-system and the finest fractional monodromy, as the latter may involve a permutation. When this happens the rational transformation corresponding to the finest fractional monodromy will simply permute the conserved charges of the $Q$-system between themselves. We conclude that we can identify the Wilson line defects with the conserved charges of the $Q$-system.

In the next Sections we will show several examples of this phenomenon. The problem of computing the framed BPS spectra for Wilson lines is reduced to finding the conserved charges of a discrete integrable system. This problem was solved in \cite{kedem2} for the case of $A_r$ as follows. One begins by constructing the matrix whose entries are $(M_{\alpha , n})_{i,j} = \mathcal{R}_{1,n+i+j-1-\alpha}$; explicitly
\begin{equation} \label{matrixM}
M_{\alpha , n}= \left(
\begin{matrix}
\mathcal{R}_{1,n-\alpha+1} && \mathcal{R}_{1,n-\alpha +2} && \cdots && \mathcal{R}_{1,n} \\
\mathcal{R}_{1,n-\alpha+2} && \mathcal{R}_{1,n-\alpha +3} && \cdots && \mathcal{R}_{1,n+1} \\
\vdots && \vdots && \ddots && \vdots  \\
\mathcal{R}_{1,n} && \mathcal{R}_{1,n+1} && \cdots && \mathcal{R}_{1,n+\alpha-1}
\end{matrix}
\right) \ .
\end{equation}
Define the discrete Wronskian determinant $W_{\alpha,n} = \det M_{\alpha , n}$. Then one can show that $\mathcal{R}_{\alpha , n} = W_{\alpha , n}$; the rationale behind the argument is that the Pl\"ucker relations between the minors of a higher rank version of (\ref{matrixM}) coincide with the relations (\ref{Rrelations}). 

The boundary condition $\mathcal{R}_{r+1,n} =1$ therefore implies that the determinant $W_{r+1,n}=1$ and in particular is a conserved quantity, that is independent on $n$. This boundary condition, together with the relations (\ref{Rrelations}) and the fact that $W_{r+2,n} = 0$, implies that the minors
\begin{equation} \label{conservedminors}
c_i = \det (M_{r+2,n})^{r+2-i}_{r+2} \ \qquad i= 0,\dots , r \, ,
\end{equation}
which are polynomials in the variables $\mathcal{R}_{\alpha,n}$,  are conserved quantities, independent on the time variable $n$. The notation $M_{i_1,\dots,i_k}^{j_1,\dots,j_l}$ means that the rows $i_1 , \dots , i_k$ and the columns $j_1 , \dots , j_l$ have been removed from the matrix $M$. The boundary conditions imply that $c_0 = c_{r+1} = 1$. The other constants $\{ c_1 , \dots , c_r \}$ are non trivial polynomials, independent on $n$ and completely fixed by the initial data. They furthermore determine the linear recursion relation for the variable $\{ \mathcal{R}_{1,n} \}_{n \in \mathbb{Z}}$
\begin{equation} \label{conservedrecursion}
\sum_{m=0}^{r+1} (-1)^m \, c_{r+1-m} \, \mathcal{R}_{1 , n+m} = 0 \ , \qquad n \in \mathbb{Z} \, .
\end{equation}

As we have mentioned before, Wilson line defects will correspond to conserved charges of the $Q$-system. We have just outlined an algorithm to find $r$ conserved charges. Of course, sum and products of conserved charges are still conserved. Therefore any Wilson line defect vev can be generated from the set of $r$ ``elementary" conserved charges of the $Q$-system, by multiplication and sum. These abstract ideas have a very concrete physical counterpart. Indeed let us consider a certain representation $\mathbf{k}$ of a simply laced group $G$. Our arguments concerning the conserved charges of the $Q$-systems were special to $A$-type Lie algebras, but we believe these ideas are more general. We would like to study the generating function $\langle \, \widehat{\mathfrak{w}}_{\zeta}^{\mathbf{k}} \, \rangle$ of a Wilson line in the representation $\mathbf{k}$ in full generality, but we can still learn a lot about them by thinking first about their classical limit first. Classically all quantum effects are suppressed and the Wilson lines literally correspond to the classical holonomy in the representation $\mathbf{k}$. By going to the Coulomb branch the gauge group $G$ is reduced to its maximal torus $U(1)^r$ and any irreducible representation $\mathbf{k}$ of $G$ decomposes into sums of $U(1)$ irreducible representations (with precisely $\text{dim} \ \mathbf{k}$ factors). Similarly the representation space of $\mathbf{k}$, $V$, decomposes into its weight spaces $V = \bigoplus_{w} \, V_w$, with $w \in \Lambda_w$ an element of the weight lattice corresponding to the representation $\mathbf{k}$. Note that the core charge is directly given by the highest weight. Therefore in the semi-classical approximation we can immediately write
\begin{equation} \label{Wexpweights}
\langle \, \widehat{\mathfrak{w}}_{\zeta}^{\mathbf{k}}  \, \rangle  = \sum_{\mathbf{w} \, \text{weight}} \ \underline{\overline{\Omega}} (\gamma ; \mathfrak{w}_\zeta^{\mathbf{k}}) \ Y_{\mathbf{e} \cdot B^{-1} \cdot \mathbf{w}} + \text{quantum effects} \, ,
\end{equation}
where the sum is over the weights of the representation $\mathbf{k}$, and we have expressed the charges $\gamma$ in terms of the weight vectors $\mathbf{w}$ using the adjacency matrix $B$. Here $\mathbf{e}$ is the standard basis of $\mathbb{R}^{\text{dim} \, \mathbf{k}}$. The degeneracies  $\underline{\overline{\Omega}} (\gamma  ; \mathfrak{w}_\zeta^{\mathbf{k}})$ can now be interpreted as the multiplicities in the weight decomposition. Physically these terms are color states of the core charge. Note that we can also turn our arguments the other way around and predict that every conserved charge of an $ADE$ $Q$-system will contain terms associated with the weights of an irreducible representation of the corresponding simply laced group $G$. 

Our general formalism predicts that these statements hold quantum mechanically. Consider the $r$ conserved charges of the $Q$-system constructed above $\{ c_1 , \dots , c_r \}$. In the next Sections we can will identify them with Wilson lines explicitly, by identifying the lowed order monomial with the highest weight of an irreducible representation of $G$. We will see that these operators correspond to the irreducible representations associated to the fundamental weights of $G$.
Assume that we have two conserved charges $c_i$ and $c_j$ computed as above and let us multiply them together. We find a result of the form
\begin{equation} \label{Cope}
c_i \cdot c_j = \sum_{k} N_{ij}^k \, \tilde{c}_k \, ,
\end{equation}
where the charges $\tilde{c}_k$ are conserved, though a priori not elements of the set $\{ c_1 , \dots , c_r \}$. At the semi-classical level this OPE is just the decomposition of the tensor product of two representations into irreducible representations, a fact that is physically clear from the weak coupling description of the Wilson line operators. But the result is much stronger: it holds for the full \textit{quantum} generating functions of framed degeneracies. 

In principle the identification of $\tilde{c}_k$ follows by restricting attention to simple line defects, that is line defects which cannot be written as sum of other line defects. But how do we proceed in practice? We can do so by recalling that line defects are identified by their core charge, by looking at the monomial whose overall power is the lowest. Multiplying the two conserved charges $c_i \cdot c_j $ together, allows us to predict the lowest core charge among all the line operators appearing on the right hand side of (\ref{Cope}), which simply results by direct multiplication of the two monomials corresponding to the core charges of $c_i$ and $c_j$ respectively. Therefore, since the electro-magnetic charge is quantized, the conserved charges are automatically ordered. Finally to identify the conserved charges on the right hand side of (\ref{Cope}), we simply have to multiply the conserved charges on the left hand side in the appropriate order, making sure every time that there is only a new term in the right hand side of (\ref{Cope}), and that the conserved charges labelled by a lower core charge have already been computed. In this way, simply by inspection of the right hand side of (\ref{Cope}), we can identify all the new conserved charges. Note that this argument strictly holds only at a \textit{fixed} point in the Coulomb branch, since crossing an anti-wall would change the core charge.

Note that in this case the concept of a conserved charge is somewhat ambiguous, as any constant term will be conserved and could be added. However physically no such term is possible, since a neutral particle will not have any electro-magnetic interaction with the core charge and therefore will not form a bound state, unless the constant term is forced upon us by gauge invariance: the irreducible representation contains the zero weight in its decomposition. Therefore the decomposition \eqref{Cope} is unambiguous and perfectly determined by the core charges of the defects, or highest weights of their irreducible representations in the ultraviolet. 

Let us summarize our strategy. We will consider BPS quivers framed by Wilson line defects. We will then check explicitly that the quivers are invariant under the smallest fractional monodromy, including explicitly the corresponding permutation if necessary. Then we will compute the conserved charges of the corresponding Q-system and identify them with the Wilson lines using the core charge. Since we will start from the smallest core charges available, which is the highest weight of the corresponding representation, we will only deal with simple line defects and the prescription is unambiguous.

\subsection{An example: $\mathfrak{su}_2$ revisited} \label{SU2-rev}

To exemplify this formalism, let us go back to the case of $\mathfrak{su}_2$ SYM discussed in Section \ref{su2onceandforall}, and consider a Wilson line in the fundamental representation $\mathfrak{w}^\mathbf{2}_\zeta$. As we have remarked in Section \ref{conserved}, this is precisely the case where we can argue that a line defect is given by a conserved charge of the associated $Q$-system. In particular this gives us a systematic approach to compute the framed spectrum. In this case the $Q$-system has seed $((\mathcal{R}_{1,0};\mathcal{R}_{1,1}), B_{SU(2)})$, and there is only one conserved charge, which has the form
\begin{equation}
c_1 = \det(M_{3,n})_3^2 = \mathcal{R}_{1,n-2} \, \mathcal{R}_{1,n+1} - \mathcal{R}_{1,n} \, \mathcal{R}_{1,n-1} \, .
\end{equation}
Now we can use the $Q$-system relations (\ref{Rrelations}), to express this charge in terms of the original seed variables:
\begin{equation}
c_1 = \frac{\mathcal{R}_{1,1}}{\mathcal{R}_{1,0}} + \frac{1}{\mathcal{R}_{1,0} \, \mathcal{R}_{1,1}} + \frac{\mathcal{R}_{1,0}}{\mathcal{R}_{1,1}} \, .
\end{equation}
Finally the latter can be expressed in term of the cluster variables $Y_{\circ}$ and $Y_\bullet$ using (\ref{y-variables}):
\begin{eqnarray}
\mathcal{R}_{1,0} &=& Y_{\circ}^{-1/2} \, , \\
\mathcal{R}_{1,1} &=& Y_{\bullet}^{1/2} \, ,
\end{eqnarray}
where we have used our standard notation $Y_{\circ}$. Finally the corresponding line defect is
\begin{equation} \label{wsu22vev}
\langle \widehat{\mathfrak{w}}_{\zeta}^{\mathbf{2}} \rangle= F [\widehat{\mathfrak{w}}_{\zeta}^{\mathbf{2}} , \{ Y_{\bullet} , Y_{\circ } \}] =\left[ \frac{1}{(Y_{\bullet} Y_{\circ})^{\frac12}}+ (Y_{\bullet} Y_{\circ})^{\frac12} \right] + \left( \frac{Y_{\circ} }{Y_{\bullet} } \right)^{\frac12} \, .
\end{equation}
which agrees with the results of Section \ref{su2onceandforall}.

\section{$\mathfrak{su}_3$ super Yang-Mills} \label{SU3cluster}

In this Section we will study $\mathfrak{su}_3$ Yang-Mills. This theory is not complete but still of quiver type. Its BPS spectrum is known in several chambers and the theory has an intricate pattern of walls of marginal stability \cite{Alim:2011kw}. We will provide a systematic analysis of the framed BPS spectra using our methods, firstly for Wilson lines, and then for other line defects. The quiver which describes the BPS spectrum of $\mathfrak{su}_3$ can be chosen as
\begin{equation} \label{Qsu3}
\begin{matrix}
\xymatrix@C=8mm{
\bullet_1 \ar[drr] & & \bullet_2 \ar[dll] \\
\circ_1 \ar@<-0.5ex>[u]  \ar@<0.5ex>[u]  & & \circ_2  \ar@<-0.5ex>[u]  \ar@<0.5ex>[u] 
 }
\end{matrix}
\end{equation}
in a certain duality frame. Using the sequence of mutations $\mathbf{m}^+= \mu^+_{\bullet_2} \,  \mu^+_{\bullet_1} \,  \mu^+_{\circ_2} \,  \mu^+_{\circ_1} \,  \mu^+_{\bullet_2} \,  \mu^+_{\bullet_1}$  we generate the finite spectrum $\{ \gamma_{\bullet_1} , \gamma_{\bullet_2} , \gamma_{\bullet_2} + \gamma_{\circ_1} , \gamma_{\bullet_1} + \gamma_{\circ_2} , \gamma_{\circ_2} , \gamma_{\circ_1} \}$, plus anti-particles. The system admits a $1/6$-monodromy, generated by the sequence $\mathbf{r}^+ = \mu^+_{\bullet_2} \, \mu^+_{\bullet_1}$, with permutation $\sigma = \{ (\bullet_1 , \circ_1) , (\bullet_2 , \circ_2) \}$. The existence of this monodromy is a consequence of an unbroken $\mathbb{Z}_6$ $R$-symmetry. The $R^{(+)}$ transformation associated with $\sigma^{-1} \, \mathbf{s}^+$ is given by:
\begin{equation} \label{RtopSU3}
R^{(+)} \equiv
\begin{cases}
  &Y_{\bullet_1}  \to 1/ Y_{\circ_1}
     \\
   & Y_{\bullet_2} \to  1/Y_{\circ_2}
         \\
   & Y_{\circ_1} \to  \frac{Y_{\circ_1}^2 (Y_{\circ_2}+1) Y_{\bullet_1}}{(Y_{\circ_1}+1)^2}
       \\
   & Y_{\circ_2} \to  \frac{(Y_{\circ_1}+1) Y_{\circ_2}^2 Y_{\bullet_2}}{(Y_{\circ_2}+1)^2}
      \end{cases}   \, .
\end{equation}

\subsection{Wilson lines}

Now we pass to the study of the Wilson line operators. These are characterized by purely electric core charges. Consider the two defects with core charges 
\begin{align}
\gamma_{\mathbf{3}} &= -\frac23 (\gamma_{\bullet_1} + \gamma_{\circ_1}) -  \frac13 (\gamma_{\bullet_2} + \gamma_{\circ_2}) \, \\
\gamma_{\mathbf{\bar{3}}} &= -\frac13 (\gamma_{\bullet_1} + \gamma_{\circ_1}) -  \frac23 (\gamma_{\bullet_2} + \gamma_{\circ_2}) \,
\end{align}
corresponding to the framed quivers
\begin{equation}
\begin{matrix}
\xymatrix@C=8mm{
\framebox{$\gamma_{\mathbf{3}}$} \ar@{..>}[dr]  & \ar@{..>}[l] \bullet_1 \ar[drr] & & \bullet_2 \ar[dll] \\
& \circ_1 \ar@<-0.5ex>[u]  \ar@<0.5ex>[u]  & & \circ_2  \ar@<-0.5ex>[u]  \ar@<0.5ex>[u] 
 }
\end{matrix}
\ , \ \qquad 
\begin{matrix}
\xymatrix@C=8mm{
\bullet_1 \ar[drr] & & \bullet_2 \ar[dll] \ar@{..>}[r] &  \framebox{$\gamma_{\mathbf{\bar{3}}}$} \ar@{..>}[dl] \\
\circ_1 \ar@<-0.5ex>[u]  \ar@<0.5ex>[u]  & & \circ_2  \ar@<-0.5ex>[u]  \ar@<0.5ex>[u]  & 
 }
\end{matrix} \, .
\end{equation}
It is easy to see that these framed quivers are \textit{invariant} under the action of $\mathbf{r}^+$ composed with the permutation $\sigma^{-1}$. We therefore predict that the corresponding vev's are the conserved charges of the associated $Q$-system, which we will now compute explicitly. In this case the $Q$-system has seed $( (\mathcal{R}_{1,0} , \mathcal{R}_{2,0} ; \mathcal{R}_{1,1} \, \mathcal{R}_{2,1} ), B_{SU(3)})$, with $B_{SU(3)}$ being the adjacency matrix of the quiver (\ref{Qsu3}). There are therefore two conserved charges, expressed in terms of the minors of the matrix (\ref{matrixM})
\begin{eqnarray}
c_1 &=& \det (M_{4,n})^3_4 \ , \\
c_2 &=& \det (M_{4,n})^2_4 \ .
\end{eqnarray}
Using the $Q$ systems relations (\ref{Rrelations}), these charges can be expresses in terms of the original seed variables. Finally the latter can be expressed in term of the cluster algebra $Y$-variables using 
\begin{align}
\mathcal{R}_{1,0} &= Y_{\circ_1}^{-2/3} \, Y_{\circ_2}^{-1/3} \\
\mathcal{R}_{2,0} &= Y_{\circ_1}^{-1/3} \, Y_{\circ_2}^{-2/3} \\
\mathcal{R}_{1,1} &= Y_{\bullet_1}^{2/3} \, Y_{\bullet_2}^{1/3} \\
\mathcal{R}_{2,1} &= Y_{\bullet_1}^{2/3} \, Y_{\bullet_2}^{1/3} \ .
\end{align}
The result is
\begin{align} \label{SU3-3-cluster}
\langle \, \widehat{\mathfrak{w}}_{\zeta}^{\mathbf{3}} \, \rangle = &F [ \widehat{\mathfrak{w}}_{\zeta}^{\mathbf{3}}  , \{ Y_i \} ] =
\left[ \frac{1}{Y_{\bullet_1}^{2/3} Y_{\bullet_2}^{1/3} Y_{\circ_1}^{2/3} Y_{\circ_2}^{1/3}} +\frac{Y_{\bullet_1}^{1/3} Y_{\circ_1}^{1/3}}{Y_{\bullet_2}^{1/3} Y_{\circ_2}^{1/3}}+Y_{\bullet_1}^{1/3} Y_{\bullet_2}^{2/3} Y_{\circ_1}^{1/3} Y_{\circ_2}^{2/3} \right]
\nonumber \\[4pt] &
+\frac{Y_{\circ_1}^{1/3}}{Y_{\bullet_1}^{2/3} Y_{\bullet_2}^{1/3} Y_{\circ_2}^{1/3}}+\frac{Y_{\bullet_1}^{1/3} Y_{\circ_1}^{1/3} Y_{\circ_2}^{2/3}}{Y_{\bullet_2}^{1/3}}    %
   \\
\langle \, \widehat{\mathfrak{w}}_{\zeta}^{\mathbf{\bar{3}}} \, \rangle = & F [ \widehat{\mathfrak{w}}_{\zeta}^{\mathbf{\overline{3}}}  , \{ Y_i \} ] = 
\left[ \frac{1}{Y_{\bullet_1}^{1/3} Y_{\bullet_2}^{2/3} Y_{\circ_1}^{1/3} Y_{\circ_2}^{2/3}}+\frac{Y_{\bullet_2}^{1/3} Y_{\circ_2}^{1/3}}{Y_{\bullet_1}^{1/3} Y_{\circ_1}^{1/3}} +Y_{\bullet_1}^{2/3} Y_{\bullet_2}^{1/3} Y_{\circ_1}^{2/3} Y_{\circ_2}^{1/3} \right]
 \nonumber \\[4pt] & 
+\frac{Y_{\circ_2}^{1/3}}{Y_{\bullet_1}^{1/3} Y_{\bullet_2}^{2/3} Y_{\circ_1}^{1/3}}+\frac{Y_{\bullet_2}^{1/3} Y_{\circ_1}^{2/3} Y_{\circ_2}^{1/3}}{Y_{\bullet_1}^{1/3}}
\end{align}
This vevs were first computed in \cite{Williams:2014efa}. The terms in square brackets contain the semi-classical contribution; indeed one can easily see that the charges involved correspond to the weights of the irreducible representations, as in \eqref{Wexpweights}
\begin{align}
\mathbf{3} & \rightsquigarrow \, [1,0] , [-1,1], [0,-1] \, , \\
\mathbf{\overline{3}} & \rightsquigarrow \, [0,1] , [1,-1], [-1,0] \, .
\end{align}
We label a weight $\lambda$ by its Dynkin labels, defined as $a_i =  2 (\lambda , \alpha) / (\alpha_i , \alpha_i)$ for $i=1,\dots,n$ for the $A_n$ Lie algebra, where $\alpha_i$ are the simple roots. In particular $\lambda = \sum_{i=1}^n a_i \, \omega_i$ where $\omega_i$ are the fundamental weights. We will employ this notation thoroughout this paper. 

As expected, one can easily see that the two line defects are invariant under the action of the rational transformation \eqref{RtopSU3}
\begin{align}
F [  \widehat{\mathfrak{w}}_{\zeta}^{\mathbf{3}}  , \{ Y_i \} ] &= F [  \widehat{\mathfrak{w}}_{\zeta}^{\mathbf{3}}  , R^{(+)} \{ Y_i \} ]  \, , \\
F [  \widehat{\mathfrak{w}}_{\zeta}^{\mathbf{\overline{3}}}  , \{ Y_i \} ] &= F [  \widehat{\mathfrak{w}}_{\zeta}^{\mathbf{\overline{3}}}  , R^{(+)} \{ Y_i \} ]  \, , 
\end{align}
where we have taken the permutation of the nodes into account\footnote{Note that without the permutation the rational transformation would exchange the two vev's. The reason for this is the change of variables \eqref{y-variables}. After a full evolution of the $Q$-system the adjacency matrix in \eqref{Rseed} has changed sign, corresponding of the exchange of the two Cartan submatrices. The effect of this exchange is compensated by the permutation $\sigma$ on the $Y$-seed in \eqref{y-variables}}. As we have discussed, all the conserved charges of the system can be derived by products and sums of $c_1$ and $c_2$. We can impose the OPE relations for $SU(3)$
\begin{align}
\mathfrak{w}_{\zeta}^{\mathbf{3}} *\mathfrak{w}_{\zeta}^{\mathbf{\bar{3}}} &= 1 + \mathfrak{w}_{\zeta}^{\mathbf{8}}  \\
\mathfrak{w}_{\zeta}^{\mathbf{3}}  * \mathfrak{w}_{\zeta}^{\mathbf{3}} &= \mathfrak{w}_{\zeta}^{\mathbf{\bar{3}}}  + \mathfrak{w}_{\zeta}^{\mathbf{6}}  \\
\mathfrak{w}_{\zeta}^{\mathbf{3}} * \mathfrak{w}_{\zeta}^{\mathbf{6}}  &= \mathfrak{w}_{\zeta}^{\mathbf{8}}  + \mathfrak{w}_{\zeta}^{\mathbf{10}} 
\end{align}
and obtain the operators
\begin{align} \label{SU3-6-cluster}
\langle \,  \widehat{\mathfrak{w}}_{\zeta}^{\mathbf{6}} \, \rangle = & \Big[ \frac{1}{Y_{\bullet_1}^{4/3} Y_{\bullet_2}^{2/3} Y_{\circ_1}^{4/3} Y_{\circ_2}^{2/3}}+\frac{1}{Y_{\bullet_1}^{1/3} Y_{\bullet_2}^{2/3} Y_{\circ_1}^{1/3} Y_{\circ_2}^{2/3}}+\frac{Y_{\bullet_1}^{2/3} Y_{\circ_1}^{2/3}}{Y_{\bullet_2}^{2/3} Y_{\circ_2}^{2/3}}+\frac{Y_{\bullet_2}^{1/3} Y_{\circ_2}^{1/3}}{Y_{\bullet_1}^{1/3} Y_{\circ_1}^{1/3}}
\nonumber\\[4pt] \ & 
+Y_{\bullet_1}^{2/3} Y_{\bullet_2}^{1/3} Y_{\circ_1}^{2/3} Y_{\circ_2}^{1/3}+Y_{\bullet_1}^{2/3} Y_{\bullet_2}^{4/3} Y_{\circ_1}^{2/3} Y_{\circ_2}^{4/3} \Big]
+\frac{2}{Y_{\bullet_1}^{4/3} Y_{\bullet_2}^{2/3} Y_{\circ_1}^{1/3} Y_{\circ_2}^{2/3}}+\frac{Y_{\circ_1}^{2/3}}{Y_{\bullet_1}^{4/3} Y_{\bullet_2}^{2/3} Y_{\circ_2}^{2/3}}
\nonumber\\[4pt] \ & 
+\frac{2 Y_{\circ_1}^{2/3}}{Y_{\bullet_1}^{1/3} Y_{\bullet_2}^{2/3} Y_{\circ_2}^{2/3}}+\frac{Y_{\circ_2}^{1/3}}{Y_{\bullet_1}^{1/3} Y_{\bullet_2}^{2/3} Y_{\circ_1}^{1/3}}
+\frac{2 Y_{\circ_1}^{2/3} Y_{\circ_2}^{1/3}}{Y_{\bullet_1}^{1/3} Y_{\bullet_2}^{2/3}}+\frac{2 Y_{\bullet_1}^{2/3} Y_{\circ_1}^{2/3} Y_{\circ_2}^{1/3}}{Y_{\bullet_2}^{2/3}}+\frac{Y_{\bullet_2}^{1/3} Y_{\circ_1}^{2/3} Y_{\circ_2}^{1/3}}{Y_{\bullet_1}^{1/3}}
\nonumber\\[4pt] \ & 
+\frac{Y_{\bullet_1}^{2/3} Y_{\circ_1}^{2/3} Y_{\circ_2}^{4/3}}{Y_{\bullet_2}^{2/3}}+2 Y_{\bullet_1}^{2/3} Y_{\bullet_2}^{1/3} Y_{\circ_1}^{2/3} Y_{\circ_2}^{4/3} \, ,
\\
\langle \,  \widehat{\mathfrak{w}}_{\zeta}^{\mathbf{8}} \, \rangle = &  \Big[
\frac{1}{Y_{\bullet_1} Y_{\bullet_2} Y_{\circ_1} Y_{\circ_2}}+\frac{1}{Y_{\bullet_2} Y_{\circ_2}}+\frac{1}{Y_{\bullet_1} Y_{\circ_1}}+2+Y_{\bullet_1} Y_{\circ_1}+Y_{\bullet_2} Y_{\circ_2}+Y_{\bullet_1} Y_{\bullet_2} Y_{\circ_1} Y_{\circ_2}
\Big]
\nonumber\\[4pt] \ & 
+ \frac{2}{Y_{\bullet_1}}+\frac{2}{Y_{\bullet_2}}+\frac{1}{Y_{\bullet_1} Y_{\bullet_2}}+\frac{1}{Y_{\bullet_1} Y_{\bullet_2} Y_{\circ_1}}+2 Y_{\circ_1}+\frac{Y_{\circ_1}}{Y_{\bullet_1}}+\frac{1}{Y_{\bullet_1} Y_{\bullet_2} Y_{\circ_2}}+2 Y_{\circ_2}+\frac{Y_{\circ_2}}{Y_{\bullet_2}}
\nonumber\\[4pt] \ & 
+Y_{\circ_1} Y_{\circ_2}+Y_{\bullet_1} Y_{\circ_1} Y_{\circ_2}+Y_{\bullet_2} Y_{\circ_1} Y_{\circ_2} \, ,
\\ \label{SU3-10-cluster}
\langle \,  \widehat{\mathfrak{w}}_{\zeta}^{\mathbf{10}}  \, \rangle = & \Big[
\frac{1}{Y_{\bullet_1}^2 Y_{\bullet_2} Y_{\circ_1}^2 Y_{\circ_2}}+\frac{1}{Y_{\bullet_1} Y_{\bullet_2} Y_{\circ_1} Y_{\circ_2}}+\frac{1}{Y_{\bullet_2} Y_{\circ_2}}+\frac{1}{Y_{\bullet_1} Y_{\circ_1}}+\frac{Y_{\bullet_1} Y_{\circ_1}}{Y_{\bullet_2} Y_{\circ_2}}+1+Y_{\bullet_1} Y_{\circ_1}+Y_{\bullet_2} Y_{\circ_2}
\nonumber\\[4pt] \ & 
+Y_{\bullet_1} Y_{\bullet_2} Y_{\circ_1} Y_{\circ_2}+Y_{\bullet_1} Y_{\bullet_2}^2 Y_{\circ_1} Y_{\circ_2}^2
\Big]
+ \frac{2}{Y_{\bullet_1}}+\frac{2}{Y_{\bullet_2}}+\frac{4}{Y_{\bullet_1} Y_{\bullet_2}}+\frac{1}{Y_{\bullet_1} Y_{\bullet_2} Y_{\circ_1}}+2 Y_{\circ_1}+\frac{Y_{\circ_1}}{Y_{\bullet_1}}
\nonumber\\[4pt] \ & 
+\frac{6 Y_{\circ_1}}{Y_{\bullet_2}}+\frac{3 Y_{\circ_1}}{Y_{\bullet_1} Y_{\bullet_2}}+\frac{3 Y_{\bullet_1} Y_{\circ_1}}{Y_{\bullet_2}}+\frac{3}{Y_{\bullet_1}^2 Y_{\bullet_2} Y_{\circ_2}}+\frac{4}{Y_{\bullet_1} Y_{\bullet_2} Y_{\circ_2}}+\frac{3}{Y_{\bullet_1}^2 Y_{\bullet_2} Y_{\circ_1} Y_{\circ_2}}+\frac{3 Y_{\circ_1}}{Y_{\bullet_2} Y_{\circ_2}}
\nonumber\\[4pt] \ & 
+\frac{Y_{\circ_1}}{Y_{\bullet_1}^2 Y_{\bullet_2} Y_{\circ_2}}+\frac{3 Y_{\circ_1}}{Y_{\bullet_1} Y_{\bullet_2} Y_{\circ_2}}+2 Y_{\circ_2}+\frac{Y_{\circ_2}}{Y_{\bullet_2}}+4 Y_{\circ_1} Y_{\circ_2}+4 Y_{\bullet_1} Y_{\circ_1} Y_{\circ_2}+\frac{3 Y_{\circ_1} Y_{\circ_2}}{Y_{\bullet_2}}
\nonumber\\[4pt] \ & 
+\frac{3 Y_{\bullet_1} Y_{\circ_1} Y_{\circ_2}}{Y_{\bullet_2}}+Y_{\bullet_2} Y_{\circ_1} Y_{\circ_2}+3 Y_{\bullet_1} Y_{\circ_1} Y_{\circ_2}^2+\frac{Y_{\bullet_1} Y_{\circ_1} Y_{\circ_2}^2}{Y_{\bullet_2}}+3 Y_{\bullet_1} Y_{\bullet_2} Y_{\circ_1} Y_{\circ_2}^2 \, .
\end{align}
Again the charges of terms in square brackets, when multiplied by the adjacency matrix, reproduce precisely the weights of the irreducible representations of $\mathfrak{su}_3$\footnote{These weights can be easily checked using, for example, the \textsc{mathematica} package \texttt{LieArt}.}
\begin{align}
\mathbf{6} & \rightsquigarrow \, [2,0] , [0,1], [-2,2], [1,-1], [-1,0] , [0, -2] \, ,  \\ 
\mathbf{8} & \rightsquigarrow \, [1,1] , [-1,2], [2,-1], [0,0], [0,0], [-2,1], [1,-2], [-1,-1] \, , \\
\mathbf{10} & \rightsquigarrow \, [3,0], [1,1] , [-1,2], [2,-1], [-3,3], [0,0], [-2,1], [1,-2], [-1,-1], [0,-3]  \ .
\end{align}
This results can be independently verified by a direct localization computation \cite{BPSlinesDT}.

\subsection{Dyonic defects} \label{dyonicSU3}

To generate infinite series of dyonic line defects we can start with the following set of ``elementary" line defects
\begin{align}
\langle \, \widehat{\mathfrak{L}}_{\zeta , f_{\star_1}} \, \rangle &= Y_{\circ_1}^{-2} \, Y_{\circ_2}^{-1} \ , & f_{\star_1} &= -2 \, \gamma_{\circ_1} -  \gamma_{\circ_2} \, , \\
\langle \, \widehat{\mathfrak{L}}_{\zeta , f_{\star_2}} \, \rangle &= Y_{\circ_1}^{-1} \, Y_{\circ_2}^{-2} \ , &  f_{\star_1}  &= - \, \gamma_{\circ_1} - 2 \gamma_{\circ_2} \, , \\
\langle \, \widehat{\mathfrak{L}}_{\zeta , f_{\star_3}} \, \rangle &= Y_{\bullet_1}^{2} \, Y_{\bullet_2}^{1} \ , & f_{\star_1} &= 2 \, \gamma_{\bullet_1} +  \gamma_{\bullet_2} \, ,\\
\langle \, \widehat{\mathfrak{L}}_{\zeta , f_{\star_4}} \, \rangle &= Y_{\bullet_1}^{1} \, Y_{\bullet_2}^{2} \ , & f_{\star_1} &=  \, \gamma_{\bullet_1} + 2 \gamma_{\bullet_2} \ .
\end{align}
These defects have the property that their framed quivers consists in a number of arrows from the unframed BPS quiver to the framing node, the only cyclic module being the trivial one. Each of them can be used to generate a whole cluster orbit of new line defects. For example we can start with the framed BPS quiver with core charge $f_{\star_1}$ and iterate the mutation sequence $\sigma^{-1} \mathbf{s}^+$:
\begin{equation}
\begin{gathered}
\xymatrix@C=8mm{
\framebox{$ f_{\star_1} $} & \ar@{..>}[l]_{\alpha^i} \bullet_1 \ar[rr]^{\psi} & & \circ_2 \ar@<-0.5ex>[d]_{\tilde{A}_1}  \ar@<0.5ex>[d]^{\tilde{A_2}}  \\
& \circ_1 \ar@<-0.5ex>[u]_{A_1}  \ar@<0.5ex>[u]^{A_2}  & & \bullet_2  \ar[ll]_{\phi}
 } \\ \mathcal{W}=  \tilde{A_1} \psi A_1 \phi + \tilde{A_2} \psi A_2 \phi
\end{gathered}
\begin{array}{c} \sigma \circ \mathbf{r}^+ \\ \Longrightarrow \end{array}
\begin{gathered}
\xymatrix@C=8mm{
\framebox{$ f_{\star_1}^{[1]} $} \ar@{..>}[dr]_{\beta^i}  & \ar@<0.5ex>@{..>}[l]^{\alpha_1^i} \ar@<-0.5ex>@{..>}[l]_{\alpha_2^i} \bullet_1 \ar[rr]^{\psi} & & \circ_2 \ar@<-0.5ex>[d]_{\tilde{A}_1}  \ar@<0.5ex>[d]^{\tilde{A_2}} \\
& \circ_1 \ar@<-0.5ex>[u]_{A_1}  \ar@<0.5ex>[u]^{A_2}  & & \bullet_2  \ar[ll]_{\phi}
 } \\ \mathcal{W} =\sum_{i=1}^3 ( \beta^i \alpha^i_2 A_2 + \beta^i \alpha^i_1 A_1) +\tilde{A_1} \psi A_1 \phi + \tilde{A_2} \psi A_2 \phi
 \end{gathered}
\ .
\end{equation}
wiht $i=1,2,3$. An iteration of the rational transformation \eqref{RtopSU3} produces
\begin{equation} \label{dyonSU3-f1}
\langle \, \widehat{\mathfrak{L}}_{\omega_6 \zeta , f_{\star_1}^{[1]}} \, \rangle = \frac{1}{Y_{\bullet_1}^{2} Y_{\bullet_2} Y_{\circ_1}^{4} Y_{\circ_2}^{2}} (1+Y_{\circ_1})^3
\end{equation}
which indeed can be verified by a direct localization computation \cite{BPSlinesDT}.

We can go one iteration further
\begin{equation} \label{SU3fstar1[2]}
\begin{gathered}
\xymatrix@C=20mm{
\framebox{$ f_{\star_1}^{[2]} $} \ar@<0.5ex>@{..>}[dr] \ar@<-0.5ex>@{..>}[dr]_{\beta^i_1,\beta^i_2}  &\ar@{..>}[l]\ar@<-0.3pc>@{..>}[l]\ar@<0.3pc>@{..>}[l]^{\a^i_1 , \a^i_2 ,\a^i_3}\bullet_1 \ar[rr]^{\psi} & & \circ_2 \ar@<-0.5ex>[d]_{\tilde{A_1}}  \ar@<0.5ex>[d]^{\tilde{A_2}}  \\
& \circ_1 \ar@<-0.5ex>[u]_{A_1}  \ar@<0.5ex>[u]^{A_2}  & & \bullet_2  \ar[ll]_{\phi}
 } 
\\ \mathcal{W}=  \sum_{i=1}^3 ( A_1 \beta_1^i \alpha_1^i + A_2 \beta_2^i \alpha_2^i + \alpha_3^i (A_1 \beta_2^i - A_2 \beta_1^i))  +\tilde{A_1} \psi A_1 \phi + \tilde{A_2} \psi A_2 \phi
\end{gathered}
\end{equation}
where again $i=1,2,3$ everywhere in the quiver. To see how this superpotential arise, mutate at $\bullet_1$ the framed quiver for $f_{\star_1}^{[2]}$
\begin{equation}
\xymatrix@C=20mm{
\framebox{$ f_{\star_1}^{[1]} $} \ar@<-0.5ex>@{..>}[r]\ar@<0.5ex>@{..>}[r]^{\alpha_1^{i,*} , \alpha_2^{i,*}}  &  \bullet_1 \ar@<-0.5ex>[d]_{A_1^*}  \ar@<0.5ex>[d]^{A_2^*}  & &\ar[ll]^{\psi^*} \circ_2 \ar@<-0.5ex>[d]_{\tilde{A}_1}  \ar@<0.5ex>[d]^{\tilde{A_2}} \\
&   \ar@<0.5ex>@{..>}[ul]^{[\alpha^i_1 A_1],[\alpha^i_1 A_2],[\alpha^i_2 A_1],[\alpha^i_2 A_2]} \ar@<-0.5ex>@{..>}[ul]_{\beta^{i,*}}  \circ_1   \ar@<0.3ex>[urr]  \ar@<-0.3ex>[urr] & & \bullet_2  \ar[ll]_{\phi}
 }
\end{equation}
We only consider the part of the superpotential associated with the framing node:
\begin{equation}
\mathcal{W} = \sum_{i=1}^3 \beta^i [\alpha_2^i A_2] + \beta^i [\alpha_1^i A_1] + [\alpha_1^i A_1] A_1^* \alpha_1^{i,*} + [\alpha_2^i A_1] A_1^* \alpha_2^{i,*} + [\alpha_2^i A_1] A_2^* \alpha_2^{i,*} + [\alpha_1^i A_2] A_2^* \alpha_1^{i,*} + \cdots
\end{equation}
The F-term equations for $\beta^i$ set $[\alpha_2^i A_2] = - [\alpha_1^i A_1]$, so that the superpotential becomes
\begin{equation}
\mathcal{W} = \sum_{i=1}^3  [\alpha_1^i A_1] (A_1^* \alpha_1^{i,*}-A_2^* \alpha_1^{i,*}) + [\alpha_2^i A_1] A_1^* \alpha_2^{i,*} + [\alpha_2^i A_1] A_2^* \alpha_2^{i,*} + \cdots
\end{equation}
and one further mutation at $\bullet_2$ reproduces \eqref{SU3fstar1[2]} after an appropriate relabeling of the arrows. The line defect vev can be again computed from the rational transformation \eqref{RtopSU3}:
\begin{equation} \label{dyonSU3-f2}
\langle L_{\omega_6^2 \zeta , f_{\star_1}^{[2]}} \rangle =
\frac{1}{Y_{\bullet_1}^{4} Y_{\bullet_2}^{2} Y_{\circ_1}^6 Y_{\circ_2}^3} (1+2 Y_{\circ_1}+Y_{\circ_1}^2+Y_{\bullet_1} Y_{\circ_1}^2+Y_{\bullet_1} Y_{\circ_1}^2 Y_{\circ_2})^3
\end{equation} 
and $\omega_6$ is a root of unity in $\mathbb{Z}_6$. Also this result can be checked with an explicit localization computation \cite{BPSlinesDT}. 

Similar results can be obtained, say, from $f_{\star_3}$
\begin{align}
\langle \, \widehat{\mathfrak{L}}_{ \omega_6 \zeta , f_{\star_3}^{[1]}} \, \rangle = & 
\frac{1}{Y_{\circ_1}^{2} Y_{\circ_2}} \, ,
\\
\langle \, \widehat{\mathfrak{L}}_{\omega_6^2 \zeta , f_{\star_3}^{[2]}} \, \rangle = &
 \frac{(1+Y_{\circ_1})^3}{Y_{\bullet_1}^{2} Y_{\bullet_2} Y_{\circ_1}^{4} Y_{\circ_2}^{2}} \, ,
\\
\langle \, \widehat{\mathfrak{L}}_{\omega_6^3 \zeta , f_{\star_3}^{[3]}} \, \rangle = &
\frac{(1+2 Y_{\circ_1}+Y_{\circ_1}^2+Y_{\bullet_1} Y_{\circ_1}^2+Y_{\bullet_1} Y_{\circ_1}^2 Y_{\circ_2})^3}{Y_{\bullet_1}^{4} Y_{\bullet_2}^{2} Y_{\circ_1}^6 Y_{\circ_2}^3} \, ,
\\
\langle \, \widehat{\mathfrak{L}}_{\omega_6^4 \zeta , f_{\star_3}^{[4]}}  \, \rangle = &
\frac{1}{Y_{\bullet_1}^6 Y_{\bullet_2}^3 Y_{\circ_1}^{8} Y_{\circ_2}^{4}}\Big(1+3 Y_{\circ_1}+3 Y_{\circ_1}^2+2 Y_{\bullet_1} Y_{\circ_1}^2+Y_{\circ_1}^3+2 Y_{\bullet_1} Y_{\circ_1}^3+Y_{\bullet_1}^2 Y_{\circ_1}^3
\nonumber\\[4pt] & +2 Y_{\bullet_1} Y_{\circ_1}^2 Y_{\circ_2}+2 Y_{\bullet_1} Y_{\circ_1}^3 Y_{\circ_2}+2 Y_{\bullet_1}^2 Y_{\circ_1}^3 Y_{\circ_2}+Y_{\bullet_1}^2 Y_{\circ_1}^3 Y_{\circ_2}^2+Y_{\bullet_1}^2 Y_{\bullet_2} Y_{\circ_1}^3 Y_{\circ_2}^2\Big)^3 \, .
\end{align}
Overall we get the picture
\begin{align}
\cdots \longrightarrow \langle \, \widehat{\mathfrak{L}}_{ \omega_6^{-1} \zeta , f_{\star_a}^{[-1]}} \, \rangle \longrightarrow 
 \langle \, \widehat{\mathfrak{L}}_{  \zeta , f_{\star_a}^{[0]}} \, \rangle \longrightarrow 
 \langle \, \widehat{\mathfrak{L}}_{ \omega_6^{1} \zeta , f_{\star_a}^{[1]}} \, \rangle \longrightarrow 
 \langle \, \widehat{\mathfrak{L}}_{ \omega_6^{2} \zeta , f_{\star_a}^{[2]}} \, \rangle \longrightarrow \cdots
\end{align}
for $a=1,2,3,4$, where an infinite class of line defects is organized into four cluster families, and their vevs are explicitly computed by the iteration of rational transformations.

\section{$\mathfrak{su}_4$ super Yang-Mills} \label{SU4cluster}

Now we will carry on the same calculations in the case of pure $\mathfrak{su}_4$ super Yang-Mills. Its BPS quiver is \cite{Alim:2011kw}
\begin{equation} \label{Qsu4}
\begin{matrix}
\xymatrix@C=8mm{
\circ_1  \ar@<-0.5ex>[d]  \ar@<0.5ex>[d] & & \bullet_2 \ar[rr] \ar[ll] & & \circ_3 \ar@<-0.5ex>[d]  \ar@<0.5ex>[d] \\
\bullet_1 \ar[rr] & & \circ_2  \ar@<-0.5ex>[u]  \ar@<0.5ex>[u] & & \bullet_3 \ar[ll]
 }
\end{matrix}
\end{equation}
The finite BPS spectrum with charges, in decreasing phase order,
\begin{align}
\{ \gamma_{\bullet_1}, \, \gamma_{\bullet_2} , \, \gamma_{\bullet_3} , \,\gamma_{\bullet_2}+\gamma_{\circ_1}, \, \gamma_{\bullet_1}+\gamma_{\bullet_3}+\gamma_{\circ_2}, \, \gamma_{\bullet_2}+\gamma_{\circ_3} , \\ \, \gamma_{\bullet_3}+\gamma_{\circ_2} , \, \gamma_{\bullet_2}+\gamma_{\circ_1}+\gamma_{\circ_3} , \, \gamma_{\bullet_1}+\gamma_{\circ_2} , \, \gamma_{\circ_3} , \, \gamma_{\circ_2} , \, \gamma_{\circ_1} \}
\end{align}
plus CPT conjugates, is generated by the sequence 
\begin{equation} 
\mathbf{m}^+ = \mu^+_{\circ_3} \, \mu^+_{\circ_2} \, \mu^+_{\circ_1} \, \mu^+_{\bullet_3} \, \mu^+_{\bullet_2} \, \mu^+_{\bullet_1} \, \mu^+_{\circ_3} \, \mu^+_{\circ_2} \, \mu^+_{\circ_1} \, \mu^+_{\bullet_3} \, \mu^+_{\bullet_2} \, \mu^+_{\bullet_1} \, .
\end{equation}
This is a repetition of the sequence 
\be
\mathbf{s}^+ = \mu^+_{\bullet_3} \, \mu^+_{\bullet_2} \, \mu^+_{\bullet_1}
\ee 
with permutation $\sigma = \{ (\bullet_1 , \circ_1) , (\bullet_2 , \circ_2) , (\bullet_3 , \circ_3) \} $, and the system exhibit a $1/8$ fractional monodromy. The rational transformation $R^{(+)}$ corresponding to $\sigma^{-1} \, \mathbf{s}^+$ is
\begin{equation} \label{RtopSU4}
R^{(+)} \equiv
\begin{cases}
  &Y_{\bullet_1}  \to 1/ Y_{\circ_1}
     \\[2pt]
   & Y_{\bullet_2} \to  1/Y_{\circ_2}
         \\[2pt]
    & Y_{\bullet_3} \to 1/ Y_{\circ_3}
    \\[2pt]     
   & Y_{\circ_1} \to \frac{Y_{\circ_1}^2 (1+Y_{\circ_2}) Y_{\bullet_1}}{(1+Y_{\circ_1})^2}
          \\[2pt]
   & Y_{\circ_2} \to  \frac{(1+Y_{\circ_1}) Y_{\circ_2}^2 (1+Y_{\circ_3}) Y_{\bullet_2}}{(1+Y_{\circ_2})^2}
   \\[2pt]
   & Y_{\circ_3} \to \frac{(1+Y_{\circ_2}) Y_{\circ_3}^2 Y_{\bullet_3}}{(1+Y_{\circ_3})^2}
      \end{cases}   \, .
\end{equation}

\subsection{Wilson lines}

Now we consider the $Q$-system \eqref{Rrelations} for the case of $SU(4)$. In this case the $Q$-system variables $\{ \mathcal{R}_{1,n} \}_{n \in \mathbb{Z}}$ satisfy the following recursion relation
\begin{equation}
 \mathcal{R}_{1,n}- c_3 \, \mathcal{R}_{1,1+n}+ c_2  \, \mathcal{R}_{1,2+n} - c_1 \,  \mathcal{R}_{1,3+n} + \mathcal{R}_{1,4+n} = 0 \, , \qquad n \in \mathbb{Z} \ ,
\end{equation}
where the constant of motion $c_1$, $c_2$ and $c_3$ are defined as the minors of (\ref{matrixM})
\begin{equation}
c_i = \left( M_{5} \right)_{5}^{5-i} \ , \qquad  i=1,\dots,3 \, .
\end{equation}
The constants of motion can be solved for in terms of the original seed variables 
\begin{equation} 
(\mathcal{R}_{1,0} , \dots, \mathcal{R}_{3,0} ; \mathcal{R}_{1,1} , \dots, \mathcal{R}_{3,1})
\end{equation} 
using the $Q$-system relations relations (\ref{Rrelations}).

We now express them in terms of the $Y$ seed using \eqref{y-variables}, which in this case reads
\begin{eqnarray}
\mathcal{R}_{i,0} &=& \prod_{j=1}^6 Y_{\circ_j}^{B_{j,i}}  \  i =1,2,3  \, ,\\
\mathcal{R}_{i,1} &=& \prod_{j=1}^6 Y_{\bullet_i}^{B_{j,i}} \ i=1,2,3 \, .
\end{eqnarray}
Explicitly one finds
\begin{align}
c_1 = \langle \, \widehat{\mathfrak{w}}_{\zeta}^{\mathbf{4}} \, \rangle = &
\Big[
\frac{1}{Y_{\bullet_1}^{3/4} Y_{\bullet_2}^{1/2} Y_{\bullet_3}^{1/4} Y_{\circ_1}^{3/4} Y_{\circ_2}^{1/2} Y_{\circ_3}^{1/4}}+\frac{Y_{\bullet_1}^{1/4} Y_{\circ_1}^{1/4}}{Y_{\bullet_2}^{1/2} Y_{\bullet_3}^{1/4} Y_{\circ_2}^{1/2} Y_{\circ_3}^{1/4}}+\frac{Y_{\bullet_1}^{1/4} Y_{\bullet_2}^{1/2} Y_{\circ_1}^{1/4} Y_{\circ_2}^{1/2}}{Y_{\bullet_3}^{1/4} Y_{\circ_3}^{1/4}}
\nonumber \\[4pt] &
+Y_{\bullet_1}^{1/4} Y_{\bullet_2}^{1/2} Y_{\bullet_3}^{3/4} Y_{\circ_1}^{1/4} Y_{\circ_2}^{1/2} Y_{\circ_3}^{3/4}
\Big]
+\frac{Y_{\circ_1}^{1/4}}{Y_{\bullet_1}^{3/4} Y_{\bullet_2}^{1/2} Y_{\bullet_3}^{1/4} Y_{\circ_2}^{1/2} Y_{\circ_3}^{1/4}}+\frac{Y_{\bullet_1}^{1/4} Y_{\circ_1}^{1/4} Y_{\circ_2}^{1/2}}{Y_{\bullet_2}^{1/2} Y_{\bullet_3}^{1/4} Y_{\circ_3}^{1/4}}
\nonumber \\[4pt] &
+\frac{Y_{\bullet_1}^{1/4} Y_{\bullet_2}^{1/2} Y_{\circ_1}^{1/4} Y_{\circ_2}^{1/2} Y_{\circ_3}^{3/4}}{Y_{\bullet_3}^{1/4}} \, ,
 \\
c_2 = \langle \, \widehat{\mathfrak{w}}_{\zeta}^{\mathbf{6}} \rangle = \ & 
\Big[
\frac{1}{Y_{\bullet_1}^{1/2} Y_{\bullet_2} Y_{\bullet_3}^{1/2} Y_{\circ_1}^{1/2} Y_{\circ_2} Y_{\circ_3}^{1/2}}+\frac{1}{Y_{\bullet_1}^{1/2} Y_{\bullet_3}^{1/2} Y_{\circ_1}^{1/2} Y_{\circ_3}^{1/2}}+\frac{Y_{\bullet_1}^{1/2} Y_{\circ_1}^{1/2}}{Y_{\bullet_3}^{1/2} Y_{\circ_3}^{1/2}}+\frac{Y_{\bullet_3}^{1/2} Y_{\circ_3}^{1/2}}{Y_{\bullet_1}^{1/2} Y_{\circ_1}^{1/2}}
\nonumber \\[4pt] &
+Y_{\bullet_1}^{1/2} Y_{\bullet_3}^{1/2} Y_{\circ_1}^{1/2} Y_{\circ_3}^{1/2}+Y_{\bullet_1}^{1/2} Y_{\bullet_2} Y_{\bullet_3}^{1/2} Y_{\circ_1}^{1/2} Y_{\circ_2} Y_{\circ_3}^{1/2}
\big]
+   \frac{1}{Y_{\bullet_1}^{1/2} Y_{\bullet_2} Y_{\bullet_3}^{1/2} Y_{\circ_1}^{1/2} Y_{\circ_3}^{1/2}}
\nonumber \\[4pt] &
+\frac{Y_{\circ_1}^{1/2}}{Y_{\bullet_1}^{1/2} Y_{\bullet_3}^{1/2} Y_{\circ_3}^{1/2}}+\frac{Y_{\circ_3}^{1/2}}{Y_{\bullet_1}^{1/2} Y_{\bullet_3}^{1/2} Y_{\circ_1}^{1/2}}+\frac{Y_{\circ_1}^{1/2} Y_{\circ_3}^{1/2}}{Y_{\bullet_1}^{1/2} Y_{\bullet_3}^{1/2}}+\frac{Y_{\bullet_1}^{1/2} Y_{\circ_1}^{1/2} Y_{\circ_3}^{1/2}}{Y_{\bullet_3}^{1/2}}
\nonumber \\[4pt] &
+\frac{Y_{\bullet_3}^{1/2} Y_{\circ_1}^{1/2} Y_{\circ_3}^{1/2}}{Y_{\bullet_1}^{1/2}}+Y_{\bullet_1}^{1/2} Y_{\bullet_3}^{1/2} Y_{\circ_1}^{1/2} Y_{\circ_2} Y_{\circ_3}^{1/2} \, ,
      \\
c_3 = \langle \, \widehat{\mathfrak{w}}_{\zeta}^{\mathbf{\bar{4}}} \, \rangle = \ &  
\Big[
\frac{1}{Y_{\bullet_1}^{1/4} Y_{\bullet_2}^{1/2} Y_{\bullet_3}^{3/4} Y_{\circ_1}^{1/4} Y_{\circ_2}^{1/2} Y_{\circ_3}^{3/4}}+\frac{Y_{\bullet_3}^{1/4} Y_{\circ_3}^{1/4}}{Y_{\bullet_1}^{1/4} Y_{\bullet_2}^{1/2} Y_{\circ_1}^{1/4} Y_{\circ_2}^{1/2}}+\frac{Y_{\bullet_2}^{1/2} Y_{\bullet_3}^{1/4} Y_{\circ_2}^{1/2} Y_{\circ_3}^{1/4}}{Y_{\bullet_1}^{1/4} Y_{\circ_1}^{1/4}}+
\nonumber \\[4pt] &
Y_{\bullet_1}^{3/4} Y_{\bullet_2}^{1/2} Y_{\bullet_3}^{1/4} Y_{\circ_1}^{3/4} Y_{\circ_2}^{1/2} Y_{\circ_3}^{1/4}
\Big] +
\frac{Y_{\circ_3}^{1/4}}{Y_{\bullet_1}^{1/4} Y_{\bullet_2}^{1/2} Y_{\bullet_3}^{3/4} Y_{\circ_1}^{1/4} Y_{\circ_2}^{1/2}}+\frac{Y_{\bullet_3}^{1/4} Y_{\circ_2}^{1/2} Y_{\circ_3}^{1/4}}{Y_{\bullet_1}^{1/4} Y_{\bullet_2}^{1/2} Y_{\circ_1}^{1/4}}
\nonumber \\[4pt] &
+\frac{Y_{\bullet_2}^{1/2} Y_{\bullet_3}^{1/4} Y_{\circ_1}^{3/4} Y_{\circ_2}^{1/2} Y_{\circ_3}^{1/4}}{Y_{\bullet_1}^{1/4}} \, .
\end{align}
Indeed it is easy to check that these operators are invariant under the action of the rational transformation \eqref{RtopSU4}. In square brackets are the terms corresponding to the weights of the irreducible representations of $\mathfrak{su}_4$
\begin{align}
\mathbf{\overline{4}} & \rightsquigarrow \, [0,0,1], [0,1,-1], [1,-1,0], [-1,0,0] \, ,  \\ 
\mathbf{6} & \rightsquigarrow \,[0,1,0], [1,-1,1], [-1,0,1], [1,0,-1], [-1,1,-1], [0,-1,0] \, , \\
\mathbf{4} & \rightsquigarrow \, [1,0,0], [-1,1,0], [0,-1,1], [0,0,-1] \, ,
\end{align}
which can be obtained by changing basis using the adjacency matrix $B_{ij}$. The highest weights are associated with the core charges
\begin{align}
\gamma_\mathbf{\overline{4}} &= -\frac14 (\gamma_{\bullet_1} + \gamma_{\circ_1}) - \frac12 (\gamma_{\bullet_2} + \gamma_{\circ_2}) - \frac34 (\gamma_{\bullet_3} + \gamma_{\circ_3}) \, ,  \\
\gamma_{\mathbf 6} &=  -\frac12 (\gamma_{\bullet_1} + \gamma_{\circ_1})- (\gamma_{\bullet_2} + \gamma_{\circ_2}) - \frac12 (\gamma_{\bullet_3} + \gamma_{\circ_3})  \, , \\
\gamma_{\mathbf 4} &=  -\frac34 (\gamma_{\bullet_1} + \gamma_{\circ_1}) - \frac12 (\gamma_{\bullet_2} + \gamma_{\circ_2}) - \frac14 (\gamma_{\bullet_3} + \gamma_{\circ_3}) \, ,
\end{align}
in the basis of charges. These line defects correspond to the framed quivers
\begin{equation}
\begin{gathered}
\xymatrix@C=8mm{
\circ_1  \ar@<-0.5ex>[d]  \ar@<0.5ex>[d] & & \bullet_2 \ar[rr] \ar[ll] & & \circ_3 \ar@<-0.5ex>[d]  \ar@<0.5ex>[d]  & \framebox{$\gamma_{\mathbf{\bar{4}}}$} \ar@{..>}[l]  \\
\bullet_1 \ar[rr] & & \circ_2  \ar@<-0.5ex>[u]  \ar@<0.5ex>[u] & & \bullet_3 \ar[ll] \ar@{..>}[ur] & 
 }
\ , \qquad
\xymatrix@C=8mm{
\framebox{$ \gamma_{\mathbf{4}} $} \ar@{..>}[r] & \circ_1  \ar@<-0.5ex>[d]  \ar@<0.5ex>[d] & & \bullet_2 \ar[rr] \ar[ll] & & \circ_3 \ar@<-0.5ex>[d]  \ar@<0.5ex>[d] \\
& \ar@{..>}[ul]  \bullet_1 \ar[rr] & & \circ_2  \ar@<-0.5ex>[u]  \ar@<0.5ex>[u] & & \bullet_3 \ar[ll]
 }
  \\
\xymatrix@C=8mm{
 & & & \framebox{$ \gamma_{\mathbf{6}} $} \ar@{..>}[ddl] &  \\
\circ_1  \ar@<-0.5ex>[d]  \ar@<0.5ex>[d] & & \bullet_2 \ar[rr] \ar[ll] \ar@{..>}[ur] &  & \circ_3 \ar@<-0.5ex>[d]  \ar@<0.5ex>[d] \\
\bullet_1 \ar[rr] & & \circ_2  \ar@<-0.5ex>[u]  \ar@<0.5ex>[u] & & \bullet_3 \ar[ll]
 }
\end{gathered}
\end{equation}

We can use these operators to compute the full ring of Wilson lines using their operator product expansion using the strategy outlined in Section \ref{Qsystems}. For example we can define
\be
\mathfrak{w}_{\zeta}^\mathbf{15} = \mathfrak{w}_{\zeta}^{\mathbf{\bar{4}}} * \mathfrak{w}_{\zeta}^{\mathbf{4}} - 1 \, .
\ee
The result is
\begin{align}
\langle \, \widehat{\mathfrak{w}}_{\zeta}^{\mathbf{15}} \, \rangle & = 
\Big[
\frac{1}{Y_{\bullet_1} Y_{\bullet_2} Y_{\bullet_3} Y_{\circ_1} Y_{\circ_2} Y_{\circ_3}}+\frac{1}{Y_{\bullet_2} Y_{\bullet_3} Y_{\circ_2} Y_{\circ_3}}+\frac{1}{Y_{\bullet_1} Y_{\bullet_2} Y_{\circ_1} Y_{\circ_2}}
+\frac{1}{Y_{\bullet_2} Y_{\circ_2}}
\nonumber \\[4pt] &
+\frac{1}{Y_{\bullet_3} Y_{\circ_3}}+\frac{1}{Y_{\bullet_1} Y_{\circ_1}}+3+Y_{\bullet_1} Y_{\circ_1}+Y_{\bullet_3} Y_{\circ_3}+Y_{\bullet_2} Y_{\circ_2}+Y_{\bullet_1} Y_{\bullet_2} Y_{\circ_1} Y_{\circ_2}+Y_{\bullet_2} Y_{\bullet_3} Y_{\circ_2} Y_{\circ_3}
\nonumber \\[4pt] &
+Y_{\bullet_1} Y_{\bullet_2} Y_{\bullet_3} Y_{\circ_1} Y_{\circ_2} Y_{\circ_3}
\Big]
+ \frac{2}{Y_{\bullet_1}}+\frac{2}{Y_{\bullet_2}}+\frac{1}{Y_{\bullet_1} Y_{\bullet_2}}+\frac{2}{Y_{\bullet_3}}+\frac{1}{Y_{\bullet_2} Y_{\bullet_3}}+\frac{1}{Y_{\bullet_1} Y_{\bullet_2} Y_{\circ_1}}+2 Y_{\circ_1}
\nonumber \\[4pt] &
+\frac{Y_{\circ_1}}{Y_{\bullet_1}}
+\frac{1}{Y_{\bullet_1} Y_{\bullet_2} Y_{\circ_2}}+\frac{1}{Y_{\bullet_2} Y_{\bullet_3} Y_{\circ_2}}
+\frac{1}{Y_{\bullet_1} Y_{\bullet_2} Y_{\bullet_3} Y_{\circ_2}}+\frac{1}{Y_{\bullet_1} Y_{\bullet_2} Y_{\bullet_3} Y_{\circ_1} Y_{\circ_2}}+2 Y_{\circ_2}+\frac{Y_{\circ_2}}{Y_{\bullet_2}}
\nonumber \\[4pt] &
+Y_{\circ_1} Y_{\circ_2}
+Y_{\bullet_1} Y_{\circ_1} Y_{\circ_2}
+Y_{\bullet_2} Y_{\circ_1} Y_{\circ_2}+\frac{1}{Y_{\bullet_2} Y_{\bullet_3} Y_{\circ_3}}+\frac{1}{Y_{\bullet_1} Y_{\bullet_2} Y_{\bullet_3} Y_{\circ_2} Y_{\circ_3}}+2 Y_{\circ_3}+\frac{Y_{\circ_3}}{Y_{\bullet_3}}
\nonumber \\[4pt] &
+Y_{\circ_2} Y_{\circ_3}
+Y_{\bullet_2} Y_{\circ_2} Y_{\circ_3}+Y_{\bullet_3} Y_{\circ_2} Y_{\circ_3}+Y_{\bullet_2} Y_{\circ_1} Y_{\circ_2} Y_{\circ_3}+Y_{\bullet_1} Y_{\bullet_2} Y_{\circ_1} Y_{\circ_2} Y_{\circ_3}
\nonumber \\[4pt] &
+Y_{\bullet_2} Y_{\bullet_3} Y_{\circ_1} Y_{\circ_2} Y_{\circ_3} \, .
\end{align}
As a non trivial check of our construction note that the above operator is manifestly strongly positive, and the terms in square brackets corresponds to the weights of the $\mathbf{15}$ of $\mathfrak{su}_4$.

\subsection{Dyonic defects}

As in the case of $\mathfrak{su}_3$, we can write down immediately the vevs of a set of ``elementary" line defects
\begin{align}
\langle \, \widehat{\mathfrak{L}}_{  \zeta , f_{\star_1}} \, \rangle & = Y_{\circ_1}^{-3} Y_{\circ_2}^{-2} Y_{\circ_3}^{-1} \, ,
\\[4pt]
\langle \, \widehat{\mathfrak{L}}_{  \zeta , f_{\star_2}} \, \rangle & = Y_{\circ_1}^{-2} Y_{\circ_2}^{-4} Y_{\circ_3}^{-2} \, ,
\\[4pt]
\langle \, \widehat{\mathfrak{L}}_{  \zeta , f_{\star_3}} \, \rangle & = Y_{\circ_1}^{-1} Y_{\circ_2}^{-2} Y_{\circ_3}^{-3} \, ,
\\[4pt]
\langle \, \widehat{\mathfrak{L}}_{  \zeta , f_{\star_4}} \, \rangle & = Y_{\bullet_1}^{3} Y_{\bullet_2}^{2} Y_{\bullet_3} \, ,
\\[4pt]
\langle \, \widehat{\mathfrak{L}}_{  \zeta , f_{\star_5}} \, \rangle & =Y_{\bullet_1}^{2} Y_{\bullet_2}^4 Y_{\bullet_3}^{2} \, ,
\\[4pt]
\langle \, \widehat{\mathfrak{L}}_{  \zeta , f_{\star_6}} \, \rangle & = Y_{\bullet_1} Y_{\bullet_2}^{2} Y_{\bullet_3}^{3} \, ,
\end{align}
whose framed BPS quiver has a single arrow from the BPS quiver $Q$ to the framing node, and starting from those derive whole new cluster families of defects by iterating the rational transformation $R^{(+)}$ from \eqref{RtopSU4}
\begin{align}
\cdots \longrightarrow \langle \, \widehat{\mathfrak{L}}_{ \omega^{-1} \zeta , f_{\star_a}^{[-1]}} \, \rangle \longrightarrow 
 \langle \, \widehat{\mathfrak{L}}_{  \zeta , f_{\star_a}^{[0]}} \, \rangle \longrightarrow 
 \langle \, \widehat{\mathfrak{L}}_{ \omega^{1} \zeta , f_{\star_a}^{[1]}} \, \rangle \longrightarrow 
 \langle \, \widehat{\mathfrak{L}}_{ \omega^{2} \zeta , f_{\star_a}^{[2]}} \, \rangle \longrightarrow \cdots \, ,
\end{align}
for $a=1,2,3,4,5,6$. For example
\begin{align}
\langle \, \widehat{\mathfrak{L}}_{  \zeta , f_{\star_2}^{[1]}} \, \rangle & = \frac{(1+Y_{\circ_2})^4}{Y_{\bullet_1}^2 Y_{\bullet_2}^4 Y_{\bullet_3}^2 Y_{\circ_1}^4 Y_{\circ_2}^8 Y_{\circ_3}^4} \, ,
\\
\langle \, \widehat{\mathfrak{L}}_{  \zeta , f_{\star_2}^{[2]}} \, \rangle & = \frac{\left(1+2 Y_{\circ_2}+Y_{\circ_2}^2+Y_{\bullet_2} Y_{\circ_2}^2+Y_{\bullet_2} Y_{\circ_1} Y_{\circ_2}^2+Y_{\bullet_2} Y_{\circ_2}^2 Y_{\circ_3}+Y_{\bullet_2} Y_{\circ_1} Y_{\circ_2}^2 Y_{\circ_3}\right)^4}{Y_{\bullet_1}^4 Y_{\bullet_2}^8 Y_{\bullet_3}^4 Y_{\circ_1}^6 Y_{\circ_2}^{12} Y_{\circ_3}^6} \, .
\end{align}
The structure of the cluster families for these line defects is similar to the case of $\mathfrak{su}_3$. This is not a surprise, since the elementary defects we have chosen are coupled to each Kronocker sub-quiver separately. Consider now a defect such as
\begin{equation}
\langle \, \widehat{\mathfrak{L}}_{  \zeta , f_{\diamond}} \, \rangle = \frac{Y_{\bullet_1}^2 Y_{\bullet_2}^4 Y_{\bullet_3}^2}{Y_{\circ_1}^3 Y_{\circ_2}^2 Y_{\circ_3}} \, ,
\end{equation}
which comes from the OPE $\mathfrak{L}_{\zeta , f_{\star_5}} * \mathfrak{L}_{\zeta , f_{\star_1}}$ between two line defects based on different Kronecker sub-quivers. Then it is easy to see that
\begin{align}
\langle \, \widehat{\mathfrak{L}}_{  \zeta , f_{\diamond}^{[1]}} \, \rangle & = \frac{(1+Y_{\circ_1})^4}{Y_{\bullet_1}^3 Y_{\bullet_2}^2 Y_{\bullet_3} Y_{\circ_1}^8 Y_{\circ_2}^8 Y_{\circ_3}^4} \, ,
\\
\langle \, \widehat{\mathfrak{L}}_{  \zeta , f_{\diamond}^{[2]}} \, \rangle & = \frac{(1+Y_{\circ_2})^4 \left(1+2 Y_{\circ_1}+Y_{\circ_1}^2+Y_{\bullet_1} Y_{\circ_1}^2+Y_{\bullet_1} Y_{\circ_1}^2 Y_{\circ_2}\right)^4}{Y_{\bullet_1}^8 Y_{\bullet_2}^8 Y_{\bullet_3}^4 Y_{\circ_1}^{13} Y_{\circ_2}^{14} Y_{\circ_3}^7} \, ,
\end{align}
which contain non-diagonal couplings between the two Kronecker sub-quivers.

\section{$\mathfrak{su}_5$ super Yang-Mills} \label{SU5cluster}

Our last example is pure $\mathfrak{su}_5$ super Yang-Mills. Its BPS quiver has the form
\begin{equation} \label{Qsu5}
\begin{matrix}
\xymatrix@C=8mm{
\circ_1  \ar@<-0.5ex>[d]  \ar@<0.5ex>[d] & & \bullet_2 \ar[rr] \ar[ll] & & \circ_3 \ar@<-0.5ex>[d]  \ar@<0.5ex>[d] & & \bullet_4 \ar[ll] \\
\bullet_1 \ar[rr] & & \circ_2  \ar@<-0.5ex>[u]  \ar@<0.5ex>[u] & & \bullet_3 \ar[ll] \ar[rr] & & \circ_4 \ar@<-0.5ex>[u]  \ar@<0.5ex>[u] 
 }
\end{matrix}
\end{equation}
This system has a $1/10$ fractional monodromy: five iterations of the sequence 
\be
\mathbf{s}^+ =  \mu^+_{\bullet_4} \, \mu^+_{\bullet_3} \, \mu^+_{\bullet_2} \, \mu^+_{\bullet_1} \, , 
\ee 
with permutation $\sigma = \{ (\bullet_1 , \circ_1) , (\bullet_2 , \circ_2) , (\bullet_3 , \circ_3) , (\bullet_4 , \circ_4) \} $, generate the spectrum of BPS particles, in decreasing phase order
\begin{equation}
\begin{gathered}
\{ \gamma_{\bullet_1}, \, \gamma_{\bullet_2} , \, \gamma_{\bullet_3} , \, \gamma_{\bullet_4} , \, \gamma_{\bullet_2}+\gamma_{\circ_1} , \, \gamma_{\bullet_1}+\gamma_{\bullet_3}+\gamma_{\circ_2} , \, \gamma_{\bullet_2}+\gamma_{\bullet_4}+\gamma_{\circ_3} , \, \gamma_{\bullet_3}+\gamma_{\circ_4} , \, \gamma_{\bullet_3}+\gamma_{\circ_2} , \cr 
\, \gamma_{\bullet_2}+\gamma_{\bullet_4}+\gamma_{\circ_1}+\gamma_{\circ_3} , 
\, \gamma_{\bullet_1}+\gamma_{\bullet_3}+\gamma_{\circ_2}+\gamma_{\circ_4} , \, \gamma_{\bullet_2}+\gamma_{\circ_3} , \, \gamma_{\bullet_4}+\gamma_{\circ_3} , \, \gamma_{\bullet_3}+\gamma_{\circ_2}+\gamma_{\circ_4} , 
\cr
\, \gamma_{\bullet_2}+\gamma_{\circ_1}+\gamma_{\circ_3} , \, \gamma_{\bullet_1}+\gamma_{\circ_2} , \, \gamma_{\circ_4} , \, \gamma_{\circ_3} , \, \gamma_{\circ_2} , \, \gamma_{\circ_1}
\} \, ,
\end{gathered}
\end{equation}
and the full spectrum is obtained by adding their CPT conjugates. The corresponding rational transformation $R^{(+)}$ is
\begin{equation} \label{RtopSU5}
R^{(+)} \equiv
\begin{cases}
  &Y_{\bullet_1}  \to 1/ Y_{\circ_1}
     \\[2pt]
   & Y_{\bullet_2} \to  1/Y_{\circ_2}
         \\[2pt]
    & Y_{\bullet_3} \to 1/ Y_{\circ_3}
    \\[2pt]     
    & Y_{\bullet_4} \to 1/ Y_{\circ_4}
 \\[2pt]  
   & Y_{\circ_1} \to \frac{Y_{\circ_1}^2 (1+Y_{\circ_2}) Y_{\bullet_1}}{(1+Y_{\circ_1})^2}
             \\[2pt]
   & Y_{\circ_2} \to \frac{(1+Y_{\circ_1}) Y_{\circ_2}^2 (1+Y_{\circ_3}) Y_{\bullet_2}}{(1+Y_{\circ_2})^2}
   \\[2pt]
   & Y_{\circ_3} \to \frac{(1+Y_{\circ_2}) Y_{\circ_3}^2 (1+Y_{\circ_4}) Y_{\bullet_3}}{(1+Y_{\circ_3})^2}
   \\[2pt]
   & Y_{\circ_4} \to \frac{(1+Y_{\circ_3}) Y_{\circ_4}^2 Y_{\bullet_4}}{(1+Y_{\circ_4})^2}
      \end{cases}   \, .
\end{equation}

\subsection{Wilson lines}

We begin with the Wilson line defects. The conserved charges of the $Q$-system are the minors
\begin{equation}
c_i = \left( M_{5} \right)_{6}^{6-i} \ , \qquad  i=1,\dots,4 \ ,
\end{equation}
and are associated with the recursion relations
\begin{equation}
 \mathcal{R}_{1,n} - c_4 \, \mathcal{R}_{1,1+n}+ c_3 \, \mathcal{R}_{1,2+n} - 
 c_2 \, \mathcal{R}_{1,3+n}+ c_1 \, \mathcal{R}_{1,4+n} - \mathcal{R}_{1,5+n} = 0 \, , \qquad n \in \mathbb{Z} \ .
\end{equation}
We now change basis to the $Y$-seed by using \eqref{y-variables} which in this case reads
\begin{eqnarray}
\mathcal{R}_{i,0} &=& \prod_{j=1}^8 Y_{\circ_i}^{B_{j,i}}  \ , \ i =1,2,3,4 \, , \\
\mathcal{R}_{i,1} &=& \prod_{j=1}^8 Y_{\bullet_i}^{B_{j,i}} \ , \ i=1,2,3,4 \, .
\end{eqnarray}
In this basis the constant of motions are
\begin{align}
\langle \, \widehat{\mathfrak{w}}_{\zeta}^{\mathbf{5}} \, \rangle  = & 
\Big[
\frac{1}{Y_{\bullet_1}^{4/5} Y_{\bullet_2}^{3/5} Y_{\bullet_3}^{2/5} Y_{\bullet_4}^{1/5} Y_{\circ_1}^{4/5} Y_{\circ_2}^{3/5} Y_{\circ_3}^{2/5} Y_{\circ_4}^{1/5}}+\frac{Y_{\bullet_1}^{1/5} Y_{\circ_1}^{1/5}}{Y_{\bullet_2}^{3/5} Y_{\bullet_3}^{2/5} Y_{\bullet_4}^{1/5} Y_{\circ_2}^{3/5} Y_{\circ_3}^{2/5} Y_{\circ_4}^{1/5}}
\nonumber \\[4pt] &
+\frac{Y_{\bullet_1}^{1/5} Y_{\bullet_2}^{2/5} Y_{\circ_1}^{1/5} Y_{\circ_2}^{2/5}}{Y_{\bullet_3}^{2/5} Y_{\bullet_4}^{1/5} Y_{\circ_3}^{2/5} Y_{\circ_4}^{1/5}}+\frac{Y_{\bullet_1}^{1/5} Y_{\bullet_2}^{2/5} Y_{\bullet_3}^{3/5} Y_{\circ_1}^{1/5} Y_{\circ_2}^{2/5} Y_{\circ_3}^{3/5}}{Y_{\bullet_4}^{1/5} Y_{\circ_4}^{1/5}}
\nonumber \\[4pt] &
+Y_{\bullet_1}^{1/5} Y_{\bullet_2}^{2/5} Y_{\bullet_3}^{3/5} Y_{\bullet_4}^{4/5} Y_{\circ_1}^{1/5} Y_{\circ_2}^{2/5} Y_{\circ_3}^{3/5} Y_{\circ_4}^{4/5}
\Big] +
\frac{Y_{\circ_1}^{1/5}}{Y_{\bullet_1}^{4/5} Y_{\bullet_2}^{3/5} Y_{\bullet_3}^{2/5} Y_{\bullet_4}^{1/5} Y_{\circ_2}^{3/5} Y_{\circ_3}^{2/5} Y_{\circ_4}^{1/5}}
\nonumber \\[4pt] &
+\frac{Y_{\bullet_1}^{1/5} Y_{\circ_1}^{1/5} Y_{\circ_2}^{2/5}}{Y_{\bullet_2}^{3/5} Y_{\bullet_3}^{2/5} Y_{\bullet_4}^{1/5} Y_{\circ_3}^{2/5} Y_{\circ_4}^{1/5}}+\frac{Y_{\bullet_1}^{1/5} Y_{\bullet_2}^{2/5} Y_{\circ_1}^{1/5} Y_{\circ_2}^{2/5} Y_{\circ_3}^{3/5}}{Y_{\bullet_3}^{2/5} Y_{\bullet_4}^{1/5} Y_{\circ_4}^{1/5}}
\nonumber \\[4pt] &
+\frac{Y_{\bullet_1}^{1/5} Y_{\bullet_2}^{2/5} Y_{\bullet_3}^{3/5} Y_{\circ_1}^{1/5} Y_{\circ_2}^{2/5} Y_{\circ_3}^{3/5} Y_{\circ_4}^{4/5}}{Y_{\bullet_4}^{1/5}} \, ,
\end{align}
\begin{align}
\langle \, \widehat{\mathfrak{w}}_{\zeta}^{\mathbf{\overline{5}}} \,  \rangle = & 
\Big[
\frac{1}{Y_{\bullet_1}^{1/5} Y_{\bullet_2}^{2/5} Y_{\bullet_3}^{3/5} Y_{\bullet_4}^{4/5} Y_{\circ_1}^{1/5} Y_{\circ_2}^{2/5} Y_{\circ_3}^{3/5} Y_{\circ_4}^{4/5}}+\frac{Y_{\bullet_4}^{1/5} Y_{\circ_4}^{1/5}}{Y_{\bullet_1}^{1/5} Y_{\bullet_2}^{2/5} Y_{\bullet_3}^{3/5} Y_{\circ_1}^{1/5} Y_{\circ_2}^{2/5} Y_{\circ_3}^{3/5}}
\nonumber \\[4pt] &
+\frac{Y_{\bullet_3}^{2/5} Y_{\bullet_4}^{1/5} Y_{\circ_3}^{2/5} Y_{\circ_4}^{1/5}}{Y_{\bullet_1}^{1/5} Y_{\bullet_2}^{2/5} Y_{\circ_1}^{1/5} Y_{\circ_2}^{2/5}}+\frac{Y_{\bullet_2}^{3/5} Y_{\bullet_3}^{2/5} Y_{\bullet_4}^{1/5} Y_{\circ_2}^{3/5} Y_{\circ_3}^{2/5} Y_{\circ_4}^{1/5}}{Y_{\bullet_1}^{1/5} Y_{\circ_1}^{1/5}}
\nonumber \\[4pt] &
+Y_{\bullet_1}^{4/5} Y_{\bullet_2}^{3/5} Y_{\bullet_3}^{2/5} Y_{\bullet_4}^{1/5} Y_{\circ_1}^{4/5} Y_{\circ_2}^{3/5} Y_{\circ_3}^{2/5} Y_{\circ_4}^{1/5}
\Big] + 
\frac{Y_{\circ_4}^{1/5}}{Y_{\bullet_1}^{1/5} Y_{\bullet_2}^{2/5} Y_{\bullet_3}^{3/5} Y_{\bullet_4}^{4/5} Y_{\circ_1}^{1/5} Y_{\circ_2}^{2/5} Y_{\circ_3}^{3/5}}
\nonumber \\[4pt] &
+\frac{Y_{\bullet_4}^{1/5} Y_{\circ_3}^{2/5} Y_{\circ_4}^{1/5}}{Y_{\bullet_1}^{1/5} Y_{\bullet_2}^{2/5} Y_{\bullet_3}^{3/5} Y_{\circ_1}^{1/5} Y_{\circ_2}^{2/5}}+\frac{Y_{\bullet_3}^{2/5} Y_{\bullet_4}^{1/5} Y_{\circ_2}^{3/5} Y_{\circ_3}^{2/5} Y_{\circ_4}^{1/5}}{Y_{\bullet_1}^{1/5} Y_{\bullet_2}^{2/5} Y_{\circ_1}^{1/5}}
\nonumber \\[4pt] &
+\frac{Y_{\bullet_2}^{3/5} Y_{\bullet_3}^{2/5} Y_{\bullet_4}^{1/5} Y_{\circ_1}^{4/5} Y_{\circ_2}^{3/5} Y_{\circ_3}^{2/5} Y_{\circ_4}^{1/5}}{Y_{\bullet_1}^{1/5}} \, ,
\end{align}
\begin{align}
\langle \, \widehat{\mathfrak{w}}_{\zeta}^{\mathbf{10}} \, \rangle  = & 
\Big[
\frac{1}{Y_{\bullet_1}^{3/5} Y_{\bullet_2}^{6/5} Y_{\bullet_3}^{4/5} Y_{\bullet_4}^{2/5} Y_{\circ_1}^{3/5} Y_{\circ_2}^{6/5} Y_{\circ_3}^{4/5} Y_{\circ_4}^{2/5}}+\frac{1}{Y_{\bullet_1}^{3/5} Y_{\bullet_2}^{1/5} Y_{\bullet_3}^{4/5} Y_{\bullet_4}^{2/5} Y_{\circ_1}^{3/5} Y_{\circ_2}^{1/5} Y_{\circ_3}^{4/5} Y_{\circ_4}^{2/5}}
\nonumber \\[4pt] &
+\frac{Y_{\bullet_1}^{2/5} Y_{\circ_1}^{2/5}}{Y_{\bullet_2}^{1/5} Y_{\bullet_3}^{4/5} Y_{\bullet_4}^{2/5} Y_{\circ_2}^{1/5} Y_{\circ_3}^{4/5} Y_{\circ_4}^{2/5}}+\frac{Y_{\bullet_3}^{1/5} Y_{\circ_3}^{1/5}}{Y_{\bullet_1}^{3/5} Y_{\bullet_2}^{1/5} Y_{\bullet_4}^{2/5} Y_{\circ_1}^{3/5} Y_{\circ_2}^{1/5} Y_{\circ_4}^{2/5}}
\nonumber \\[4pt] &
+\frac{Y_{\bullet_1}^{2/5} Y_{\bullet_3}^{1/5} Y_{\circ_1}^{2/5} Y_{\circ_3}^{1/5}}{Y_{\bullet_2}^{1/5} Y_{\bullet_4}^{2/5} Y_{\circ_2}^{1/5} Y_{\circ_4}^{2/5}}+\frac{Y_{\bullet_3}^{1/5} Y_{\bullet_4}^{3/5} Y_{\circ_3}^{1/5} Y_{\circ_4}^{3/5}}{Y_{\bullet_1}^{3/5} Y_{\bullet_2}^{1/5} Y_{\circ_1}^{3/5} Y_{\circ_2}^{1/5}}+\frac{Y_{\bullet_1}^{2/5} Y_{\bullet_3}^{1/5} Y_{\bullet_4}^{3/5} Y_{\circ_1}^{2/5} Y_{\circ_3}^{1/5} Y_{\circ_4}^{3/5}}{Y_{\bullet_2}^{1/5} Y_{\circ_2}^{1/5}}
\nonumber \\[4pt] &
+\frac{Y_{\bullet_1}^{2/5} Y_{\bullet_2}^{4/5} Y_{\bullet_3}^{1/5} Y_{\circ_1}^{2/5} Y_{\circ_2}^{4/5} Y_{\circ_3}^{1/5}}{Y_{\bullet_4}^{2/5} Y_{\circ_4}^{2/5}}+Y_{\bullet_1}^{2/5} Y_{\bullet_2}^{4/5} Y_{\bullet_3}^{1/5} Y_{\bullet_4}^{3/5} Y_{\circ_1}^{2/5} Y_{\circ_2}^{4/5} Y_{\circ_3}^{1/5} Y_{\circ_4}^{3/5}
\nonumber \\[4pt] &
+Y_{\bullet_1}^{2/5} Y_{\bullet_2}^{4/5} Y_{\bullet_3}^{6/5} Y_{\bullet_4}^{3/5} Y_{\circ_1}^{2/5} Y_{\circ_2}^{4/5} Y_{\circ_3}^{6/5} Y_{\circ_4}^{3/5}
\Big] + 
\frac{1}{Y_{\bullet_1}^{3/5} Y_{\bullet_2}^{6/5} Y_{\bullet_3}^{4/5} Y_{\bullet_4}^{2/5} Y_{\circ_1}^{3/5} Y_{\circ_2}^{1/5} Y_{\circ_3}^{4/5} Y_{\circ_4}^{2/5}}
\nonumber \\[4pt] &
+\frac{Y_{\circ_1}^{2/5}}{Y_{\bullet_1}^{3/5} Y_{\bullet_2}^{1/5} Y_{\bullet_3}^{4/5} Y_{\bullet_4}^{2/5} Y_{\circ_2}^{1/5} Y_{\circ_3}^{4/5} Y_{\circ_4}^{2/5}}+\frac{Y_{\circ_3}^{1/5}}{Y_{\bullet_1}^{3/5} Y_{\bullet_2}^{1/5} Y_{\bullet_3}^{4/5} Y_{\bullet_4}^{2/5} Y_{\circ_1}^{3/5} Y_{\circ_2}^{1/5} Y_{\circ_4}^{2/5}}
\nonumber \\[4pt] &
+\frac{Y_{\circ_1}^{2/5} Y_{\circ_3}^{1/5}}{Y_{\bullet_1}^{3/5} Y_{\bullet_2}^{1/5} Y_{\bullet_3}^{4/5} Y_{\bullet_4}^{2/5} Y_{\circ_2}^{1/5} Y_{\circ_4}^{2/5}}+\frac{Y_{\bullet_1}^{2/5} Y_{\circ_1}^{2/5} Y_{\circ_3}^{1/5}}{Y_{\bullet_2}^{1/5} Y_{\bullet_3}^{4/5} Y_{\bullet_4}^{2/5} Y_{\circ_2}^{1/5} Y_{\circ_4}^{2/5}}+\frac{Y_{\bullet_3}^{1/5} Y_{\circ_1}^{2/5} Y_{\circ_3}^{1/5}}{Y_{\bullet_1}^{3/5} Y_{\bullet_2}^{1/5} Y_{\bullet_4}^{2/5} Y_{\circ_2}^{1/5} Y_{\circ_4}^{2/5}}
\nonumber \\[4pt] &
+\frac{Y_{\bullet_1}^{2/5} Y_{\bullet_3}^{1/5} Y_{\circ_1}^{2/5} Y_{\circ_2}^{4/5} Y_{\circ_3}^{1/5}}{Y_{\bullet_2}^{1/5} Y_{\bullet_4}^{2/5} Y_{\circ_4}^{2/5}}+\frac{Y_{\bullet_3}^{1/5} Y_{\circ_3}^{1/5} Y_{\circ_4}^{3/5}}{Y_{\bullet_1}^{3/5} Y_{\bullet_2}^{1/5} Y_{\bullet_4}^{2/5} Y_{\circ_1}^{3/5} Y_{\circ_2}^{1/5}}+\frac{Y_{\bullet_3}^{1/5} Y_{\circ_1}^{2/5} Y_{\circ_3}^{1/5} Y_{\circ_4}^{3/5}}{Y_{\bullet_1}^{3/5} Y_{\bullet_2}^{1/5} Y_{\bullet_4}^{2/5} Y_{\circ_2}^{1/5}}
\nonumber \\[4pt] &
+\frac{Y_{\bullet_1}^{2/5} Y_{\bullet_3}^{1/5} Y_{\circ_1}^{2/5} Y_{\circ_3}^{1/5} Y_{\circ_4}^{3/5}}{Y_{\bullet_2}^{1/5} Y_{\bullet_4}^{2/5} Y_{\circ_2}^{1/5}}+\frac{Y_{\bullet_3}^{1/5} Y_{\bullet_4}^{3/5} Y_{\circ_1}^{2/5} Y_{\circ_3}^{1/5} Y_{\circ_4}^{3/5}}{Y_{\bullet_1}^{3/5} Y_{\bullet_2}^{1/5} Y_{\circ_2}^{1/5}}+\frac{Y_{\bullet_1}^{2/5} Y_{\bullet_3}^{1/5} Y_{\circ_1}^{2/5} Y_{\circ_2}^{4/5} Y_{\circ_3}^{1/5} Y_{\circ_4}^{3/5}}{Y_{\bullet_2}^{1/5} Y_{\bullet_4}^{2/5}}
\nonumber \\[4pt] &
+\frac{Y_{\bullet_1}^{2/5} Y_{\bullet_2}^{4/5} Y_{\bullet_3}^{1/5} Y_{\circ_1}^{2/5} Y_{\circ_2}^{4/5} Y_{\circ_3}^{1/5} Y_{\circ_4}^{3/5}}{Y_{\bullet_4}^{2/5}}+\frac{Y_{\bullet_1}^{2/5} Y_{\bullet_3}^{1/5} Y_{\bullet_4}^{3/5} Y_{\circ_1}^{2/5} Y_{\circ_2}^{4/5} Y_{\circ_3}^{1/5} Y_{\circ_4}^{3/5}}{Y_{\bullet_2}^{1/5}}
\nonumber \\[4pt] &
+Y_{\bullet_1}^{2/5} Y_{\bullet_2}^{4/5} Y_{\bullet_3}^{1/5} Y_{\bullet_4}^{3/5} Y_{\circ_1}^{2/5} Y_{\circ_2}^{4/5} Y_{\circ_3}^{6/5} Y_{\circ_4}^{3/5} \, ,
\end{align}
\begin{align}
\langle \, \widehat{\mathfrak{w}}_{\zeta}^{\mathbf{\overline{10}}} \, \rangle  = &
\Big[
\frac{1}{Y_{\bullet_1}^{2/5} Y_{\bullet_2}^{4/5} Y_{\bullet_3}^{6/5} Y_{\bullet_4}^{3/5} Y_{\circ_1}^{2/5} Y_{\circ_2}^{4/5} Y_{\circ_3}^{6/5} Y_{\circ_4}^{3/5}}+\frac{1}{Y_{\bullet_1}^{2/5} Y_{\bullet_2}^{4/5} Y_{\bullet_3}^{1/5} Y_{\bullet_4}^{3/5} Y_{\circ_1}^{2/5} Y_{\circ_2}^{4/5} Y_{\circ_3}^{1/5} Y_{\circ_4}^{3/5}}
\nonumber \\[4pt] &
+\frac{Y_{\bullet_4}^{2/5} Y_{\circ_4}^{2/5}}{Y_{\bullet_1}^{2/5} Y_{\bullet_2}^{4/5} Y_{\bullet_3}^{1/5} Y_{\circ_1}^{2/5} Y_{\circ_2}^{4/5} Y_{\circ_3}^{1/5}}+\frac{Y_{\bullet_2}^{1/5} Y_{\circ_2}^{1/5}}{Y_{\bullet_1}^{2/5} Y_{\bullet_3}^{1/5} Y_{\bullet_4}^{3/5} Y_{\circ_1}^{2/5} Y_{\circ_3}^{1/5} Y_{\circ_4}^{3/5}}+\frac{Y_{\bullet_1}^{3/5} Y_{\bullet_2}^{1/5} Y_{\circ_1}^{3/5} Y_{\circ_2}^{1/5}}{Y_{\bullet_3}^{1/5} Y_{\bullet_4}^{3/5} Y_{\circ_3}^{1/5} Y_{\circ_4}^{3/5}}
\nonumber \\[4pt] &
+\frac{Y_{\bullet_2}^{1/5} Y_{\bullet_4}^{2/5} Y_{\circ_2}^{1/5} Y_{\circ_4}^{2/5}}{Y_{\bullet_1}^{2/5} Y_{\bullet_3}^{1/5} Y_{\circ_1}^{2/5} Y_{\circ_3}^{1/5}}+\frac{Y_{\bullet_1}^{3/5} Y_{\bullet_2}^{1/5} Y_{\bullet_4}^{2/5} Y_{\circ_1}^{3/5} Y_{\circ_2}^{1/5} Y_{\circ_4}^{2/5}}{Y_{\bullet_3}^{1/5} Y_{\circ_3}^{1/5}}+\frac{Y_{\bullet_2}^{1/5} Y_{\bullet_3}^{4/5} Y_{\bullet_4}^{2/5} Y_{\circ_2}^{1/5} Y_{\circ_3}^{4/5} Y_{\circ_4}^{2/5}}{Y_{\bullet_1}^{2/5} Y_{\circ_1}^{2/5}}
\nonumber \\[4pt] &
+Y_{\bullet_1}^{3/5} Y_{\bullet_2}^{1/5} Y_{\bullet_3}^{4/5} Y_{\bullet_4}^{2/5} Y_{\circ_1}^{3/5} Y_{\circ_2}^{1/5} Y_{\circ_3}^{4/5} Y_{\circ_4}^{2/5}+Y_{\bullet_1}^{3/5} Y_{\bullet_2}^{6/5} Y_{\bullet_3}^{4/5} Y_{\bullet_4}^{2/5} Y_{\circ_1}^{3/5} Y_{\circ_2}^{6/5} Y_{\circ_3}^{4/5} Y_{\circ_4}^{2/5}
\Big]
\nonumber \\[4pt] &
+ \frac{1}{Y_{\bullet_1}^{2/5} Y_{\bullet_2}^{4/5} Y_{\bullet_3}^{6/5} Y_{\bullet_4}^{3/5} Y_{\circ_1}^{2/5} Y_{\circ_2}^{4/5} Y_{\circ_3}^{1/5} Y_{\circ_4}^{3/5}}+\frac{Y_{\circ_2}^{1/5}}{Y_{\bullet_1}^{2/5} Y_{\bullet_2}^{4/5} Y_{\bullet_3}^{1/5} Y_{\bullet_4}^{3/5} Y_{\circ_1}^{2/5} Y_{\circ_3}^{1/5} Y_{\circ_4}^{3/5}}
\nonumber \\[4pt] &
+\frac{Y_{\bullet_2}^{1/5} Y_{\circ_1}^{3/5} Y_{\circ_2}^{1/5}}{Y_{\bullet_1}^{2/5} Y_{\bullet_3}^{1/5} Y_{\bullet_4}^{3/5} Y_{\circ_3}^{1/5} Y_{\circ_4}^{3/5}}+\frac{Y_{\circ_4}^{2/5}}{Y_{\bullet_1}^{2/5} Y_{\bullet_2}^{4/5} Y_{\bullet_3}^{1/5} Y_{\bullet_4}^{3/5} Y_{\circ_1}^{2/5} Y_{\circ_2}^{4/5} Y_{\circ_3}^{1/5}}
\nonumber \\[4pt] &
+\frac{Y_{\circ_2}^{1/5} Y_{\circ_4}^{2/5}}{Y_{\bullet_1}^{2/5} Y_{\bullet_2}^{4/5} Y_{\bullet_3}^{1/5} Y_{\bullet_4}^{3/5} Y_{\circ_1}^{2/5} Y_{\circ_3}^{1/5}}+\frac{Y_{\bullet_2}^{1/5} Y_{\circ_2}^{1/5} Y_{\circ_4}^{2/5}}{Y_{\bullet_1}^{2/5} Y_{\bullet_3}^{1/5} Y_{\bullet_4}^{3/5} Y_{\circ_1}^{2/5} Y_{\circ_3}^{1/5}}
+\frac{Y_{\bullet_4}^{2/5} Y_{\circ_2}^{1/5} Y_{\circ_4}^{2/5}}{Y_{\bullet_1}^{2/5} Y_{\bullet_2}^{4/5} Y_{\bullet_3}^{1/5} Y_{\circ_1}^{2/5} Y_{\circ_3}^{1/5}}
\nonumber \\[4pt] &
+\frac{Y_{\bullet_2}^{1/5} Y_{\circ_1}^{3/5} Y_{\circ_2}^{1/5} Y_{\circ_4}^{2/5}}{Y_{\bullet_1}^{2/5} Y_{\bullet_3}^{1/5} Y_{\bullet_4}^{3/5} Y_{\circ_3}^{1/5}}+\frac{Y_{\bullet_1}^{3/5} Y_{\bullet_2}^{1/5} Y_{\circ_1}^{3/5} Y_{\circ_2}^{1/5} Y_{\circ_4}^{2/5}}{Y_{\bullet_3}^{1/5} Y_{\bullet_4}^{3/5} Y_{\circ_3}^{1/5}}+\frac{Y_{\bullet_2}^{1/5} Y_{\bullet_4}^{2/5} Y_{\circ_1}^{3/5} Y_{\circ_2}^{1/5} Y_{\circ_4}^{2/5}}{Y_{\bullet_1}^{2/5} Y_{\bullet_3}^{1/5} Y_{\circ_3}^{1/5}}
\nonumber \\[4pt] &
+\frac{Y_{\bullet_2}^{1/5} Y_{\bullet_4}^{2/5} Y_{\circ_2}^{1/5} Y_{\circ_3}^{4/5} Y_{\circ_4}^{2/5}}{Y_{\bullet_1}^{2/5} Y_{\bullet_3}^{1/5} Y_{\circ_1}^{2/5}}+\frac{Y_{\bullet_2}^{1/5} Y_{\bullet_4}^{2/5} Y_{\circ_1}^{3/5} Y_{\circ_2}^{1/5} Y_{\circ_3}^{4/5} Y_{\circ_4}^{2/5}}{Y_{\bullet_1}^{2/5} Y_{\bullet_3}^{1/5}}
\nonumber \\[4pt] &
+\frac{Y_{\bullet_1}^{3/5} Y_{\bullet_2}^{1/5} Y_{\bullet_4}^{2/5} Y_{\circ_1}^{3/5} Y_{\circ_2}^{1/5} Y_{\circ_3}^{4/5} Y_{\circ_4}^{2/5}}{Y_{\bullet_3}^{1/5}}+\frac{Y_{\bullet_2}^{1/5} Y_{\bullet_3}^{4/5} Y_{\bullet_4}^{2/5} Y_{\circ_1}^{3/5} Y_{\circ_2}^{1/5} Y_{\circ_3}^{4/5} Y_{\circ_4}^{2/5}}{Y_{\bullet_1}^{2/5}}
\nonumber \\[4pt] &
+Y_{\bullet_1}^{3/5} Y_{\bullet_2}^{1/5} Y_{\bullet_3}^{4/5} Y_{\bullet_4}^{2/5} Y_{\circ_1}^{3/5} Y_{\circ_2}^{6/5} Y_{\circ_3}^{4/5} Y_{\circ_4}^{2/5} \, .
\end{align}
These polynomials are indeed invariant under the operation \eqref{RtopSU5}. From these expressions we can read directly the core charges and therefore write down the corresponding framed BPS quivers
\begin{equation}
\begin{gathered}
\xymatrix@C=5mm{
\framebox{$ \gamma_{\mathbf{5}} $} \ar@{..>}[r] & \circ_1  \ar@<-0.5ex>[d]  \ar@<0.5ex>[d] & & \bullet_2 \ar[rr] \ar[ll] & & \circ_3 \ar@<-0.5ex>[d]  \ar@<0.5ex>[d] & & \bullet_4 \ar[ll] \\
&\ar@{..>}[ul] \bullet_1 \ar[rr] & & \circ_2  \ar@<-0.5ex>[u]  \ar@<0.5ex>[u] & & \bullet_3 \ar[ll] \ar[rr] & & \circ_4 \ar@<-0.5ex>[u]  \ar@<0.5ex>[u] 
 }
\  \qquad
\xymatrix@C=5mm{
\circ_1  \ar@<-0.5ex>[d]  \ar@<0.5ex>[d] & & \bullet_2 \ar[rr] \ar[ll] & & \circ_3 \ar@<-0.5ex>[d]  \ar@<0.5ex>[d] & & \bullet_4 \ar[ll]  \ar@{..>}[r]& \framebox{$ \gamma_{\mathbf{\bar{5}}} $} \ar@{..>}[dl] \\
\bullet_1 \ar[rr] & & \circ_2  \ar@<-0.5ex>[u]  \ar@<0.5ex>[u] & & \bullet_3 \ar[ll] \ar[rr] & & \circ_4 \ar@<-0.5ex>[u]  \ar@<0.5ex>[u] & 
 }
\\
\xymatrix@C=5mm{ & \framebox{$ \gamma_{\mathbf{10}} $} \ar@{..>}[ddr] & & & & & \\
\circ_1  \ar@<-0.5ex>[d]  \ar@<0.5ex>[d] & & \bullet_2 \ar@{..>}[ul] \ar[rr] \ar[ll] & & \circ_3 \ar@<-0.5ex>[d]  \ar@<0.5ex>[d] & & \bullet_4 \ar[ll] \\
\bullet_1 \ar[rr] & & \circ_2  \ar@<-0.5ex>[u]  \ar@<0.5ex>[u] & & \bullet_3 \ar[ll] \ar[rr] & & \circ_4 \ar@<-0.5ex>[u]  \ar@<0.5ex>[u] 
 }
\  \qquad
\xymatrix@C=5mm{ & & & & & \framebox{$ \gamma_{\mathbf{\overline{10}}} $} \ar@{..>}[ddl]  & \\
\circ_1  \ar@<-0.5ex>[d]  \ar@<0.5ex>[d] & & \bullet_2 \ar[rr] \ar[ll] & & \ar@{..>}[ur] \circ_3 \ar@<-0.5ex>[d]  \ar@<0.5ex>[d] & & \bullet_4 \ar[ll] \\
\bullet_1 \ar[rr] & & \circ_2  \ar@<-0.5ex>[u]  \ar@<0.5ex>[u] & & \bullet_3 \ar[ll] \ar[rr] & & \circ_4 \ar@<-0.5ex>[u]  \ar@<0.5ex>[u] 
 }
\end{gathered} \, .
\end{equation}
It is easy to check that these are indeed the charges of the Wilson line operators and that the above framed quivers have the expected properties under the sequence of mutations corresponding to the transformation $R^{(+)}$ of \eqref{RtopSU5}.

As before we can now obtain all the other Wilson line defects by imposing the operator product expansion relations
\begin{align}
\mathfrak{w}_\zeta^{\mathbf{24}} &= \mathfrak{w}_\zeta^{\mathbf{5}} * \mathfrak{w}_\zeta^{\mathbf{\overline{5}}} - 1 \label{24su5} \, , \\
\mathfrak{w}_\zeta^{\mathbf{15}} & = \mathfrak{w}_\zeta^{\mathbf{5}} * \mathfrak{w}_\zeta^{\mathbf{5}} - \mathfrak{w}_\zeta^{\mathbf{10}} \label{15su5} \, , \\
\mathfrak{w}_\zeta^{\mathbf{\overline{45}}} &= \mathfrak{w}_\zeta^{\mathbf{\overline{10}}} * \mathfrak{w}_\zeta^{\mathbf{5}} - \mathfrak{w}_\zeta^{\mathbf{\overline{5}}} \, . \label{45su5}
\end{align} 
For example
\begin{align}
\langle \, \widehat{\mathfrak{w}}_\zeta^{\mathbf{15}} \, \rangle = & 
\frac{1}{ Y_{\bullet_1}^{8/5} Y_{\bullet_2}^{6/5} Y_{\bullet_3}^{4/5} Y_{\bullet_4}^{2/5} Y_{\circ_1}^{8/5} Y_{\circ_2}^{6/5} Y_{\circ_3}^{4/5} Y_{\circ_4}^{2/5} }
\Big(1+2 Y_{\circ_1}+Y_{\bullet_1} Y_{\circ_1}+Y_{\circ_1}^2+2 Y_{\bullet_1} Y_{\circ_1}^2+Y_{\bullet_1}^2 Y_{\circ_1}^2
\nonumber\\[4pt] \ &
+Y_{\bullet_1} Y_{\circ_1} Y_{\circ_2}+Y_{\bullet_1} Y_{\bullet_2} Y_{\circ_1} Y_{\circ_2}+2 Y_{\bullet_1} Y_{\circ_1}^2 Y_{\circ_2}+2 Y_{\bullet_1}^2 Y_{\circ_1}^2 Y_{\circ_2}+Y_{\bullet_1} Y_{\bullet_2} Y_{\circ_1}^2 Y_{\circ_2}+Y_{\bullet_1}^2 Y_{\bullet_2} Y_{\circ_1}^2 Y_{\circ_2}
\nonumber\\[4pt] \ &
+Y_{\bullet_1}^2 Y_{\circ_1}^2 Y_{\circ_2}^2+2 Y_{\bullet_1}^2 Y_{\bullet_2} Y_{\circ_1}^2 Y_{\circ_2}^2+Y_{\bullet_1}^2 Y_{\bullet_2}^2 Y_{\circ_1}^2 Y_{\circ_2}^2+Y_{\bullet_1} Y_{\bullet_2} Y_{\circ_1} Y_{\circ_2} Y_{\circ_3}+Y_{\bullet_1} Y_{\bullet_2} Y_{\bullet_3} Y_{\circ_1} Y_{\circ_2} Y_{\circ_3}
\nonumber\\[4pt] \ &
+Y_{\bullet_1} Y_{\bullet_2} Y_{\circ_1}^2 Y_{\circ_2} Y_{\circ_3}+Y_{\bullet_1}^2 Y_{\bullet_2} Y_{\circ_1}^2 Y_{\circ_2} Y_{\circ_3}+Y_{\bullet_1} Y_{\bullet_2} Y_{\bullet_3} Y_{\circ_1}^2 Y_{\circ_2} Y_{\circ_3}+Y_{\bullet_1}^2 Y_{\bullet_2} Y_{\bullet_3} Y_{\circ_1}^2 Y_{\circ_2} Y_{\circ_3}
\nonumber\\[4pt] \ &
+2 Y_{\bullet_1}^2 Y_{\bullet_2} Y_{\circ_1}^2 Y_{\circ_2}^2 Y_{\circ_3}+2 Y_{\bullet_1}^2 Y_{\bullet_2}^2 Y_{\circ_1}^2 Y_{\circ_2}^2 Y_{\circ_3}+Y_{\bullet_1}^2 Y_{\bullet_2} Y_{\bullet_3} Y_{\circ_1}^2 Y_{\circ_2}^2 Y_{\circ_3}+Y_{\bullet_1}^2 Y_{\bullet_2}^2 Y_{\bullet_3} Y_{\circ_1}^2 Y_{\circ_2}^2 Y_{\circ_3}
\nonumber\\[4pt] \ &
+Y_{\bullet_1}^2 Y_{\bullet_2}^2 Y_{\circ_1}^2 Y_{\circ_2}^2 Y_{\circ_3}^2+2 Y_{\bullet_1}^2 Y_{\bullet_2}^2 Y_{\bullet_3} Y_{\circ_1}^2 Y_{\circ_2}^2 Y_{\circ_3}^2+Y_{\bullet_1}^2 Y_{\bullet_2}^2 Y_{\bullet_3}^2 Y_{\circ_1}^2 Y_{\circ_2}^2 Y_{\circ_3}^2
\nonumber\\[4pt] \ &
+Y_{\bullet_1} Y_{\bullet_2} Y_{\bullet_3} Y_{\circ_1} Y_{\circ_2} Y_{\circ_3} Y_{\circ_4}+Y_{\bullet_1} Y_{\bullet_2} Y_{\bullet_3} Y_{\bullet_4} Y_{\circ_1} Y_{\circ_2} Y_{\circ_3} Y_{\circ_4}+Y_{\bullet_1} Y_{\bullet_2} Y_{\bullet_3} Y_{\circ_1}^2 Y_{\circ_2} Y_{\circ_3} Y_{\circ_4}
\nonumber\\[4pt] \ &
+Y_{\bullet_1}^2 Y_{\bullet_2} Y_{\bullet_3} Y_{\circ_1}^2 Y_{\circ_2} Y_{\circ_3} Y_{\circ_4}+Y_{\bullet_1} Y_{\bullet_2} Y_{\bullet_3} Y_{\bullet_4} Y_{\circ_1}^2 Y_{\circ_2} Y_{\circ_3} Y_{\circ_4}+Y_{\bullet_1}^2 Y_{\bullet_2} Y_{\bullet_3} Y_{\bullet_4} Y_{\circ_1}^2 Y_{\circ_2} Y_{\circ_3} Y_{\circ_4}
\nonumber\\[4pt] \ &
+Y_{\bullet_1}^2 Y_{\bullet_2} Y_{\bullet_3} Y_{\circ_1}^2 Y_{\circ_2}^2 Y_{\circ_3} Y_{\circ_4}+Y_{\bullet_1}^2 Y_{\bullet_2}^2 Y_{\bullet_3} Y_{\circ_1}^2 Y_{\circ_2}^2 Y_{\circ_3} Y_{\circ_4}+Y_{\bullet_1}^2 Y_{\bullet_2} Y_{\bullet_3} Y_{\bullet_4} Y_{\circ_1}^2 Y_{\circ_2}^2 Y_{\circ_3} Y_{\circ_4}
\nonumber\\[4pt] \ &
+Y_{\bullet_1}^2 Y_{\bullet_2}^2 Y_{\bullet_3} Y_{\bullet_4} Y_{\circ_1}^2 Y_{\circ_2}^2 Y_{\circ_3} Y_{\circ_4}+2 Y_{\bullet_1}^2 Y_{\bullet_2}^2 Y_{\bullet_3} Y_{\circ_1}^2 Y_{\circ_2}^2 Y_{\circ_3}^2 Y_{\circ_4}+2 Y_{\bullet_1}^2 Y_{\bullet_2}^2 Y_{\bullet_3}^2 Y_{\circ_1}^2 Y_{\circ_2}^2 Y_{\circ_3}^2 Y_{\circ_4}
\nonumber\\[4pt] \ &
+Y_{\bullet_1}^2 Y_{\bullet_2}^2 Y_{\bullet_3} Y_{\bullet_4} Y_{\circ_1}^2 Y_{\circ_2}^2 Y_{\circ_3}^2 Y_{\circ_4}+Y_{\bullet_1}^2 Y_{\bullet_2}^2 Y_{\bullet_3}^2 Y_{\bullet_4} Y_{\circ_1}^2 Y_{\circ_2}^2 Y_{\circ_3}^2 Y_{\circ_4}+Y_{\bullet_1}^2 Y_{\bullet_2}^2 Y_{\bullet_3}^2 Y_{\circ_1}^2 Y_{\circ_2}^2 Y_{\circ_3}^2 Y_{\circ_4}^2
\nonumber\\[4pt] \ &
+2 Y_{\bullet_1}^2 Y_{\bullet_2}^2 Y_{\bullet_3}^2 Y_{\bullet_4} Y_{\circ_1}^2 Y_{\circ_2}^2 Y_{\circ_3}^2 Y_{\circ_4}^2+Y_{\bullet_1}^2 Y_{\bullet_2}^2 Y_{\bullet_3}^2 Y_{\bullet_4}^2 Y_{\circ_1}^2 Y_{\circ_2}^2 Y_{\circ_3}^2 Y_{\circ_4}^2\Big) \, .
\end{align}
Note that this operator is manifestly strongly positive, and its core charge 
\begin{equation}
\gamma_\mathbf{15} = -\frac85 (\gamma_{\bullet_1} + \gamma_{\circ_1}) -\frac65 (\gamma_{\bullet_2} + \gamma_{\circ_2}) -\frac45 (\gamma_{\bullet_3} + \gamma_{\circ_3}) - \frac25 (\gamma_{\bullet_4} + \gamma_{\circ_4}) 
\end{equation}
indeed corresponds to the highest weight associated with the $\mathbf{15}$ irreducible representation of $\mathfrak{su}_5$. We stress that the positivity property of the generating functions (\ref{24su5})-(\ref{45su5}) is highly non trivial, since it is the consequence of delicate cancellations between very complicated polynomials.

\subsection{Dyonic defects}

Again we can generate cluster families of defects starting from a set of elementary operators
\begin{align}
\langle \, \widehat{\mathfrak{L}}_{\zeta , f_{\star_1}} \, \rangle & =  \frac{1}{Y_{\circ_1}^{4} Y_{\circ_2}^{3} Y_{\circ_3}^{2} Y_{\circ_4}^{1}}
\ , \qquad
\langle \, \widehat{\mathfrak{L}}_{\zeta , f_{\star_2}} \, \rangle  =  \frac{1}{Y_{\circ_1}^{3} Y_{\circ_2}^{6} Y_{\circ_3}^{4} Y_{\circ_4}^{2}} \, ,
\\
\langle \, \widehat{\mathfrak{L}}_{\zeta , f_{\star_3}} \, \rangle & =  \frac{1}{Y_{\circ_1}^{2} Y_{\circ_2}^{4} Y_{\circ_3}^{6} Y_{\circ_4}^{3}}
\ , \qquad
\langle \, \widehat{\mathfrak{L}}_{\zeta , f_{\star_4}} \, \rangle  =  \frac{1}{Y_{\circ_1}^{1} Y_{\circ_2}^{2} Y_{\circ_3}^{3} Y_{\circ_4}^{4}} \, ,
\\
\langle \, \widehat{\mathfrak{L}}_{\zeta , f_{\star_5}} \, \rangle & =  Y_{\bullet_1}^{4} Y_{\bullet_2}^{3} Y_{\bullet_3}^{2} Y_{\bullet_4}^{1}
\ , \qquad
\langle \, \widehat{\mathfrak{L}}_{\zeta , f_{\star_6}} \, \rangle  =  Y_{\bullet_1}^{3} Y_{\bullet_2}^{6} Y_{\bullet_3}^{4} Y_{\bullet_4}^{2} \, ,
\\
\langle \, \widehat{\mathfrak{L}}_{\zeta , f_{\star_7}}  \, \rangle & =  Y_{\bullet_1}^{2} Y_{\bullet_2}^{4} Y_{\bullet_3}^{6} Y_{\bullet_4}^{3}
\ , \qquad
\langle \, \widehat{\mathfrak{L}}_{\zeta , f_{\star_8}} \, \rangle  =  Y_{\bullet_1} Y_{\bullet_2}^{2} Y_{\bullet_3}^{3} Y_{\bullet_4}^{4} \, .
\end{align}
The computations are somewhat technical and the results rather lengthy; therefore instead of writing down the full operators we will limit ourselves to listing only the core charges of the defects of these families
\begin{align}
\gamma_{f_{\star_1}^{[n]}} =&  -4 n \gamma_{\bullet_1} - 3 n \gamma_{\bullet_2} - 2 n \gamma_{\bullet_3} -  n \gamma_{\bullet_4} 
\nonumber\\ & \ 
- 4 (n+1) \gamma_{\circ_1} - 3 (n+1) \gamma_{\circ_2} - 2 (n+1) \gamma_{\circ_3} -  (n+1) \gamma_{\circ_4}  \, ,
\\[4pt]
\gamma_{f_{\star_2}^{[n]}} =&  -3 n \gamma_{\bullet_1} - 6 n \gamma_{\bullet_2} - 4 n \gamma_{\bullet_3} - 2 n \gamma_{\bullet_4} 
\nonumber\\ & \ 
- 3 (n+1) \gamma_{\circ_1} - 6 (n+1) \gamma_{\circ_2} - 4 (n+1) \gamma_{\circ_3} - 2 (n+1) \gamma_{\circ_4}  \, ,
\\[4pt]
\gamma_{f_{\star_3}^{[n]}} =&  -2 n \gamma_{\bullet_1} - 4 n \gamma_{\bullet_2} - 6 n \gamma_{\bullet_3} - 3 n \gamma_{\bullet_4} 
\nonumber\\ & \ 
- 2 (n+1) \gamma_{\circ_1} - 4 (n+1) \gamma_{\circ_2} - 6 (n+1) \gamma_{\circ_3} - 3 (n+1) \gamma_{\circ_4} \, ,
\\[4pt]
\gamma_{f_{\star_4}^{[n]}} =&  - n \gamma_{\bullet_1} - 2 n \gamma_{\bullet_2} - 3 n \gamma_{\bullet_3} - 4 n \gamma_{\bullet_4} 
\nonumber\\ & \ 
-  (n+1) \gamma_{\circ_1} - 2 (n+1) \gamma_{\circ_2} - 3 (n+1) \gamma_{\circ_3} - 4 (n+1) \gamma_{\circ_4} \, ,
\\[4pt]
\gamma_{f_{\star_5}^{[n]}} =&  - 4 (n-1) \gamma_{\bullet_1} - 3 (n-1) \gamma_{\bullet_2} - 2 (n-1) \gamma_{\bullet_3} - (n-1) \gamma_{\bullet_4} 
\nonumber\\ & \ 
- 4 n \gamma_{\circ_1} - 3 n \gamma_{\circ_2} - 2 n \gamma_{\circ_3} - n \gamma_{\circ_4}  \, ,
\\[4pt]
\gamma_{f_{\star_6}^{[n]}} =&  - 3 (n-1) \gamma_{\bullet_1} - 6 (n-1) \gamma_{\bullet_2} - 4 (n-1) \gamma_{\bullet_3} - 2 (n-1) \gamma_{\bullet_4} 
\nonumber\\ & \ 
- 3 n \gamma_{\circ_1} - 6 n \gamma_{\circ_2} - 4 n \gamma_{\circ_3} - 2 n \gamma_{\circ_4} \, ,
\\[4pt]
\gamma_{f_{\star_7}^{[n]}} =&  - 2 (n-1) \gamma_{\bullet_1} - 4 (n-1) \gamma_{\bullet_2} - 6 (n-1) \gamma_{\bullet_3} - 3 (n-1) \gamma_{\bullet_4} 
\nonumber\\ & \ 
- 2 n \gamma_{\circ_1} - 4 n \gamma_{\circ_2} - 6 n \gamma_{\circ_3} - 3 n \gamma_{\circ_4} \, ,
\\[4pt]
\gamma_{f_{\star_8}^{[n]}} =&  - (n-1) \gamma_{\bullet_1} - 2 (n-1) \gamma_{\bullet_2} - 3 (n-1) \gamma_{\bullet_3} - 4 (n-1) \gamma_{\bullet_4} 
\nonumber\\ & \ 
- n \gamma_{\circ_1} - 2 n \gamma_{\circ_2} - 3 n \gamma_{\circ_3} - 4 n \gamma_{\circ_4} \, .
\end{align}
With our formalism we can compute explicitly the vev's for each of these operators, the only limit being computational power.


\section{Conclusions and future directions}

In this paper we have discussed a connection between line defects in $\mathcal{N}=2$ QFT and certain discrete integrable systems. These systems arise naturally when studying the problem with quiver methods. The existence of certain discrete symmetries for the theory without defects is reflected in the presence of fractional quantum monodromies. These operators are associated with Seiberg dualities of the supersymmetric quantum mechanics which describes the effective dynamics of BPS particles. We have shown that all these statements have a clear counterpart when the theory is defined in the presence of line defects. Fractional monodromies naturally act on the set of line defects. This action identifies the Wilson lines in asymptotically free theories with conserved charges of the $Q$-system, which arises from the interplay between TBA coordinates on the Hitchin moduli space and cluster algebras. For other kind of defects, our methods allow to compute explicitly infinite series of vevs just by iteration of certain rational transformations.

Notably our formalism is powerful enough to distinguish theories which have the same local description but inequivalent sets of line defects. The action of the appropriate monodromy operator descends to the framed quiver and to the corresponding line operator vev and maps the set of line operators of the same theory into itself. The physical reason for this is that the existence of fractional monodromies is a consequence of unbroken $R$-symmetries which can distinguish quantum field theories based on different gauge groups via global effects, such as the periodicity of the $\theta$-angles or the lattice of instanton charges.

A forthcoming paper will address these issue in more detail in the case of SCFT. A natural open problem is to clarify the relation between our formalism and the S-duality of $\mathcal{N}=2$ QFT. In this case particular sequences of mutations relate different quiver SQM characterizations of S-dual systems \cite{Alim:2011kw,Cecotti:2013lda, Cecotti:2015hca,Caorsi:2016ebt}. For an explicit example, see the discussion about the first historical example of Argyres-Seiberg duality in \cite{Alim:2011kw}. From the perspective of our formalism these sequences of mutations now can be lifted to transformations acting on the space of BPS line defects. In the part II of this work we plan to investigate in depth these actions. Of course for theories of class $\cs$ we expect to find a matching among them and the modular transformations of the UV curve \cite{Gaiotto:2009we}. Also, extending this analysis to a more broad class of defects for the $\cn=2$ theory would be interesting. A natural question, for instance, is about the action of the broken discrete symmetries as well as S-dualties on the space of surface defects, perhaps along the lines of \cite{Gaiotto:2008ak}.


\section*{Acknowledgements}
MdZ thanks Andy Royston for explaining him several interesting aspects of the physics of framed BPS states at weak coupling. We also thank D. Allegretti, S. Cecotti, C. Cordova, T. Dumitrescu, V. Pestun, B. Pioline and C. Vafa for discussions. The work of MC was partially supported by FCT/Portugal and IST-ID through EXCL/MAT-GEO/0222/2012 and the program Investigador FCT IF2014, contract IF/01426/2014/CP1214/CT0001. MC is thankful to the Theory Division of CERN for the hospitality and support during the last stages of this project. MC acknowledges support by the Action MP1405 QSPACE from the European Cooperation in Science and Technology (COST). The author also acknowledges the support of IH\'ES during a visit.  The research of M.C. on this project has received funding from the European Research Council (ERC) under the European Union's Horizon 2020 research and innovation programme (QUASIFT grant agreement 677368).

\appendix

\section{Mutation of the Cordova-Neitzke superpotential}\label{failarmy}
Though very appealing and natural from the point of view of an analogous problem in quantum gravity \cite{Andriyash:2010qv,Andriyash:2010yf}, in \cite{Cordova:2013bza} at the level of the Wilson lines of pure $SU(2)$ SYM it was noticed an apparent paradox. Let us proceed by explaining it: the framed BPS degeneracies for a Wilson 
line in the $n$--dimensional representation of $SU(2)$ computed using the Coulomb branch 
\cite{Manschot:2010qz,Manschot:2011xc,Sen:2011aa,Manschot:2012rx,Manschot:2013sya,Manschot:2013dua,Manschot:2014fua} of the corresponding quiver SQM do not agree with those 
computed quantizing  \cite{Witten:1996qb, Denef:2002ru} the SQM Higgs branch moduli spaces for 
$n>2$.

Now, whenever 
a given quiver has non-trivial loops, the corresponding quiver SQM admits superpotential terms. 
There are quivers which admit several distinct superpotentials which are not equivalent, each 
corresponding to a different physics. For reproducing correctly the physics of a given model, it is of 
crucial importance to choose the correct superpotential, which is a subtle point. The generic 
superpotential proposed in \cite{Cordova:2013bza} for the SQM describing a Wilson line in the $n$-dimensional representation is mutation equivalent to the null superpotential on the same quiver for all 
$n>2$, and therefore it is not DWZ-generic. We believe that this explains the paradox discussed in \cite{Cordova:2013bza}. Of course, the choice of a correct superpotential must reconcile the Higgs 
branch analysis with the Coulomb branch one. 

The BPS quiver for pure $SU(2)$ SYM with a Wilson line in the $N$ dimensional representation of $SU(2)$ is \cite{Cordova:2013bza}
\be
\begin{gathered}
\xymatrix{
\bullet\ar@{..>}[drr]^{B_j}\\
&&\ast\ar@{..>}[dll]^{C_j}\\
\circ \ar@<-0.5ex>[uu]_{A_1}\ar@<0.5ex>[uu]^{A_2}
}
\end{gathered} \quad j=1,\dots,N.
\ee
The superpotential proposed in Equation (5.41) of \cite{Cordova:2013bza} is
\be
\cw = \sum_{j=1}^N C_j B_j A_1 + \sum_{j=1}^N \lambda_j C_j B_j A_2.
\ee
Mutating the above quiver on $\bullet$, we obtain:
\be
\begin{gathered}
\xymatrix{
\bullet\ar@<-0.5ex>[dd]_{A^*_1}\ar@<0.5ex>[dd]^{A^*_2}\ar@/^0.7pc/@{<..}[drr]^{B^*_j}\\
&&\ast\ar@{..>}[dll]^{C_j}\\
\circ\ar@/_1.5pc/@{..>}[urr]_{[B_jA_i]}\
}
\end{gathered} \quad j=1,\dots,N, \quad i =1,2.
\ee
with
\be
\tilde{\mu}_\bullet \cw = \sum_{j=1}^N C_j [B_j A_1] + \sum_{j=1}^N \lambda_j C_j [B_j A_2] + \sum_{j=1}^N\sum_{i=1}^2 [B_j A_i]  A_i^*B_j^*.
\ee
Notice that
\begin{align}
&\partial_{[B_j A_1]}  \mu_\bullet \cw = C_j + A_1^*B_j^* = 0\\
&\partial_{[B_j A_2]}  \mu_\bullet \cw = \lambda_j C_j + A_2^*B_j^* = 0.
\end{align}
Therefore, integrating out the fields which appear quadratically in the superpotential, as we are instructed to do by the definition of mutation \cite{DWZ}, we obtain
\begin{align}
\mu_\bullet \cw &= - \sum_{j=1}^N A_1^*B_j^* [B_j A_1] - \sum_{j=1}^N A_2^*B_j^* [B_j A_2] + \sum_{j=1}^N\sum_{i=1}^2 [B_j A_i]  A_i^*B_j^*\\
&= 0
\end{align}
by cyclicity and linearity of the trace. Therefore, the superpotential is not DWZ-generic.

\section{$SU(2)$ Wilson lines and cyclic modules}\label{Wsu23vevk}
Consider the case of a Wilson line in the Adjoint representation of $SU(2)$. Let us first consider the superpotential
\be\label{adjojo}
\cw_{2,1} = \b_1 \a_1 A + \b_2 \a_2 B,
\ee
with relations
\be
\b_1 \a_1 = \a_1 A = A \b_1 = \b_2 \a_2 = \a_2 B = B \b_2 = 0.
\ee
In this case the skeleton diagram of eqn.\eqref{skelwsu2} reduces to
\be
\xymatrix{
&\b_1 v\ar[r]^B& B \b_1 v\ar[r]^{\a_1\quad} & \a_1 B \b_1 v = b v & \cdots \\
v\ar[ur]^{\b_1}\ar[dr]_{\b_2}&&\\
&\b_2 v\ar[r]^A& A \b_2 v\ar[r]^{\a_2\quad} & \a_2 A \b_2 v = a v & \cdots \\
}
\ee
However,
\be
a \b_2 v = \b_2 \a_2 A \b_2 v = 0  \qquad b \b_1 v = \b_1 \a_1 B \b_1 v = 0 
\ee
Therefore, if $\b_1 v$ (resp. $\b_2 v$) is nonzero, $a = 0$ (resp. $b=0$),
\be
\a_1 B \b_1 v = 0 \quad \text{and}\quad \a_2 A \b_2 v = 0,
\ee
and the skeleton diagram reduces to an $A_5$ Dynkin graph:
\begin{equation}
\begin{gathered}
\xymatrix{
&\b_1 v\ar[r]^B& B \b_1 v\\
v\ar[ur]^{\b_1}\ar[dr]_{\b_2}&&\\
&\b_2 v\ar[r]^A& A \b_2 v\\
}
\end{gathered}
\end{equation}
Let us proceed by determining the allowed cyclic modules. Let us first consider the case in which both arrows $A$ and $B$ are zero. There are three options to be considered: {\it 1.)} $0 = \b_1 v = \b_2 v $: this correspond to the trivial cyclic module; {\it 2.)} $0 \neq \b_1 v \neq \b_2 v $: this correspond to a rigid representation with dimension vector $(1,2,0)$, indeed in this case $\dim X_\circ=2$ and we can choose its basis to set $\b_1 = (1,0)^t$ and $\b_2 = (0,1)^t$; {\it 3.)} $0 \neq \b_1 v = \b_2 v$: in this case $\dim X_\circ=1$, and we obtain the minimal imaginary root of the Kronecker quiver obtained by folding the above skeleton diagram according to the identification $\b_1 v = \b_2 v$, it has a $\mathbb{P}^1$ moduli space. Let us now consider the case in which the arrows $A$ and $B$ are nonzero. This rules out automatically options {\it 1.)} and {\it 3.)} above: indeed, if $\b_1 v = \b_2 v$, we have that $A \b_2 v = A \b_1 v = 0$ by the e.o.m.,  and $B \b_1 v = B \b_2 v = 0$. We are left with option {\it 2.)} above plus the two options: $0 = \b_1 v \neq \b_2 v$ and $0 = \b_2 v \neq \b_1 v$. These two options are special points of the $\PP^1$ moduli space of solutions of {\it 3.)} that play an interesting role here. In this case we obtain two representations:
\be
\begin{gathered}
\xymatrix{
&\b_1 v\ar[r]^B& B \b_1 v\\
v\ar[ur]^{\b_1}\ar[dr]_{\b_2}&&\\
& 0\ar[r]^A & 0 \\
}
\end{gathered}\quad\text{and}\quad\begin{gathered}
\xymatrix{
&0\ar[r]^B& 0\\
v\ar[ur]^{\b_1}\ar[dr]_{\b_2}&&\\
&\b_2 v\ar[r]^A& A \b_2 v\\
}
\end{gathered}
\ee
Both representations are rigid and give rise to the dimension vector $(1,1,1)$. Notice that the superpotential $\cw_{2,1}$ has a discrete symmetry $S_2$ that acts only on the arrows of the quiver leaving the nodes invariant:
\be
S_2 \colon (A,\a_1,\b_1) \longleftrightarrow (B,\a_2,\b_2),
\ee
As remarked in \cite{Cecotti:2012va} this is a crucial property of this system: it introduces a superselection rule on the Hilbert space of the SQM. We claim that the ground states correctly capturing the framed BPS excitations lie in the singlet subsector of the Hilbert space. With this caveat in mind it is clear that the two representations above are indistinguishable and give rise to a single $S_2$ symmetric ground state. Let us consider now option {\it 2.)}. We have two cases: {\it 2a.)} $A \b_2 v = B \b_1 v$, and {\it 2b.)} $A \b_2 v \neq B \b_1 v$. In case {\it 2a.)} by folding the skeleton diagram according to the identification $A \b_2 v = B \b_1 v$ we obtain an affine $\hat{A}(2,2)$ quiver, and the corresponding dimension vector is again the minimal imaginary root: we obtain again a representation with a $\PP^1$ moduli space with dimension vector $(1,2,1)$. In case {\it 2b.)} we obtain again a rigid representation with dimension vector $(1,2,2)$ corresponding to the root $(1,1,1,1,1)$ of the $A_5$ Dynkin graph. Therefore, we are lead to claim that the superpotential $\cw_{2,1}$ correctly reproduces the framed BPS excitations to the adjoint Wilson line of $SU(2)$, indeed we obtain:
\bea
\langle W_{\mathbf{3}} \rangle &= \frac{1}{Y_\circ Y_\bullet} (1 + 2 Y_\circ + Y^2_\circ + Y_\circ Y_\bullet + 2 Y_\circ^2 Y_\bullet  +Y_\circ^2 Y_\bullet^2)\\
&=\frac{1}{Y_\circ Y_\bullet} + \frac{2}{Y_\bullet} + \frac{Y_\circ }{Y_\bullet} + 1 + 2 Y_\circ + Y_\circ Y_\bullet\\
&= \left[\frac{1}{Y_\circ Y_\bullet} + 1 +  Y_\circ Y_\bullet\right] + \frac{2}{Y_\bullet} + \frac{Y_\circ }{Y_\bullet} + 2 Y_\circ.
\eea
This result agrees with the Coulomb branch computation carried over in \cite{Cordova:2013bza}: the discrete $S_2$ symmetry is crucial to obtain $c_{11}=1$.

Let us now exclude the other superpotential:
\be
\cw_{2,2} = (\b_1\a_1 + \b_2 \a_2) A
\ee
with relations
\be
A \b_1 = A \b_2 = \a_1 A = \a_2 A = 0 = \b_1\a_1 + \b_2 \a_2.
\ee
The corresponding skeleton diagram is
\begin{equation}
\begin{gathered}
\xymatrix{
&&&\a_1 B \b_1 v=b_{11}v&\cdots\\
&\b_1 v\ar[r]& B \b_1 v\ar[ur]\ar[r]& \a_2 B \b_1 v = b_{21} v & \cdots \\
v\ar[ur]\ar[dr]&&\\
&\b_2 v\ar[r]& B \b_2 v\ar[r]\ar[dr]& \a_1 B \b_2 v = b_{12} v & \cdots \\
&&&\a_2 B \b_2 v=b_{22}v&\cdots\\
}
\end{gathered}
\end{equation}
For $\a_1$ and $\a_2$ generic we obtain the relations:
\be
b_{11} \b_1 v + b_{21} \b_2 v = 0 \qquad b_{12} \b_1 v + b_{22} \b_2 v = 0,
\ee
these needs to be modified according to the relations at nongeneric values for $\a_1$ or $\a_2$. Notice that whenever the dimension of $X_\circ$ equals 2, $\a_1 = \a_2 = 0$, as there are no linear relations in between $\b_1 v$ and $\b_2 v$. Therefore the modules with dimension vectors $(1,2,0)$, $(1,2,1)$ and $(1,2,2)$ have the same skeleton diagram as in the previous case (up to relabeling some arrows). Similarly for the modules with $\dim X_\bullet = 0$: the modules with dimensions $(1,0,0)$, $(1,1,0)$ are also equivalent. We are left with only one case to consider: the module with dimension vector $(1,1,1)$. Before doing that let us notice that the SQM with superpotential $\cw_{2,2}$ has an $S_2$ discrete symmetry too: it acts by switching $\a_1 \leftrightarrow \a_2$ and $\b_1 \leftrightarrow \b_2$ simultaneously. In this case, by gauge fixing away $\b_1$ and $B$, we obtain the following representation:
\be
\xymatrix{&&\C\ar@<-0.2pc>[dll]_{\a}\ar@<0.2pc>[dll]^{-\a \b }\\
\C\ar@<-0.2pc>[drr]_{\b}\ar@<0.2pc>[drr]^{1}&\\
&&\C\ar@<-0.2pc>[uu]_1\ar@<0.2pc>[uu]^0
}
\ee
Instead, by gauge fixing away $\b_2$ and $B$, we obtain the following representation:
\be
\xymatrix{&&\C\ar@<-0.2pc>[dll]_{-\a\b}\ar@<0.2pc>[dll]^{\a }\\
\C\ar@<-0.2pc>[drr]_{1}\ar@<0.2pc>[drr]^{\b}&\\
&&\C\ar@<-0.2pc>[uu]_1\ar@<0.2pc>[uu]^0
}
\ee
these two combine into an $S_2$ singlet that clearly has $c_{11}\neq 1$.


\bibliography{ithaca.bib}

\end{document}